\documentclass[twocolumn,floatfix,showpacs,prd,aps,tightenlines]{revtex4}
\usepackage{amsmath}
\usepackage{graphicx}
\usepackage{psfrag}
\usepackage{color}
\usepackage{dcolumn}
\usepackage{bm}
\usepackage{longtable}
\usepackage[mathscr]{eucal}
\usepackage{mathrsfs}

\newcommand{\bea}{\begin{eqnarray}}
\newcommand{\eea}{\end{eqnarray}}
\newcommand{\beq}{\begin{equation}}
\newcommand{\eeq}{\end{equation}}
\newcommand{\KMS}{\rm km\ s^{-1}}

\newcommand{\scri}{\mathscr{I}}

\begin{document}

\def\fun#1#2{\lower3.6pt\vbox{\baselineskip0pt\lineskip.9pt
  \ialign{$\mathsurround=0pt#1\hfil##\hfil$\crcr#2\crcr\sim\crcr}}}
\def\lap{\mathrel{\mathpalette\fun <}}
\def\gap{\mathrel{\mathpalette\fun >}}
\def\kms{{\rm km\ s}^{-1}}
\def\vk{V_{\rm recoil}}

\title{Intermediate-mass-ratio black hole binaries II:\\
Modeling Trajectories and Gravitational Waveforms}

\author{
Hiroyuki Nakano,
Yosef Zlochower,
Carlos O. Lousto,
Manuela Campanelli
}

\affiliation{Center for Computational Relativity and Gravitation,\\
and School of Mathematical Sciences, Rochester Institute of
Technology, 85 Lomb Memorial Drive, Rochester, New York 14623}

\begin{abstract}
We revisit the scenario of small-mass-ratio ($q$) black-hole binaries;
performing new, more accurate, simulations of
mass ratios 10:1 and 100:1 for initially nonspinning black holes.
We propose fitting functions for the trajectories of the
two black holes as a function of time and mass ratio (in the range
 $1/100\leq q\leq1/10$) that combine
aspects of post-Newtonian trajectories at smaller orbital frequencies
and plunging geodesics at larger frequencies.
We then use these trajectories to compute waveforms 
via black hole perturbation theory. 
Using the advanced 
LIGO noise curve, we see a match of $\sim 99.5\%$ for the leading 
$(\ell,\,m)=(2,\,2)$ mode between the numerical relativity and 
perturbative waveforms. Nonleading modes have similarly high
matches. We thus prove the feasibility of efficiently generating 
a bank of gravitational waveforms in the intermediate-mass-ratio 
regime using only
a sparse set of full numerical simulations.
\end{abstract}

\pacs{04.25.dg, 04.30.Db, 04.25.Nx, 04.70.Bw} \maketitle

\section{Introduction}\label{sec:Introduction}

Numerical relativity (NR) has come a long way since the breakthroughs
of 2005~\cite{Pretorius:2005gq, Campanelli:2005dd, Baker:2005vv}
that  allowed, for the first time, long-term evolution of
black hole binaries (BHBs). Among NR's significant achievements are 
its contributions towards the modeling of astrophysical gravitational 
wave sources that will be relevant for the first direct detection and parameter
estimation by gravitational wave observatories~\cite{Aylott:2009ya}.
NR has also made contributions to the modeling of astrophysical sources,
notably with the discovery of very large recoil
velocities~\cite{Campanelli:2007ew, Gonzalez:2007hi,
Campanelli:2007cga, Lousto:2011kp}, and
the application of the numerical techniques to combined systems
of black holes and neutron stars~\cite{Sekiguchi:2010ja}. More mathematical 
aspects of relativity have also recently been investigated, including the
evolution of N-black holes~\cite{Lousto:2007rj}, the exploration of the no-hair
theorem~\cite{Campanelli:2008dv, Owen:2010vw}, and
cosmic~\cite{Campanelli:2006uy} and topological
censorship~\cite{Ponce:2010fq}, 
as well as BHBs in dimensions higher than four~\cite{Shibata:2010wz}.

Among the remaining challenges are the exploration of the extremes
of the BHB parameter space. The current state of the art simulations
can
simulate  BHBs with mass ratios as small as $q=1/100$~\cite{Lousto:2010ut, Sperhake:2011ik}
 and highly spinning BHBs with intrinsic spins
$\alpha=S_H/M_H^2$ up to (at least) $0.97$~\cite{Lovelace:2010ne}.
Currently these runs
are very costly and it is hard to foresee the possibility of completely
covering the
 parameter space densely enough for match filtering the data coming
from advanced laser interferometric detectors by the time they become
operational. It is therefore imperative to develop
interpolation techniques that
allow for astrophysical parameter estimations, at a reasonable level
of accuracy, based on a sparse set
of numerical simulations.

In this spirit, we designed a set of prototypical runs for initially
non spinning BHBs with small mass ratios $q$ in the range 
$0.1\leq q\leq0.01$. Since we expect that, for small enough mass
ratios,
this BHB system will be describable by perturbation theory, we compare the
full numerical waveforms with those produced by perturbative
evolutions via the Regge-Wheeler~\cite{Regge:1957td} 
and Zerilli~\cite{Zerilli:1971wd} equations,
supplemented by linear corrections in the spin of the large black
hole~\cite{Lousto:2010qx}. The key ingredients perturbation theory needs is the
relative trajectory of the small black hole with respect to the larger
one and the background mass and spin. In our previous tests we used the full numerical tracks and
proved the perturbative waveforms agree reasonably well with
the full numerical ones~\cite{Lousto:2010qx}. 
In this paper we develop a model with free fitting parameters for these
trajectories based on post-Newtonian and geodesic input,
 and fit to full numerical tracks. For these fits, we use
full numerical
evolutions of  $q=1/10$ and $q=1/100$ BHBs~\cite{Lousto:2010qx,Lousto:2010ut}
and perform new, more accurate simulations. We  compare the
waveforms for the modeled tracks with the full numerical waveforms to
confirm matching agreement within a fraction of a percent. Hence
this method paves the way for further generalizations and simulations 
to provide an approximate, yet accurate, bank for waveforms
for second generation gravitational wave detectors.

This paper is organized as follows. In Sec.~\ref{Sec:Numerical} we
review
the numerical methodology to perform the small mass ratio runs.
In Sec.~\ref{SubSec:Runs} we  describe the runs and results. In Sec.~\ref{Sec:Perturbations} we describe the track modeling
on post-Newtonian expansions and fits to the full
numerical results. In Sec.~\ref{sec:Results} we give the
results of using those tracks to generate perturbative
waveforms and compares them with those extracted from the full numerical
evolutions. We discuss the consequences and future application of
this techniques in Sec.~\ref{sec:Discussion}.
We also include an appendix~\ref{app:Trumpet} to briefly discuss
perturbative theory in numerical coordinates (1+log, trumpet coordinates).

\section{Numerical Relativity Techniques}\label{Sec:Numerical}

To compute the numerical initial data, we use the puncture
approach~\cite{Brandt97b} along with the {\sc
TwoPunctures}~\cite{Ansorg:2004ds} thorn.  In this approach the
3-metric on the initial slice has the form $\gamma_{a b} = (\psi_{BL}
+ u)^4 \delta_{a b}$, where $\psi_{BL}$ is the Brill-Lindquist
conformal factor, $\delta_{ab}$ is the Euclidean metric, and $u$ is
(at least) $C^2$ on the punctures.  The Brill-Lindquist conformal
factor is given by $ \psi_{BL} = 1 + \sum_{i=1}^n m_{i}^p / (2 |\vec r
- \vec r_i|), $ where $n$ is the total number of `punctures',
$m_{i}^p$ is the mass parameter of puncture $i$ ($m_{i}^p$ is {\em
not} the horizon mass associated with puncture $i$), and $\vec r_i$ is
the coordinate location of puncture $i$.  We evolve these
black-hole-binary data-sets using the {\sc
LazEv}~\cite{Zlochower:2005bj} implementation of the moving puncture
approach~\cite{Campanelli:2005dd,Baker:2005vv} with the conformal
function $W=\sqrt{\chi}=\exp(-2\phi)$ suggested by Ref.~\cite{Marronetti:2007wz}.
For the runs presented here,
we use centered, eighth-order finite differencing in
space~\cite{Lousto:2007rj} and a fourth-order Runge-Kutta time integrator. (Note that we do
not upwind the advection terms.)

Our code uses the {\sc Cactus}/{\sc EinsteinToolkit}~\cite{cactus_web,
einsteintoolkit} infrastructure.
We use the {\sc Carpet}~\cite{Schnetter-etal-03b} mesh refinement driver to
provide a ``moving boxes'' style of mesh refinement. In this approach
refined grids of fixed size are arranged about the coordinate centers
of both holes.  The {\sc Carpet} code then moves these fine grids about the
computational domain by following the trajectories of the two black
holes.

We use {\sc AHFinderDirect}~\cite{Thornburg2003:AH-finding} to locate
apparent horizons.  We measure the magnitude of the horizon spin using
the Isolated Horizon algorithm detailed in Ref.~\cite{Dreyer02a}.
Note that once we have the
horizon spin, we can calculate the horizon mass via the Christodoulou
formula
\begin{equation}
{m^H} = \sqrt{m_{\rm irr}^2 +
 S^2/(4 m_{\rm irr}^2)},
\end{equation}
where $m_{\rm irr} = \sqrt{A/(16 \pi)}$ and $A$ is the surface area of
the horizon.
We measure radiated energy, linear momentum, and angular momentum, in
terms of $\psi_4$, using the formulae provided in
Refs.~\cite{Campanelli:1998jv,Lousto:2007mh}. However, rather than using
the full $\psi_4$, we decompose it into $\ell$ and $m$ modes and solve
for the radiated linear momentum, dropping terms with $\ell \geq 5$.
The formulae in Refs.~\cite{Campanelli:1998jv,Lousto:2007mh} are valid at
$r=\infty$.
Typically, we would extract the radiated energy-momentum at finite
radius and extrapolate to $r=\infty$. However, for the smaller mass
ratios examined here, noise in the waveform introduces spurious
effects that make these extrapolations inaccurate. We therefore use
the average of these quantities extracted at radii $r=70$, $80$,
$90$, $100$ and use the difference between these quantities at
different radii as a measure of the error.

We extrapolate the waveform to $r\to\infty$ using the perturbative
formula~\cite{Lousto:2010qx}
\begin{eqnarray}
&&\lim_{r\to\infty}[r \,\psi_{4}^{\ell m}(r,t)]  \nonumber \\
&&=\left[r \,\psi_{4}^{\ell m}(r,t)
- \frac{(\ell -1)(\ell +2)}{2} \int_0^t dt \, \psi_{4}^{\ell m}(r,t)
\right]_{r=r_{\rm Obs}}\nonumber \\
&& 
 + O(R_{\rm Obs}^{-2}) \,,
\label{eq:asymtpsi4ext}
\end{eqnarray}
where $r_{\rm Obs}$ is the approximate areal radius of the
sphere $R_{\rm Obs}=const$ [Add a factor $(1/2 - M/r)$ multiplying
the square bracket to correct for a difference in normalization
between the Psikadelia (numerical) and Kinnersley tetrads at large
distances.]
We have found that this formula gives reliable extrapolations
for $R_{\rm Obs}\gtrsim100M$.

Recent Cauchy-Characteristic extraction (CCE)
studies~\cite{Babiuc:2010ze}
(see also Ref.~\cite{Reisswig:2009us}) showed that this perturbative extrapolation
formula can be more accurate than a linear extrapolation
of the waveforms at finite radius to infinite resolution (provided
that
the observer $R$ is in the far zone). Those
studies compared the extrapolated waveforms with the gauge invariant
waveform on $\scri^+$ obtained using a nonlinear characteristic
evolution from a finite radius to $\scri^+$ along
outgoing null slices. 
The authors of Ref.~\cite{Babiuc:2010ze} measured the errors in
$r \psi_4$ for an equal-mass BHB simulation when
extracting at $R=50M$ and $R=100M$, the corresponding extrapolation
to $\infty$, as well as the error in $r\psi_4$ obtained by applying
the perturbative extrapolation formula~(\ref{eq:asymtpsi4ext})
 to the $R=100M$ waveform. The
errors in the perturbative extrapolation where the smallest over the
entire waveform (both in amplitude and phase).

We note that multi-patch~\cite{Pollney:2009yz} and
pseudospectral~\cite{Szilagyi:2009qz}
techniques allow extraction radii very far from the source, leading to
very small extrapolation errors.

\subsection{Gauge}

We obtain accurate, convergent waveforms and horizon parameters by
evolving this system in conjunction with a modified 1+log lapse and a
modified Gamma-driver shift
condition~\cite{Alcubierre02a,Campanelli:2005dd}, and an initial lapse
$\alpha(t=0) = 2/(1+\psi_{BL}^{4})$.  The lapse and shift are evolved
with
\begin{subequations}
\label{eq:gauge}
  \begin{eqnarray}
(\partial_t - \beta^i \partial_i) \alpha &=& - 2 \alpha K,\\
 \partial_t \beta^a &=& (3/4) \tilde \Gamma^a - \eta(x^a,t) \beta^a,
 \label{eq:Bdot}
 \end{eqnarray}
 \end{subequations}
where different functional dependences for $\eta(x^a,t)$ have been
proposed in
Refs.~\cite{Alcubierre:2004bm, Zlochower:2005bj, Mueller:2009jx, Muller:2010zze, Schnetter:2010cz,Alic:2010wu}. 
Here we use a modification of the form proposed
in Ref.~\cite{Mueller:2009jx},
\begin{equation}
  \eta(x^a,t) =  R_0 \frac{\sqrt{\partial_i W \partial_j W \tilde
\gamma^{ij}}}{ \left(1 - W^a\right)^b},
\end{equation}
where we chose $R_0=1.31$.
The above gauge condition is inspired by, but differs from Ref.~\cite{Mueller:2009jx}
between the BHs and in the outer zones when $a\neq1$ and $b\neq2$.
Once the conformal factor settles down to its asymptotic
$\psi=C/\sqrt{r} + O(1)$ form near the puncture, $\eta$ will have the
form  $\eta = (R_0/C^2) ( 1+ b (r/C^2)^a)$ near the puncture and
$\eta= R_0 r^{b-2} M/(a M)^b$ as $r\to \infty$. In practice we used
$a=2$ and $b=2$, which reduces $\eta$ by a factor of $4$ at infinity
when compared to the original version of this gauge proposed
by Ref.~\cite{Mueller:2009jx}.
 We note that if we set $b=1$ then $\eta$
will have a $1/r$ falloff at $r=\infty$ as suggested
by Ref.~\cite{Schnetter:2010cz}. Our tests indicate that the choices
$(a=2$, $b=1)$ and $(a=1, b=1)$ lead to more noise in the waveform
than $(a=2,b=2)$.

\section{Simulations and results}\label{SubSec:Runs}

\begin{table}[t]
   \caption{Initial data parameters for the numerical simulations.
    Note that the $q=1/15$ simulations are older and used a CFL factor
    twice as large. Here $m^p_1$ and $m^p_2$ are the two puncture mass
parameters, the puncture were located at $(x_1,0,0)$ and $(x_2,0,0)$
with momentum $\pm (P_r, P_t,0)$ and zero spin. The measured horizon
masses and total ADM mass are also provided.
   }
   \label{tab:ID}
\begin{ruledtabular}
\begin{tabular}{l|lll}
 Param & $q=1/10$ & $q=1/15$ & $q=1/100$\\
\hline
$m^p_1$ & 0.085237276   & 0.057566227   & 0.0086894746 \\
$m^p_2$ & 0.907396855   & 0.936224183   & 0.9896192142 \\
$x_1$ & 7.633129115     & 6.806172805   & 4.952562636 \\
$x_2$ & -0.7531758055   & -0.4438775230 & -0.04743736368 \\
$P_t$ & 0.0366988        & 0.0290721     & 0.00672262416584 \\
$P_r$ & -0.000168519     & -0.000160518  & -0.00001026521884 \\
$m_{H1}$ & 0.09129      & 0.06254       & 0.99065 \\
$m_{H2}$ & 0.91255      & 0.94044       & 0.00990841 \\
$M_{\rm ADM}$ & 1.00004 & 1.00005       & 1.00000000 \\
\end{tabular}
\end{ruledtabular}
\end{table}

The initial data parameters for the $q=1/10$, $q=1/15$, and $q=1/100$
simulations are given in Table~\ref{tab:ID}. Note that the $q=1/15$
simulations are older and suffer from the mass loss error discussed
below.

We used a base (coarsest) resolution of $h_0=4M$ with 
11 levels of refinement for $q=1/10$ simulations and
15 levels of refinement for $q=1/100$ for the low resolution 
simulations. The outer boundaries were at $400M$. The higher resolution simulations were all based on
these grids, but with correspondingly more gridpoints per level.
The grid structure was chosen 
by studying the behavior  of the background
potential for the propagation of perturbations~\cite{Lousto:2010ut}.
In the appendix \ref{app:Trumpet} we show this potential, both
in the isotropic coordinates of the initial data
and final 'trumpet' coordinates.

\begin{figure}
\includegraphics[width=3in]{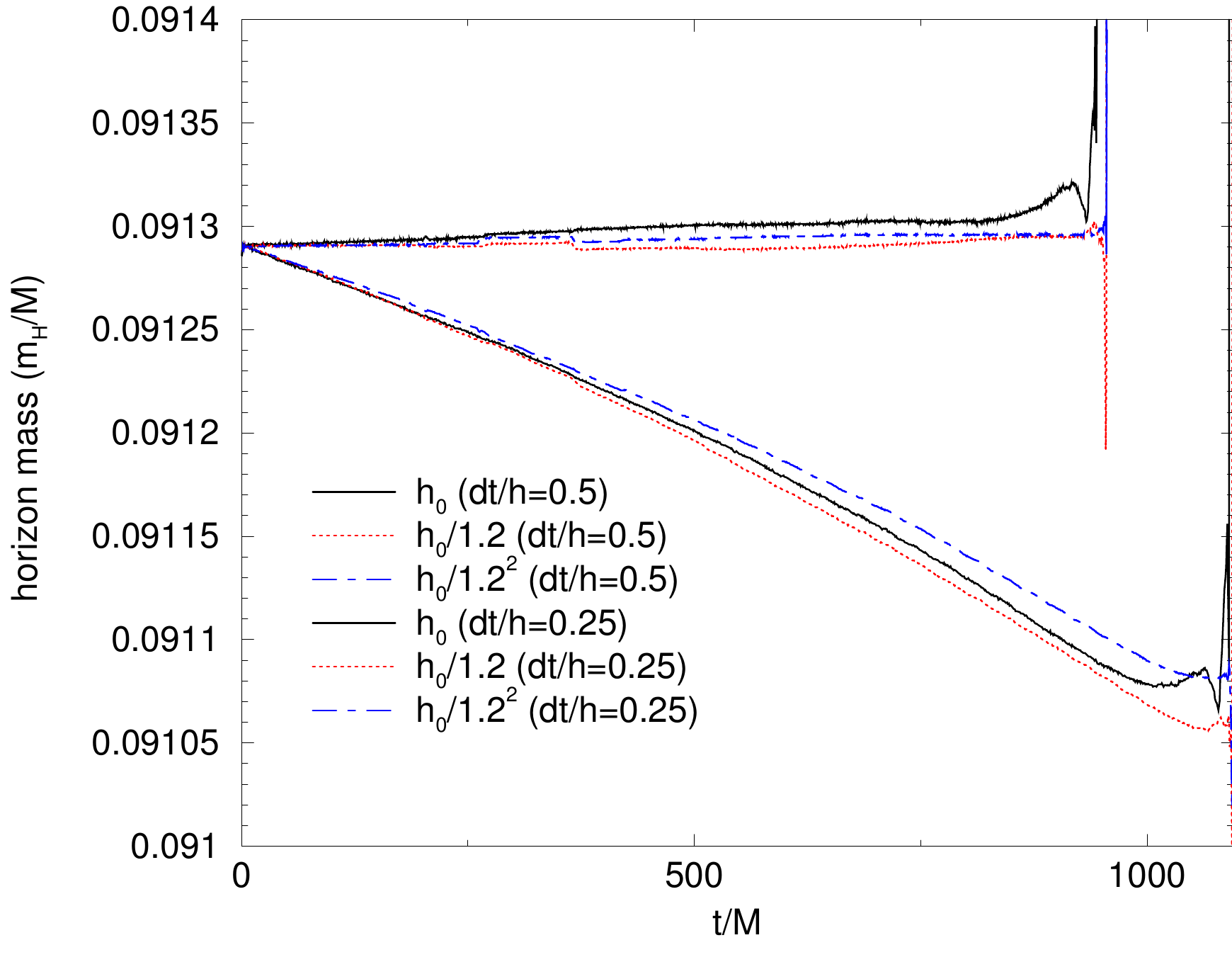}
\caption{The horizon mass conservation using CFL factors of 
$0.5$ and $0.25$ for the $q=1/10$ simulations.
 A factor of 10 better results are obtained using
a CFL factor of $0.25$. The effect on the trajectory is significant because the
mass loss is dynamically important (it significantly delays the merger).
The mass changes post merger (sharp spikes near $t\sim950M$
 and $t\sim 1080M$) are not
dynamically important.}
\label{fig:mass_cons}
\end{figure}

\begin{figure}
\includegraphics[width=3in]{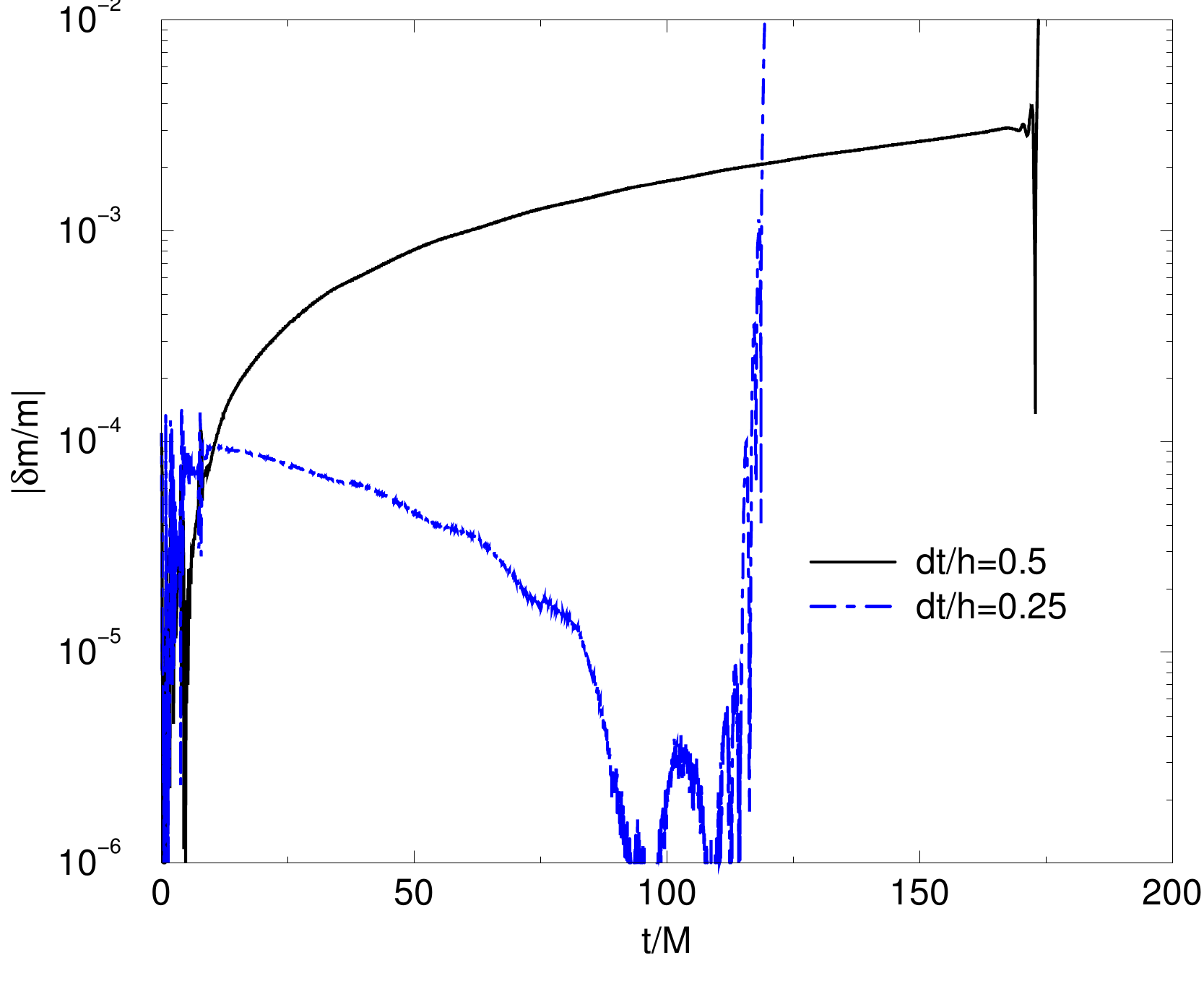}
\caption{The horizon mass conservation using CFL factors of 
$0.5$ and $0.25$ for the $q=1/100$ simulations. Here $|\delta m|$ is
the absolute value of the difference between the expected horizon mass 
and the measured
horizon mass for the smaller BH. In the case of $q=1/100$, as in the
case of $q=1/10$, the horizon mass decreases (hence $|\delta m|$
increases) with time when using
a CFL factor of $0.5$.
 A factor of 20  better results are obtained using
a CFL factor of $0.25$. The effect on the trajectory is significant because the
mass loss is dynamically important (it delays the merger). The mass
changes post merger (sharp spikes near $t=115M$ and $t=172M$) are {\em not}
dynamically important.}
\label{fig:h2o_mass_cons}
\end{figure}

\begin{figure}
\includegraphics[width=3in]{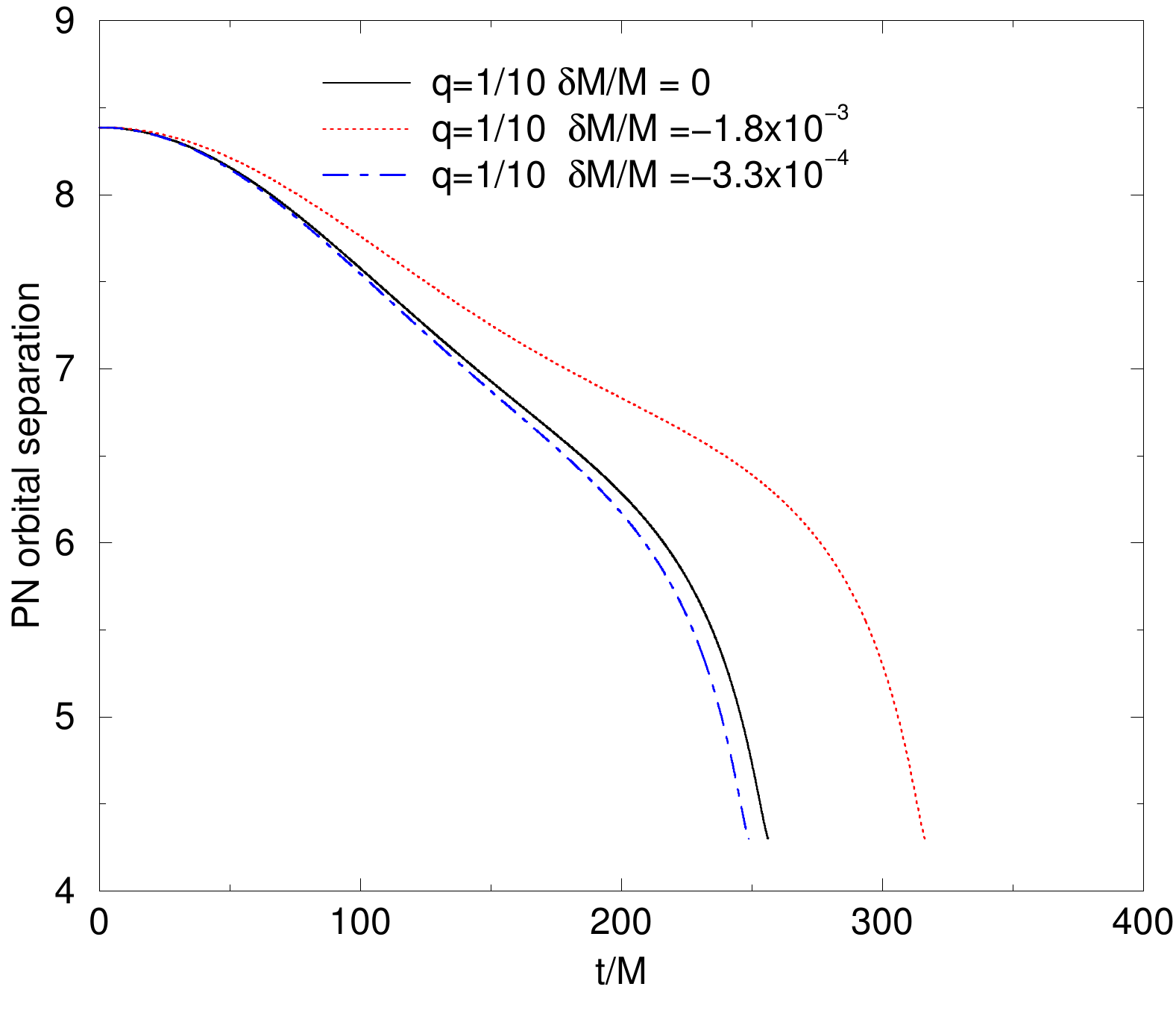}
\caption{The 3.5PN orbital separations for the $q=1/10$ binary
(the eccentricity is due to the fact that we use identical
parameters as the full numerical simulation, which are not
quasi-circular PN parameters) for slightly different small BH masses.
A mass loss of 1 part in $10^{4}$ is dynamically important, leading
to a significantly delayed merger. Reducing the mass change by
a factor of 10 significantly reduces this error.}
\label{fig:pn}
\end{figure}

We previously evolved a set of $q=1/10$,
$q=1/15$~\cite{Lousto:2010qx}, and
$q=1/100$~\cite{Lousto:2010ut} BHBs using the standard choice of
Courant-Friedrichs-Lewy (CFL) factor $dt/h\sim0.5$. Although we
found the waveform at lower resolution appear to converge, an
unphysical mass loss led to incorrect dynamics at later times
(see Fig.~\ref{fig:mass_cons}). This, in turn, led to oscillations
in the errors as a function of grid resolution $h$. We found that
reducing the CFL factor significantly reduces these unphysical
effects~\footnote{Marcelo Ponce, private communication}. In
Figs.~\ref{fig:mass_cons} and \ref{fig:h2o_mass_cons}
 we plot the horizon mass as
a function of time for three resolutions using both the old and
new CFL factor for both the $q=1/10$ and $q=1/100$ simulations.  Note the much better conservation of the mass and
the corresponding reduction in the lifetime of binary.
Post-Newtonian trajectory evolutions (see
Fig.~\ref{fig:pn}) indicate that the mass losses as small as 1 part in
$10^4$ are dynamically important, and the error introduced by this
mass loss is significantly reduced by the new integration. The effects of
the time integrator on the numerical error, and in particular the mass
loss and constraint violation errors, will be the subject of an
upcoming paper by Ponce, Lousto, and Zlochower.

For the $q=1/10$ nonspinning BHB, we performed simulations with
central resolutions of $h = M/256$, $h=M/307.2$, $M/337.9$,
$M/368.64$,
and $M/404.48$, with 11 levels of refinement in each case (the
coarsest grid resolutions where $h_0$, $h_0/1.2$, $h_0/1.32$,
$h_0/1.44$, $h_0/1.58$, respectively).
 We note that the waveform showed
oscillations in error as a function of resolution. We are therefore
using
the most widely spaced (in resolution) runs for the convergence plot.
In particular, we use the $h = M/256$,  $M/337.9$, and $M/404.48$
simulations.

\begin{figure}
  \includegraphics[width=3in]{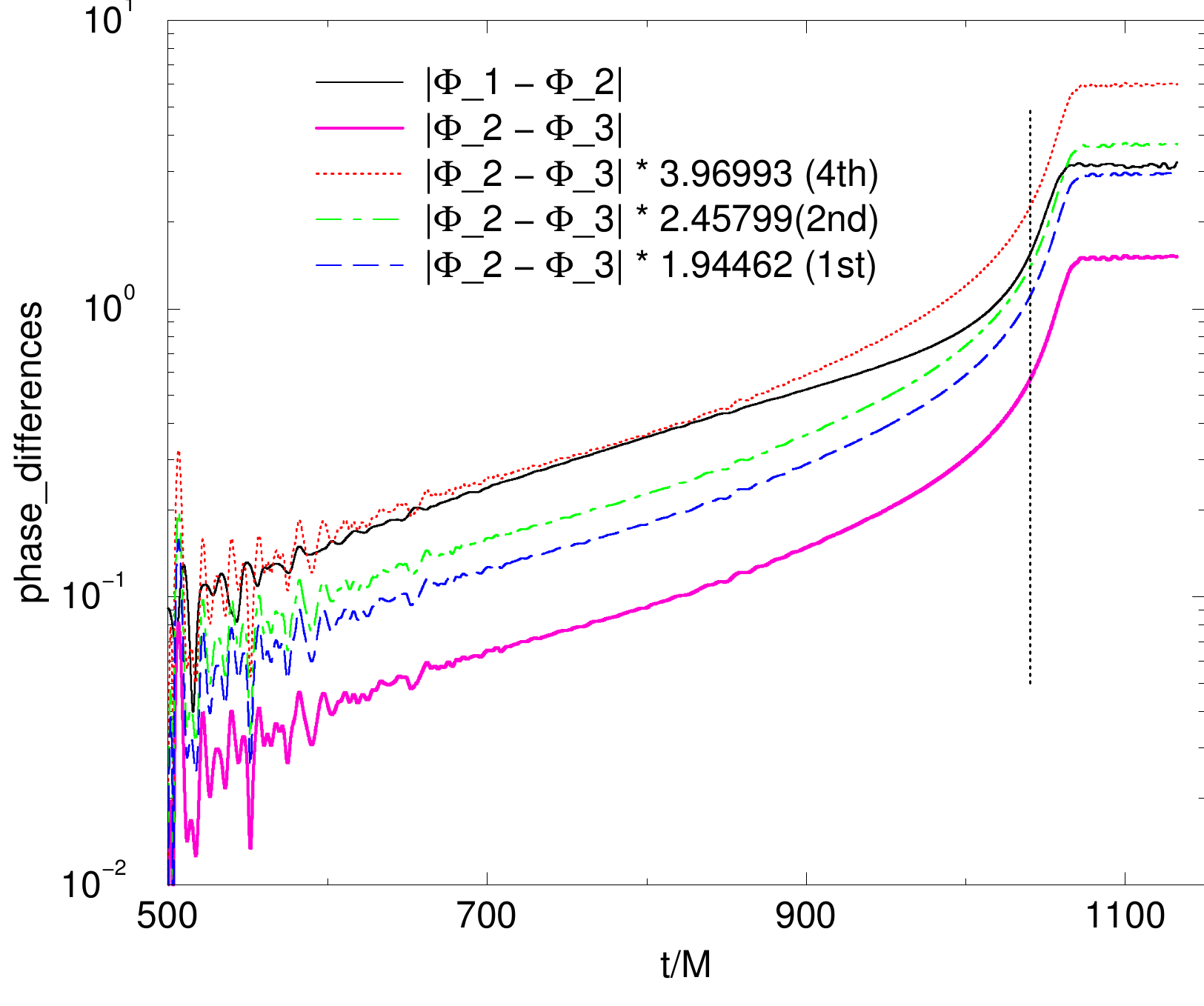}
  \includegraphics[width=3in]{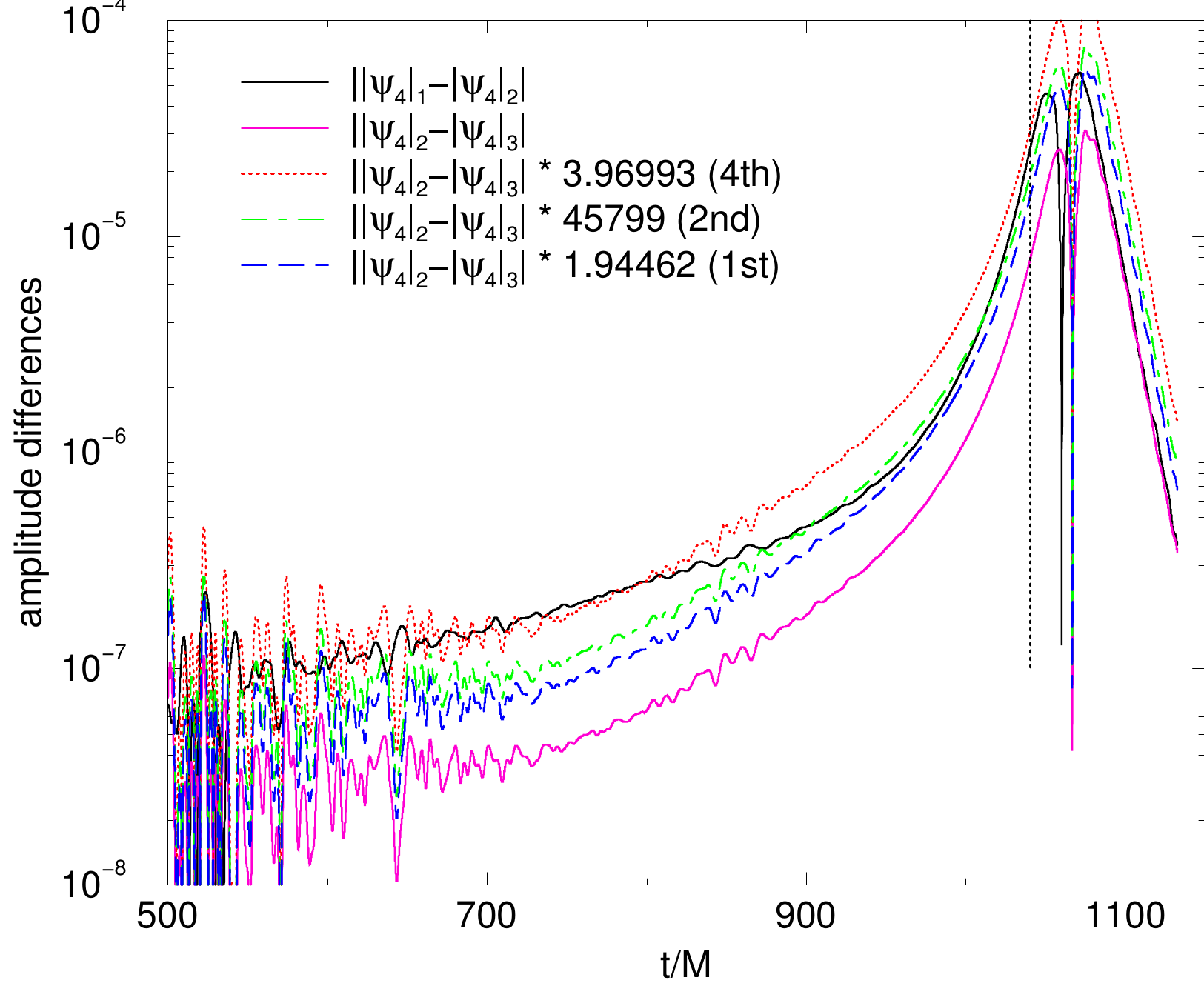}
  \caption{Convergence of the $(\ell=2,m=2)$ mode of $\psi_4$ for
 the $q=1/10$ waveform. The waveform starts out fourth-order accurate
(due to
our fourth-order waveform extraction algorithm), but then reduces to
second
order during the late inspiral, and dropping to first-order during the
ringdown.
The vertical line indicates the time when the frequency is
$\omega=0.2/M$.
The central resolutions for the three cases were
$h_1=M/256$, $h_2=M/337.9$, and $h_3=M/404.48$, respectively,
and each configuration used 11 levels of refinement.
}
  \label{fig:tentoone_conv}
\end{figure}

\begin{figure}
  \includegraphics[width=3in]{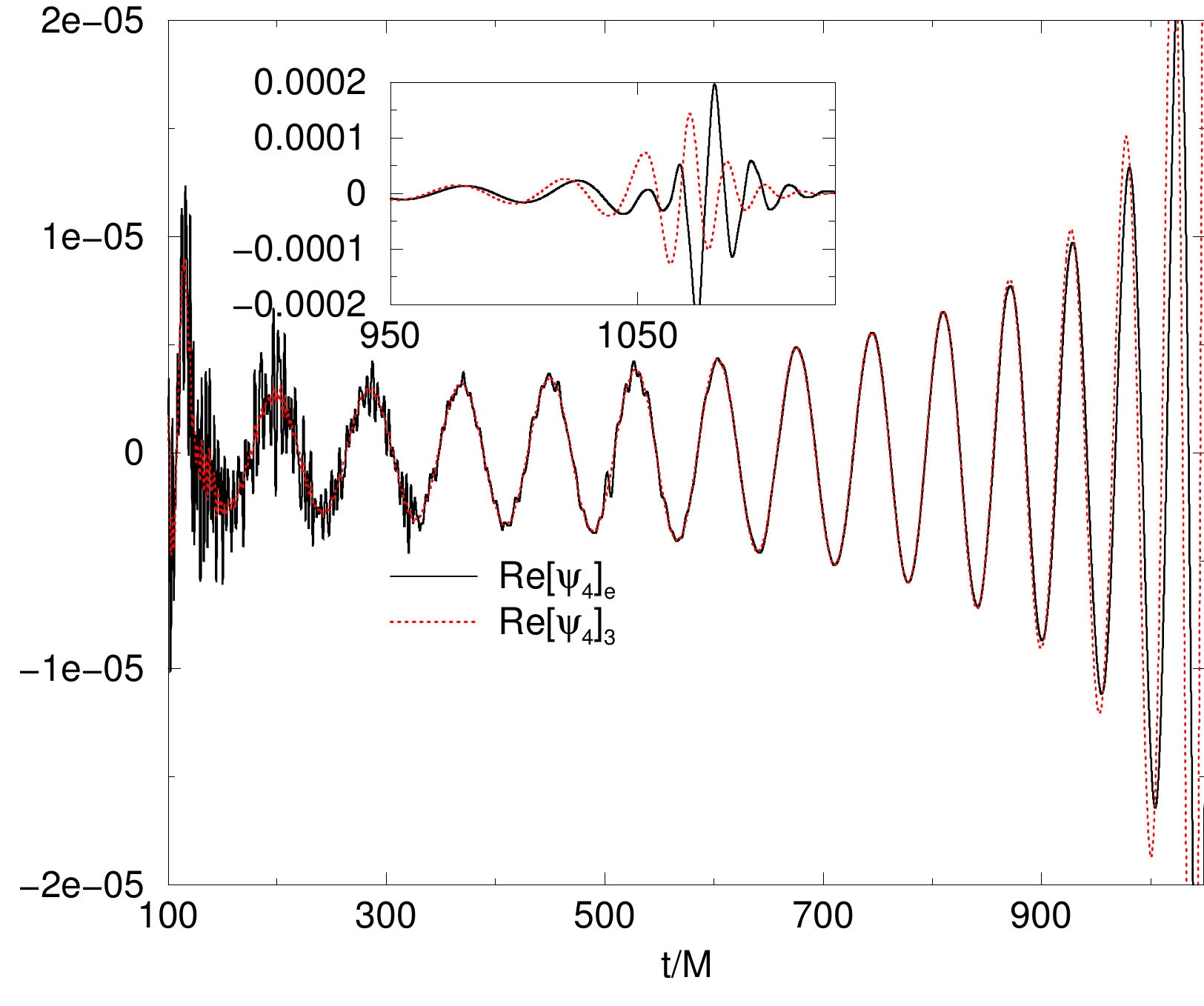}
  \includegraphics[width=3in,height=2.5in]{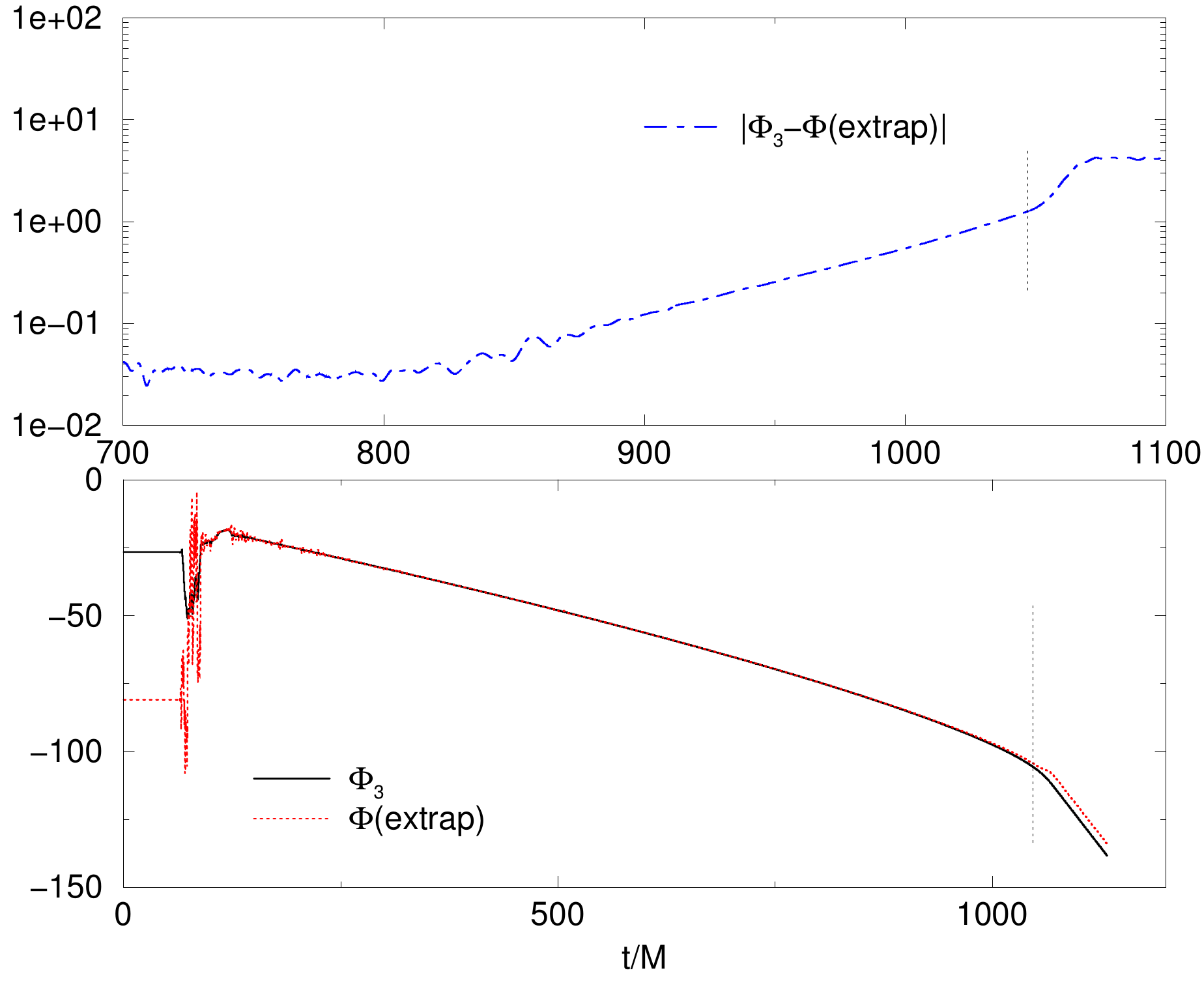}
  \caption{(TOP) A comparison of the highest resolution run with the
Richardson extrapolated waveform. At the plunge, the phase error in
the
waveform is sufficiently large that the extrapolation give erroneous
results (note the unphysical oscillation near $t=1050M$).
The extrapolation breaks down at $\omega=0.19/M$. (BOTTOM) The phase
difference between the Richardson extrapolated waveform and
the highest resolution.
}
  \label{fig:tentoone_extrap_comp}
\end{figure}

In Fig.~\ref{fig:tentoone_conv} we show the convergence of the amplitude
and
phase of $\psi_4$ for these three resolutions.
Convergence appears to break just at the start of the final plunge.
Figure~\ref{fig:tentoone_extrap_comp} shows the highest resolution waveform
and the Richardson extrapolation (assuming a second-order error).
The extrapolation error becomes large just near the plunge, at
a frequency of $\omega=0.19/M$.
We also plot the phase deference between the highest resolution run
and the Richardson extrapolated waveform.
The vertical line shows the point when the frequency is
$\omega=0.2/M$. The phase error, up to $t=850M$ is within $0.04$
radian, but increases exponentially
to $1.2$ radians when $\omega=0.2/M$.

\begin{figure}
\includegraphics[width=3in]{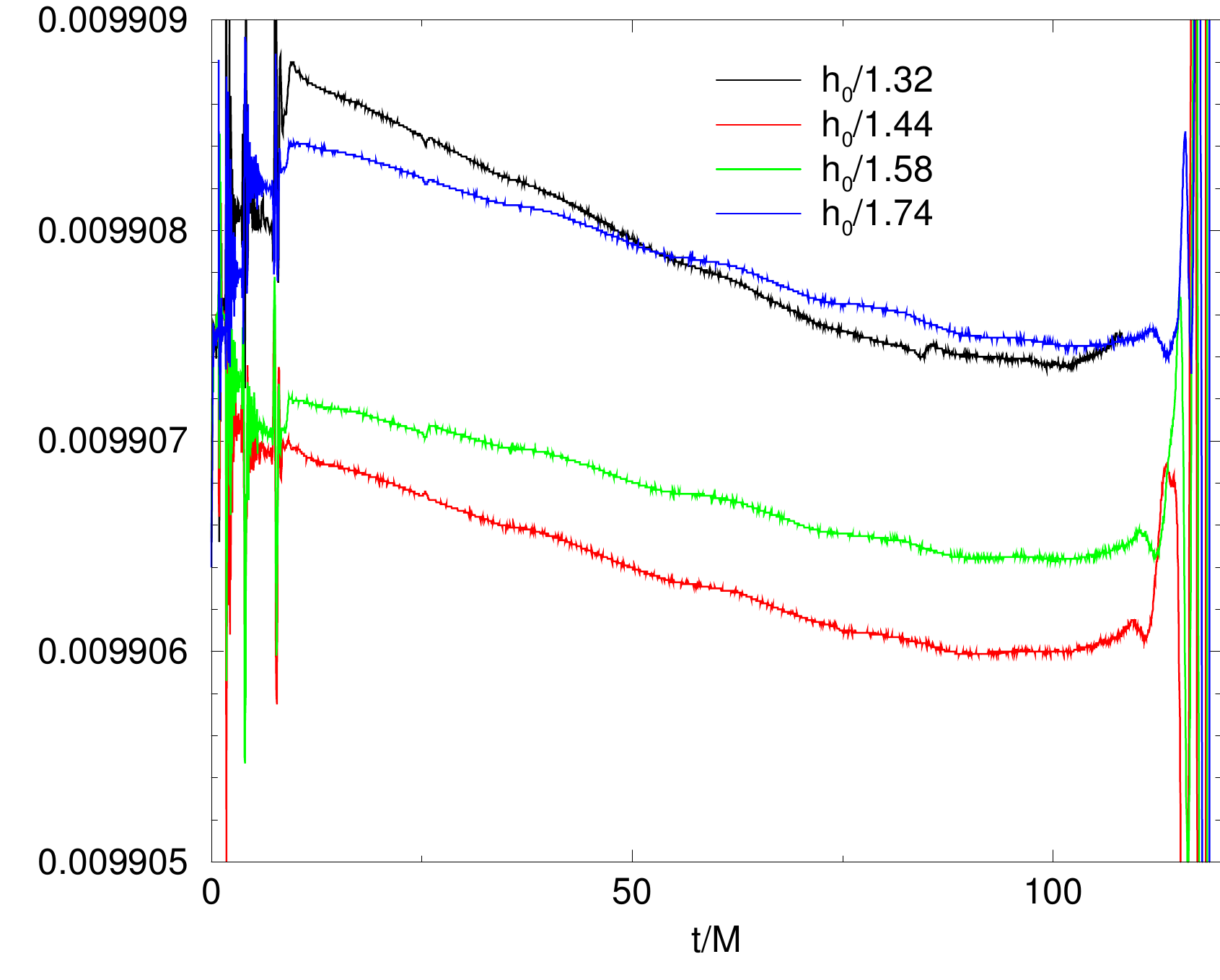}
\caption{The horizon mass for the $q=1/100$ configurations. Jumps in
the horizon mass of order $1/10^4$ are apparent between resolutions.
this has a small but noticeable dynamic effect.}
\label{fig:h2o_mass_comp}
\end{figure}

\begin{figure}
\includegraphics[width=3in]{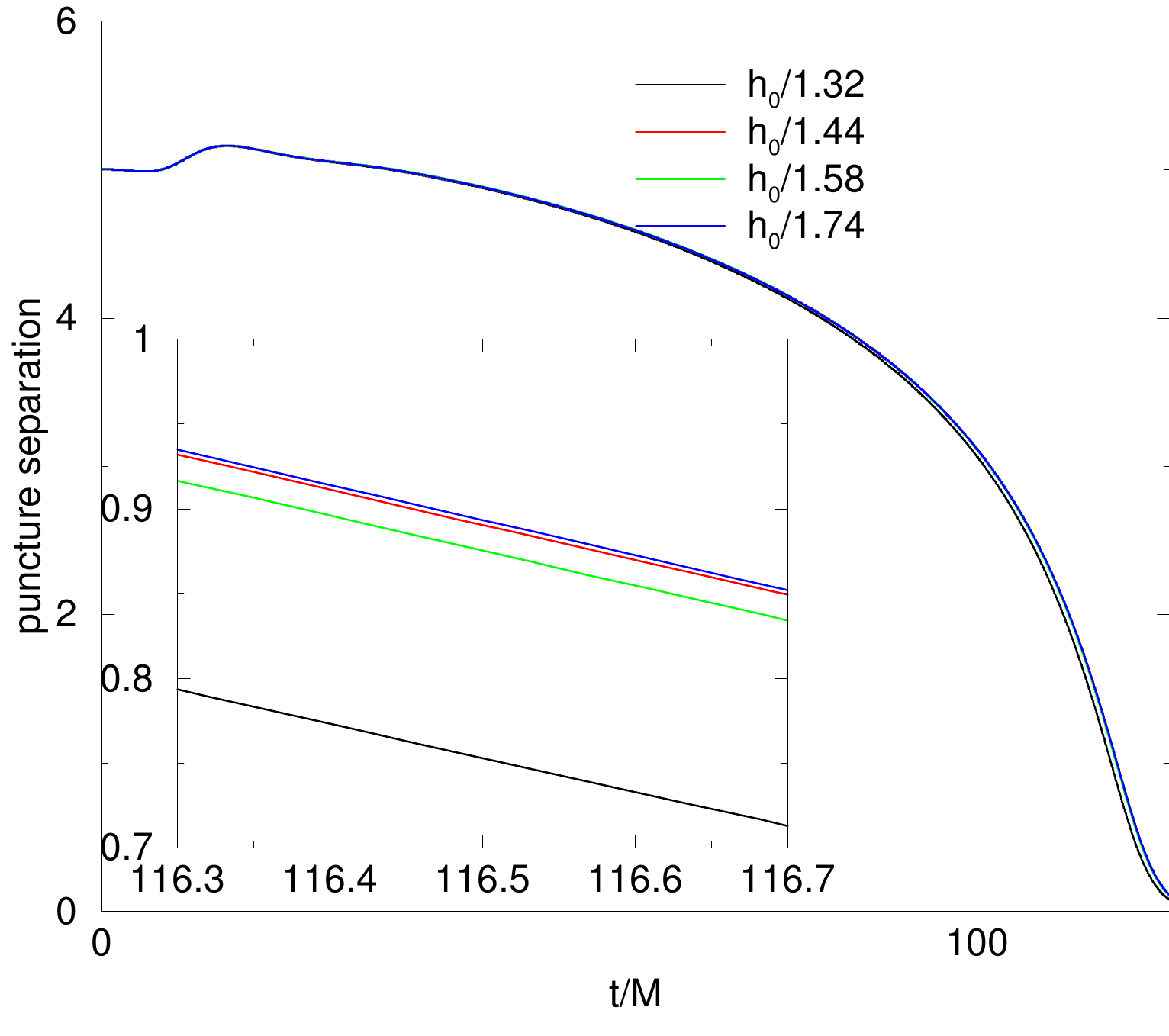}
\caption{The puncture separation as a function of time for the $q=1/100$ 
configurations. The second lowest resolution run appears to merge
later than expected, consistent with its reduced mass (apparent in
Fig.~\ref{fig:h2o_mass_comp})}
\label{fig:h2o_r_comp}
\end{figure}

\begin{figure}
\includegraphics[width=3in]{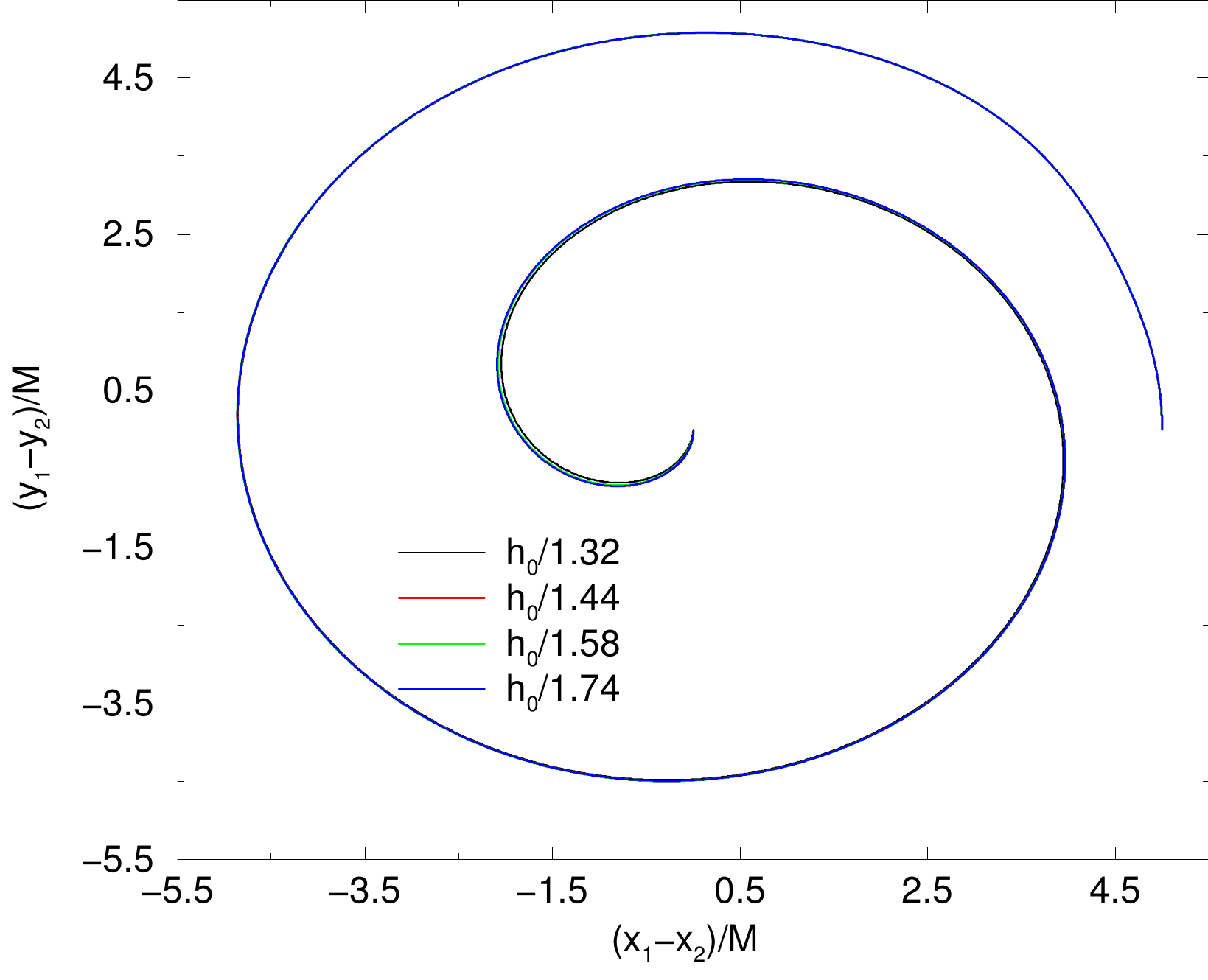}
\caption{The puncture trajectories for the $q=1/100$ configuration.}
\label{fig:h2o_track_comp}
\end{figure}

\begin{figure}
\includegraphics[width=3in]{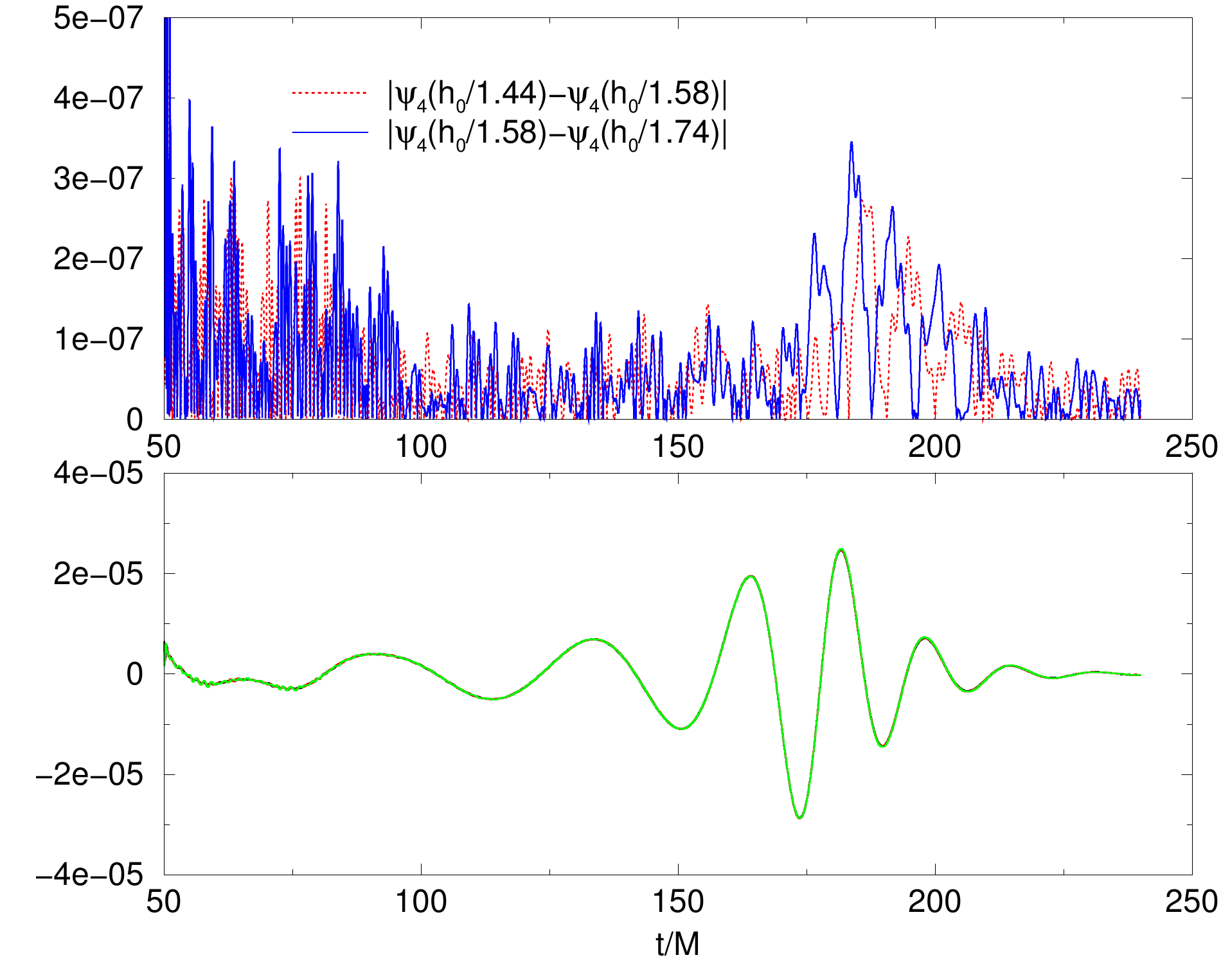}
\caption{The imaginary part (which is much less noisy than the real
part) of the $(\ell=2,m=2)$ mode of $\psi_4$ for the $q=1/100$
configurations. The differences in waveform between resolutions are
dominated by grid noise at high resolutions. At the peak, the
differences in the waveforms between resolutions is about $1\%$ of
the amplitude of the waveform. The real part of the waveform is 
roughly a factor of 10 more noisy.} 
\label{fig:h2o_wave_comp}
\end{figure}

\begin{figure}
\includegraphics[width=3in]{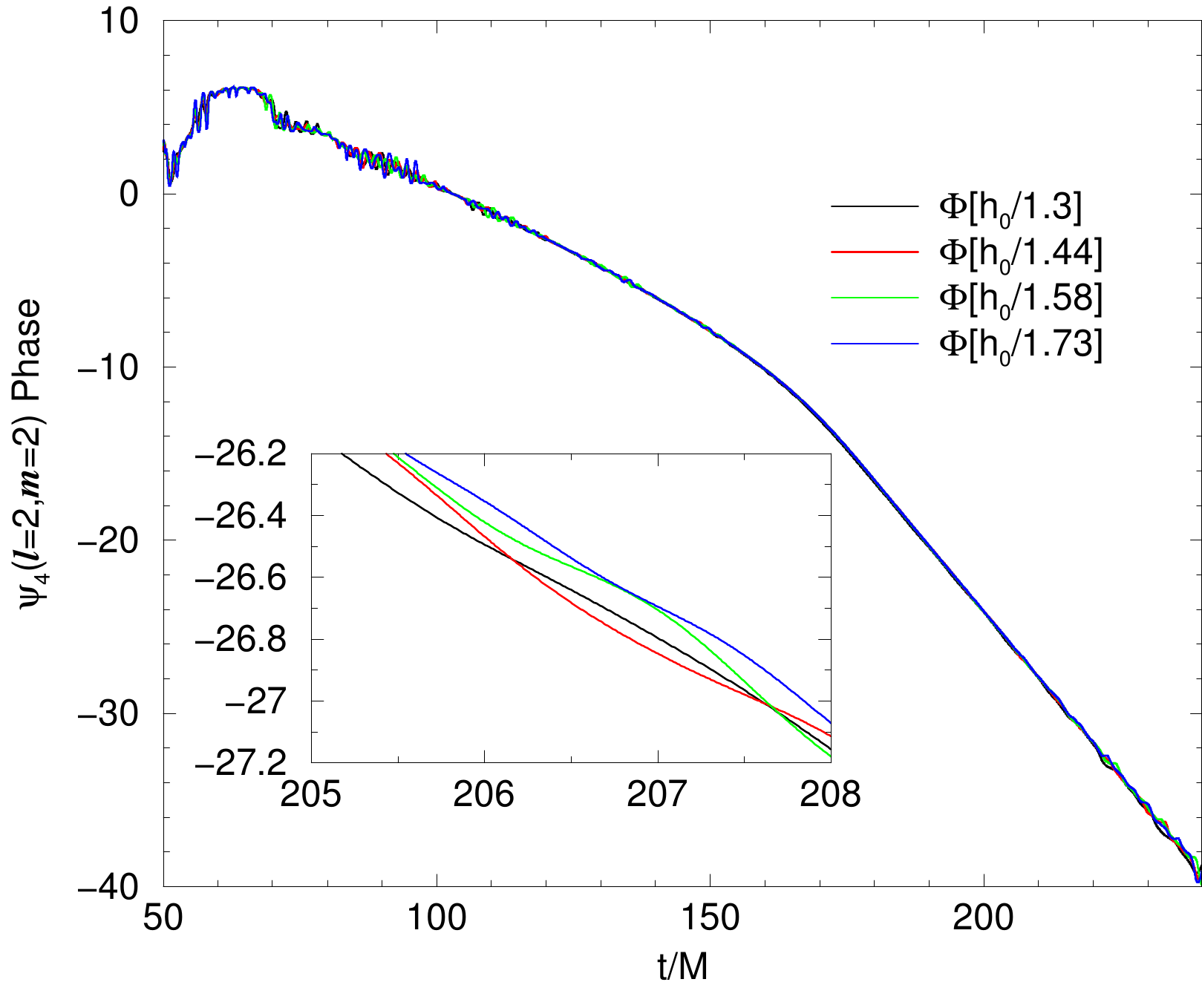}
\caption{The phase of the $(\ell=2,m=2)$ mode of $\psi_4$ for the $q=1/100$
configurations. The differences in phases between resolutions 
are
dominated by grid noise.}
\label{fig:h2o_phase_comp}
\end{figure}

Determining convergence for the $q=1/100$ simulations proved
challenging for two reasons. There was a small random jump in the mass of
the smaller BH between resolutions that had a small effect on the
trajectory. From Fig.~\ref{fig:h2o_mass_comp} we can see a difference
of one part in $10^4$ in the small BH mass. The two medium resolution
have lower masses, and therefore merge slightly later than expected.
The net effect is to confuse the order when the mergers happens (we
expected the higher resolutions runs to merge slightly later than
the lower resolutions runs). We see from Fig.~\ref{fig:h2o_r_comp}
that the second-lowest resolution merges out of order with the
other runs. Similarly, the third lowest resolution appears to merge
late, as well.
Another source of confusion for our convergence study has to do with 
lower-order errors (not
necessarily associated with the mass) becoming important at higher
resolutions. This leads to oscillations in the error as a function
of resolution that can cause the merger time to oscillate with
resolution.
 Nevertheless, the actual deviations in the trajectories, for the higher
resolutions, is small. The improvement over the original $q=1/100$
simulation is substantial. The mass conservation is a factor of
20 better. The old simulations had an unphysical extra orbit inside
the ISCO due to significant mass loss. Convergence measurements of the
waveform do not appear to suffer much from the above mentioned mass
loss. However, due to low amplitude of the waveform compared to the
grid noise (induced by the refinement boundaries), the differences
in the waveforms at the various resolution were smaller than the
grid-noise fluctuations.
Figures~\ref{fig:h2o_wave_comp}~and~\ref{fig:h2o_phase_comp} show the
waveform and phase for the three highest resolutions. Note that the
gridnoise apparent in Fig.~\ref{fig:h2o_wave_comp} is due to reflections of
the initial data pulse at the refinement boundaries. For our
configuration, the imaginary part of the $(\ell=2,m=2)$ components of
this pulse is small compared to the real part. This leads to a
significant reduction (about a factor of 10) in the noise in the
imaginary part when compared to the real part.
Figure~\ref{fig:h2o_track_comp} shows the trajectories for the
four highest resolutions. Note that good agreement between the curves
when using the new smaller CFL factor.

For reference, we report the final remnant properties for the three configurations
in Table~\ref{tab:remnant}.
\begin{table}
  \caption{Final remnant intrinsic spin, radiated energy, and recoil
velocity for the
$q=1/10$, $q=1/15$, and $q=1/100$ configurations.
For reference we also include the predicted recoil velocity based 
on~\cite{Gonzalez:2006md}.}
   \label{tab:remnant}
\begin{ruledtabular}
\begin{tabular}{l|lll}
 Param             & $q=1/10$          &  $q=1/15$ & $q=1/100$\\
\hline 
 $\alpha$          & $0.261\pm0.002$   &  $0.189\pm 0.006$  &
$0.0332\pm0.0001$\\ 
 $E_{\rm rad}/M$     & $0.0044\pm0.0001$ &  $0.0022 \pm 0.0001$ &
$(5.5\pm1.0) \times10^{-5}$ \\
 $V_{\rm kick}$    & $60\pm2\ \KMS$    &  $34\pm2\ \KMS$          &
$1.0\pm0.1\ \KMS$ \\
 $V_{\rm pred} $ & $62\ \KMS$ & $34\ \KMS$ & $1.1\ \KMS$
\end{tabular}
\end{ruledtabular}

\end{table}

\begin{figure}
\includegraphics[width=3in]{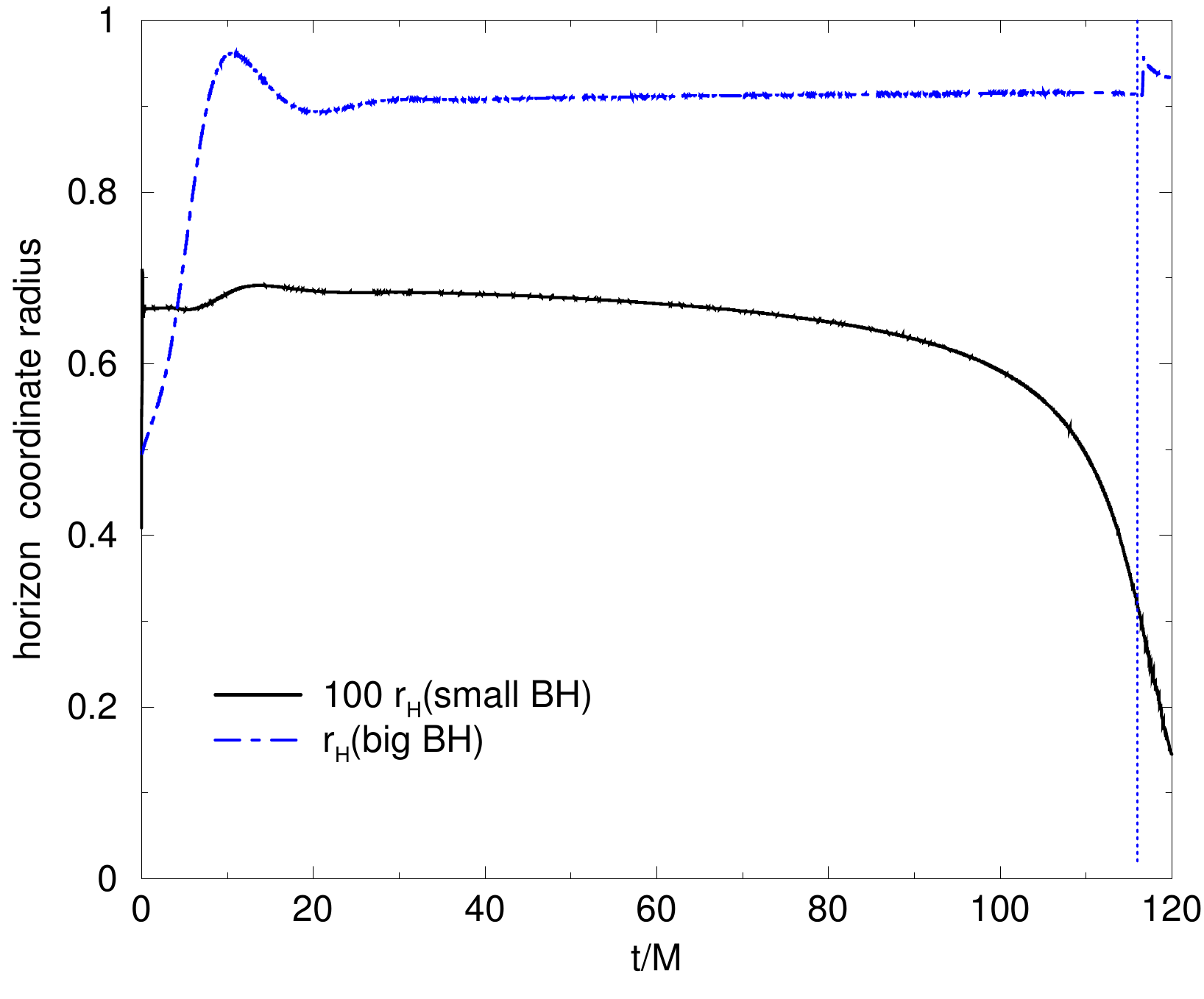}
\caption{The horizon radius of both BHs for the new $q=1/100$
simulations. The horizon radius of the larger BH barely changes, while
the radius of the smaller BH decreases sharply at merger.}
\label{fig:h2o_horizon_radius}
\end{figure}

\begin{figure}
\includegraphics[width=3in]{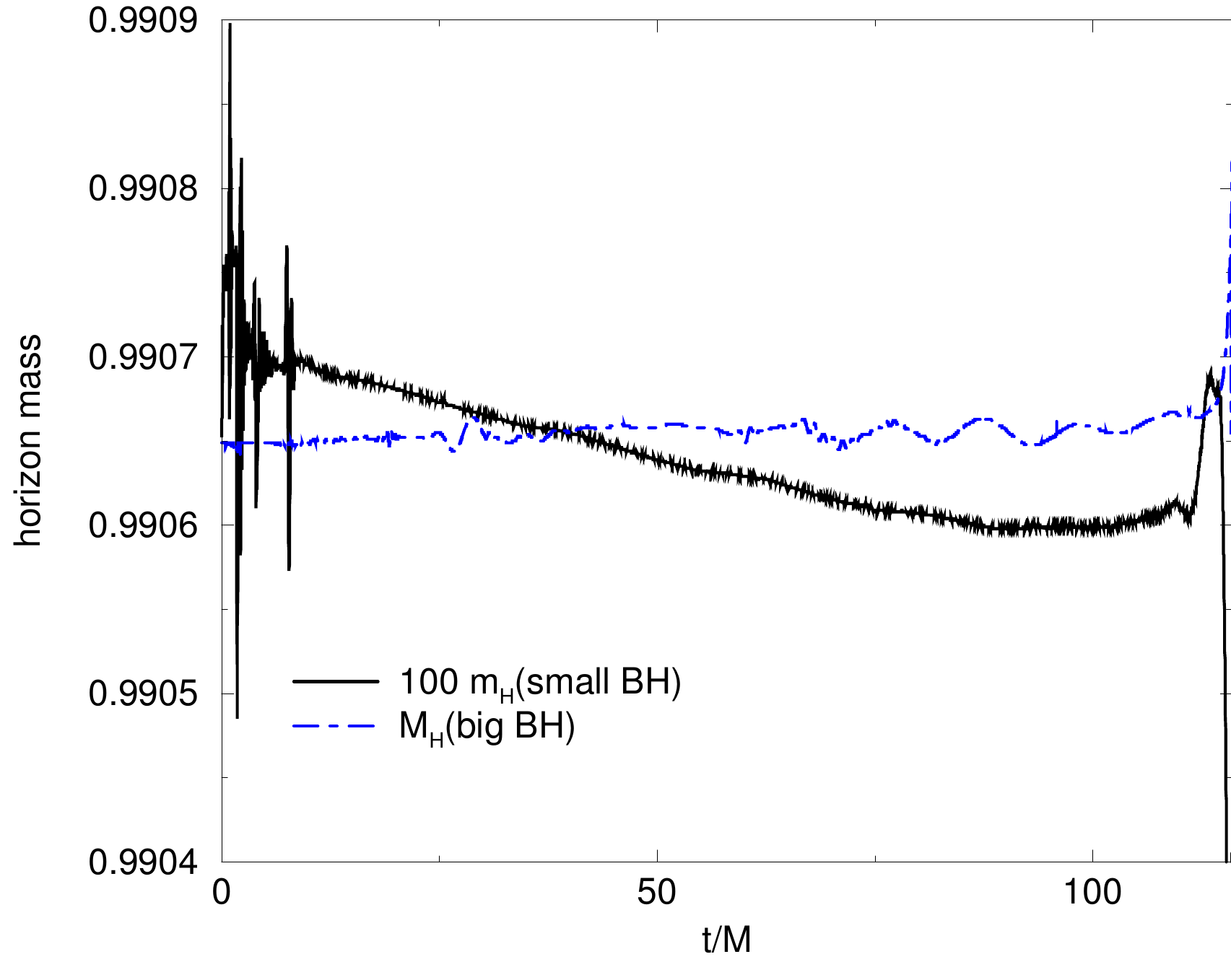}
\caption{The horizon mass of both BHs for the new $q=1/100$
simulations. The horizon mass of the larger BH is conserved to a
higher degree than that of the smaller BH. This is the case even
though the number of gridpoints inside the smaller BH is actually
larger than the number of gridpoints in the larger BH.}
\label{fig:h2o_horizon_mass}
\end{figure}

The above gauge conditions~(\ref{eq:gauge}) have a drawback near merger. The conformal
function $W$ goes to zero at the two punctures and there is a local
maximum between the two punctures. Due to this local maximum,
$\eta\to0$ at some point between the two punctures. As the two
punctures approach each other, this zero in $\eta$ get progressively
closer to the smaller BH. This, in turn, causes the coordinate radius
of the horizon to shrink near merger. As seen in
Fig.~\ref{fig:h2o_horizon_radius}, the horizon radius for the
new $q=1/100$ simulations decreases at merger, leading to a loss in
resolution of the smaller BH. Another issue related to this gauge is
that the relative effective size (i.e.\ $r_{H}/m_H$)
 of the smaller BH is less than that of the larger BH. So the required
resolution (gridpsacing) for each hole does not scale exactly with the
BH's mass. This lower effective relative size of the smaller BH may
be partially responsible for the reduced conservation of the horizon
mass of the smaller BH when compared to the larger on (see
Fig.~\ref{fig:h2o_horizon_mass}).

\section{Perturbative Techniques}\label{Sec:Perturbations}

In Ref.~\cite{Lousto:2010qx}, we extended the Regge-Wheeler-Zerilli
(RWZ) formalism~\cite{Regge:1957td,Zerilli:1971wd} of black hole perturbation theory
to include, perturbatively, a term linear in the spin of the larger black hole.
This Spin-Regge-Wheeler-Zerilli (SRWZ) formalism has second-order perturbations
with linear dependence on the spin
in the wave equations for Schwarzschild perturbations.

In the SRWZ formalism, the wave functions for the even and odd parity
perturbations are expressed as a  combined wave function of the first
and second perturbative orders,
\bea
\Psi_{\ell m} \left( t,r \right) &=& \Psi_{\ell m}^{(1)} \left( t,r \right)
+ \Psi_{\ell m}^{(2)} \left( t,r \right) \,,
\nonumber \\ 
\Psi_{\ell m}^{\rm (o)} 
&=& \Psi_{\ell m}^{\rm (o,1)} + 2\,\int dt \, \Psi_{\ell m}^{\rm (o,Z,2)}
\,,
\eea
where the superscript $1$ and $2$ denote the perturbative order,
$\Psi_{\ell m}$ denotes the even parity Zerilli function, and
$\Psi_{\ell m}^{\rm (o,1)}$ and $\Psi_{\ell m}^{\rm (o,Z,2)}$ denote
the Regge-Wheeler (Cunningham et al.) and the Zerilli functions for
the odd parity perturbations
(see Ref.~\cite{Lousto:2005ip}
for the relation between the wave functions).
These functions satisfy the following three equations.
\bea
&& 
-{\frac {\partial ^{2}}{\partial {t}^{2}}}\Psi_{\ell m} \left( t,r \right) 
+{\frac {\partial ^{2}}{\partial {r^*}^{2}}}
\Psi_{\ell m} \left( t,r \right) 
\nonumber \\ && \qquad 
-V_{\ell}^{\rm (even)}(r) 
\Psi_{\ell m} \left( t,r \right) 
+ i\,m\,\alpha\,{\hat P}_{\ell}^{\rm (even)} \Psi_{\ell m} \left( t,r \right)
\nonumber \\ 
&& =  S_{\ell m}^{\rm (even)}(t,r; r_p(t), \phi_p(t))
\,, \nonumber\\
&&
-{\frac {\partial ^{2}}{\partial {t}^{2}}}\Psi_{\ell m}^{\rm (o,1)} \left( t,r \right) 
+{\frac {\partial ^{2}}{\partial {r^*}^{2}}}
\Psi_{\ell m}^{\rm (o,1)} \left( t,r \right) 
\nonumber \\ && \qquad 
-V_{\ell}^{\rm (odd)}(r) 
\Psi_{\ell m}^{\rm (o,1)} \left( t,r \right) \nonumber\\
&& = S_{\ell m}^{\rm (odd,1)} \left(t,r; r_p(t), \phi_p(t) \right) \,,
\nonumber \\
&&
-{\frac {\partial ^{2}}{\partial {t}^{2}}}\Psi_{\ell m}^{\rm (o,Z,2)} \left( t,r \right) 
+{\frac {\partial ^{2}}{\partial {r^*}^{2}}}
\Psi_{\ell m}^{\rm (o,Z,2)} \left( t,r \right) 
\nonumber \\ && \qquad 
-V_{\ell}^{\rm (odd)}(r) 
\Psi_{\ell m}^{\rm (o,Z,2)} \left( t,r \right)  \nonumber\\
&& = S_{\ell m}^{\rm (odd,Z,2)} \left(t, r;  r_p(t), \phi_p(t)\right)
\,,
\label{eq:SRWZeqs}
\eea
where $r^*=r+2M\ln[r/(2M)-1]$ is a characteristic coordinate, 
$\alpha$ is the non-dimensional spin parameter,
$V_{\ell}^{\rm (even/odd)}$ denotes the Zerilli and Regge-Wheeler potential,
respectively, $S_{\ell m}$ is the source term, and $(r_p(t),
\phi_p(t))$ are the particle separation and orbital phase as a
function of time.
The differential operator ${\hat P}_{\ell}^{\rm (even)}$ arises from the coupling
between the BH's spin and the first-order wave functions.
We note that  $S_{\ell m}^{\rm (even)}$ and $S_{\ell m}^{\rm (odd,Z,2)}$
include local and extended source terms
arising from the second perturbative order associated with
products of terms linear in $\alpha$ and the first-order perturbation
functions.
See Appendix A in Ref.~\cite{Lousto:2010qx} for more details.

In order to evolve the perturbative equations~(\ref{eq:SRWZeqs}), we need 
the particle trajectory
$(r_p(t), \phi_p(t))$ as a function of time.
Here we use a post-Newtonian
inspired  model for the
BH trajectories, which will approximate the numerical trajectories
from intermediate-mass-ratio
BHB inspirals.
In order to model the numerical trajectories
for the late inspiral phase of a BHB merger using a relatively
straightforward  fitting function, we start with an adiabatic
quasicircular evolution, i.e., ignoring the eccentricity of the orbit
and the component of radial velocity not due to the radiation
reaction, and then parametrize deviations from a 3.5PN-TaylorT4 adiabatic
inspiral. We
model the time derivative of the orbital frequency $d\Omega/dt$ and
the orbital separation $R$ as a function of the orbital frequency
$\Omega$ based on an extension of standard post-Newtonian
calculations,
described below.

\subsection{The orbital frequency $\Omega$ evolution of the NR trajectories}

First, we focus on the $d\Omega/dt$ evolution with respect to the orbital frequency $\Omega$.
In post-Newtonian (PN) calculations for the quasicircular case,
the evolution for $\Omega$ is obtained from the energy loss $dE/dt$ 
and the relation between energy and frequency $E(\Omega)$ given
by
\bea
\frac{d\Omega}{dt} &=& \frac{dE}{dt} \left(\frac{dE}{d\Omega}\right)^{-1} \,,
\label{eq:basicOt}
\eea
where $dE/dt$ and $dE/d\Omega$ are obtained by appropriate PN
expansions.
In the TaylorT1 waveform, we simply substitute these quantities into Eq.~(\ref{eq:basicOt}).
By expanding the TaylorT1 in the PN series again, we obtain 
the TaylorT4 waveform, which has been shown to give good agreement 
with numerical simulations~\cite{Boyle:2007ft}.
Therefore, we develop a fitting function based on the TaylorT4 evolution.
These Taylor series waveforms are summarized in Ref.~\cite{Brown:2007jx}.

We note here that for Schwarzschild geodesics, 
$d\Omega/dt$ diverges at the innermost stable circular orbit (ISCO), 
$R_{\rm Sch}=6M$ in Schwarzschild coordinates.
This is because $dE/d\Omega$ becomes zero at the ISCO frequency 
$M\Omega = (1/6)^{3/2} \sim 0.0680413817$.
Although we need to develop techniques for the transition to 
plunge~\cite{Ori:2000zn,Kesden:2011ma} in the above situation,
there is no divergence in the TaylorT4 waveform
because the Taylor series expansion always has finite $d\Omega/dt$ .

When fitting of the trajectories in the NR coordinates 
for various mass ratios,
we use the following modified TaylorT4 evolution.
\begin{widetext}
\bea
\frac{d\Omega}{dt} &=& 
{\frac {96}{5}}\,{\Omega}^{11/3}{M}^{5/3}\eta\, 
 \left( 1+B \left( \Omega/\Omega_0 \right) ^{\beta/3} \right) ^{-1}
\Biggl( 1
+ \left( -{\frac {743}{336}}-\frac{11}{4}\,\eta \right)  \left( M\Omega \right) ^{2/3}
+4\,\pi \,M\Omega
\nonumber \\ &&
+ \left( {\frac {34103}{18144}}+{\frac {13661}{2016}}\,\eta+{\frac {59}{18}}\,{\eta}^{2} \right) 
 \left( M\Omega \right) ^{4/3}
+ \left( -{\frac {4159}{672}}\,\pi -{\frac {189}{8}}\,\eta\,\pi 
 \right)  \left( M\Omega \right) ^{5/3}
\nonumber \\ &&
+
 \left( {\frac {16447322263}{139708800}}+\frac{16}{3}\,{\pi }^{2}-{\frac {1712
}{105}}\,\gamma -{\frac {
1712}{315}}\,\ln  \left( 64\,M\Omega \right) -{\frac {56198689}{217728}}\,\eta
\right. \nonumber \\ && \left. \quad 
+{\frac {451}{48}}\,\eta\,{\pi }^{2}
+{\frac {541}{896}}\,{\eta}^{2
}-{\frac {5605}{2592}}\,{\eta}^{3} \right) \left( M\Omega \right) ^{2}
\nonumber \\ && 
+ \left( -{\frac {4415}{4032}}\,\pi +{\frac {358675}{6048}}\,\eta\,\pi +{\frac {
91495}{1512}}\,{\eta}^{2}\pi  \right)  \left( M\Omega \right) ^{7/3}
+A \left( \Omega/\Omega_0 \right) ^{\alpha/3} 
\Biggr) 
\,,
\label{eq:OOt_PNP}
\eea
\end{widetext}
where $d\Omega/dt$ and $\Omega$ are obtained from the NR trajectories,
and $M$ and $\eta$ denote the total mass and symmetric mass ratio
of the binary system, 
\bea
M &=& m_1 + m_2 \,,
\nonumber \\ 
\eta &=& \frac{m_1 m_2}{M^2} \,,
\eea
respectively. 
Here, $A$, $\alpha$, $B$ and $\beta$ are the fitting parameters,
and the power in the parameters should be $\alpha>7$ and $\beta>7$
in order to be consistent with this ($7/2$)PN formula.
Here $\Omega_0$ is the frequency at $R_{\rm Sch}=3M$ for circular
orbits, 
i.e.\ $M\Omega_0=(1/3)^{3/2} \sim 0.19245009$. 

In Table~\ref{tab:dOdt_AalpBbet_mT4},
we summarize the parameters obtained from fitting for the three cases,
$q=1/10$,
$q=1/15$ (from Ref.~\cite{Lousto:2010qx}), 
and $q=1/100$.

\begin{table}[t]
   \caption{
   The summary of the fitting parameters
   for the ``PN$+$'' (a modification of the TaylorT4 evolution) in Eq.~(\ref{eq:OOt_PNP}).
   The first line for the $q=1/100$ case ($*$) gives the fitting
   parameters obtained by allowing all 4 constants to vary, 
   the second line ($\dag$) for the  $q=1/100$ case 
   gives the parameters from a fitting of $A$ and $B$ only  by assuming $\alpha=8$ and $\beta=15$,
   and the third line ($\ddag$) gives the parameters extrapolated
   from the $q=1/10$ and $q=1/15$ results using Eq.~(\ref{eq:c0c1}).
   }
   \label{tab:dOdt_AalpBbet_mT4}
\begin{ruledtabular}
\begin{tabular}{c|cccc}
 Mass ratio & $A$ & $\alpha$ & $B$ & $\beta$ \\ \hline
 $q=1/10$ & $17.0500$ &  $7.21975$ & $8.18920$ & $12.5197$
 \\
 $q=1/15$ & $26.0150$ &  $7.54047$ & $8.65525$ & $13.6168$
 \\
 $q=1/100^*$ & $93.0650$ &  $4.32071$ & $5.42457$ & $14.9711$
 \\
\hline
 $q=1/100^\dag$ & $288.888$ &  $8$ & $16.2033$ & $15$
 \\
 $q=1/100^\ddag$ & $233.985$ &  $9.45214$ & $11.5411$ & $21.0731$
\end{tabular}
\end{ruledtabular}
\end{table}

\begin{figure}[!h]
  \caption{
  The fitting curve for the $q=1/10$ case using Eq.~(\ref{eq:OOt_PNP})
and orbital frequencies  up to $M\Omega = 0.175$.
  The solid (black), dashed (red) and dotted (blue)
  curves show the NR, fitting and TaylorT4 PN evolutions, respectively.
  In the NR evolution, the end point of the solid (black) curve corresponds to $R_{\rm Sch}=2M$,
  and $M\Omega = 0.175$ is the NR orbital frequency around $R_{\rm Sch}=3M$,
  assuming the NR coordinate system can be approximated by a ``trumpet'' 
  stationary $1+\log$ slice of the Schwarzschild spacetime.
  The inset shows the zoom-in of the initial part (around $M\Omega = 0.05$)
  } 
  \includegraphics[width=3in]{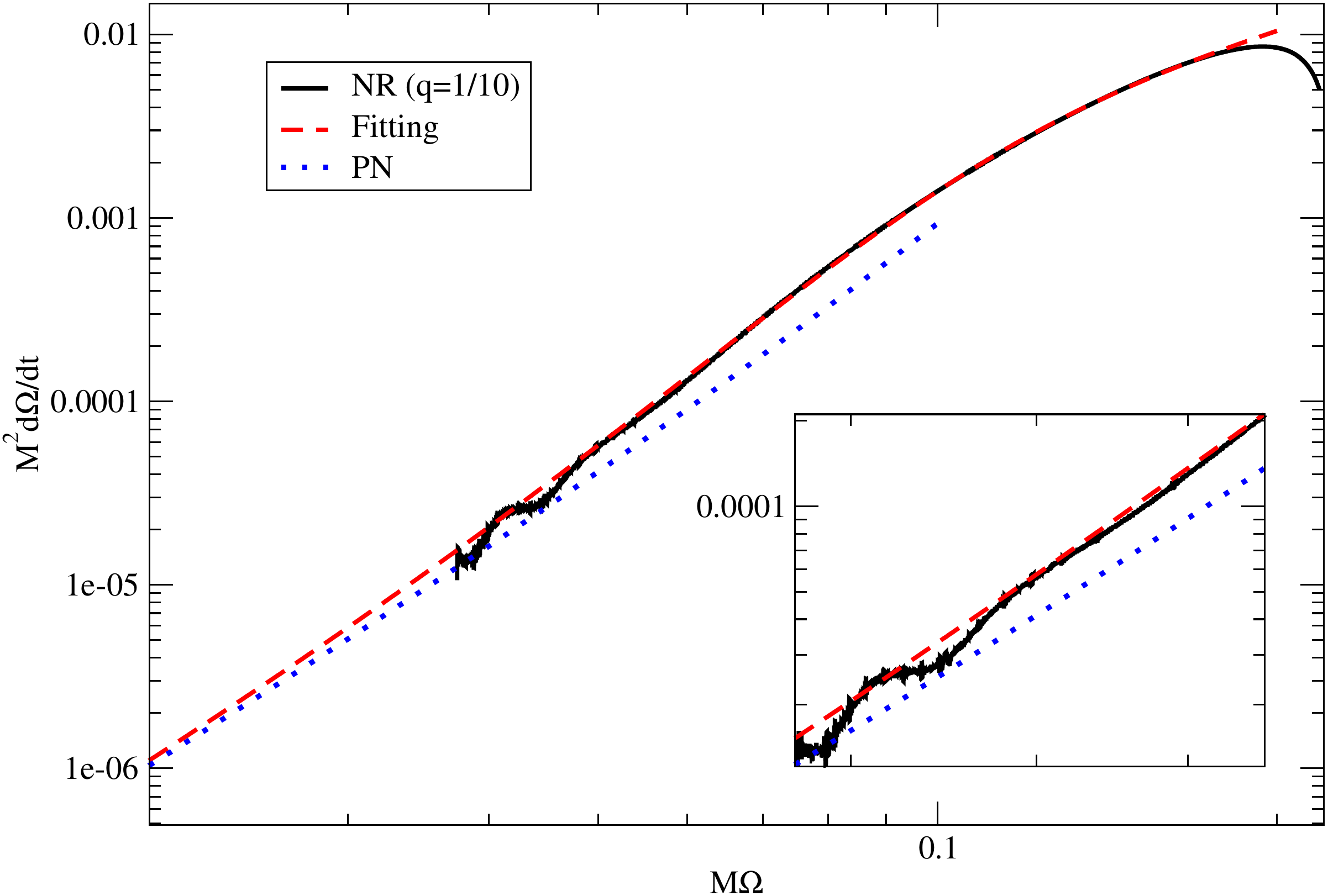}  
  \label{fig:OOt_10to1_PNP}
\end{figure}

The fitting curve for the $q=1/10$ case, 
which is valid up to $M\Omega = 0.175$,
is shown in Fig.~\ref{fig:OOt_10to1_PNP}.
The fitting parameters, $\alpha$ and $\beta$, for this case 
are consistent with the current PN formula,
and we see that the fitting curve is close to the TaylorT4 evolution 
for small $M\Omega$.

\begin{figure}[!h]
  \caption{
  The fitting curve for the $q=1/15$ case using Eq.~(\ref{eq:OOt_PNP})  up to
orbital frequencies of $M\Omega = 0.17$.
  The solid (black), dashed (red) and dotted (blue)
  curves show the NR, fitting and TaylorT4 PN evolutions, respectively.
  In the NR evolution, the end point of the solid (black) curve corresponds to $R_{\rm Sch}=2M$,
  and $M\Omega = 0.17$ is the NR orbital frequency around $R_{\rm Sch}=3M$,
assuming the NR coordinate system can be approximated by a ``trumpet'' 
  stationary $1+\log$ slice of the Schwarzschild spacetime.
  } 
  \includegraphics[width=3in]{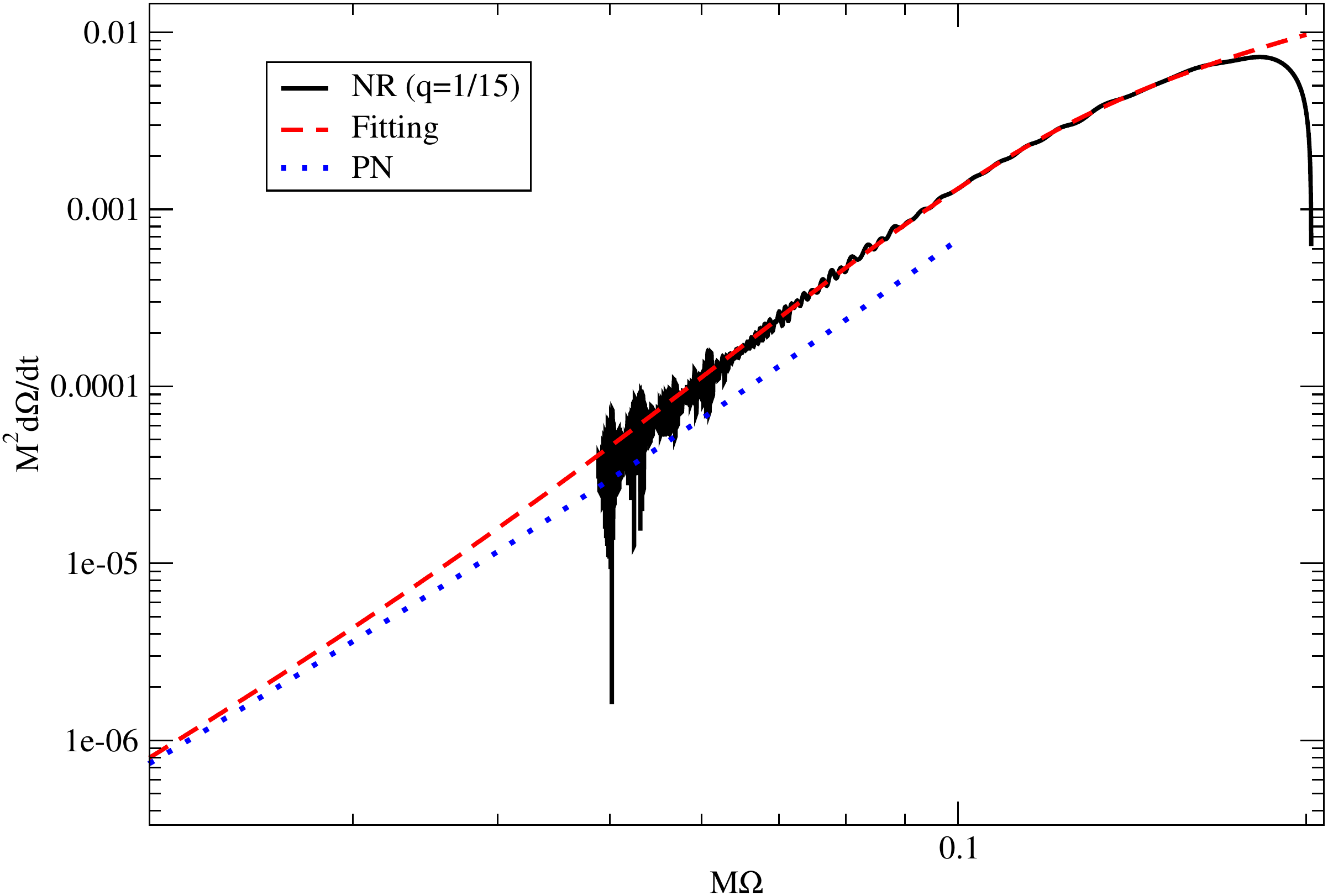}
  \label{fig:OOt_15to1_PNP}
\end{figure}

We show the fitting curve for the $q=1/15$ case,
which is valid up to $M\Omega = 0.17$, in Fig.~\ref{fig:OOt_15to1_PNP}.
Again, the fitting parameters, $\alpha$ and $\beta$, for this case are
consistent with the PN formula. 

\begin{figure}[!h]
  \caption{
  The fitting curve for the $q=1/100$ case using
  Eq.~(\ref{eq:OOt_PNP}) with orbital frequencies 
  up to $M\Omega = 0.15$.
  The solid (black) and dotted (blue)
  curves show the NR and TaylorT4 PN evolutions, respectively.
  The Fitting ($*$, dashed (red)), ($1$, dot-dashed (green)) and
  ($2$, dot-dot-dashed (magenta)) curves show the fitting curve
  with the ($*$), ($\dag$) and ($\ddag$) cases in Table~\ref{tab:dOdt_AalpBbet_mT4}.
  In the NR evolution,
  $M\Omega = 0.15$ is the NR orbital frequency around $R_{\rm Sch}=3M$,
  assuming the NR coordinate system can be approximated by a ``trumpet'' 
  stationary $1+\log$ slice of the Schwarzschild spacetime.
} 
  \includegraphics[width=3in]{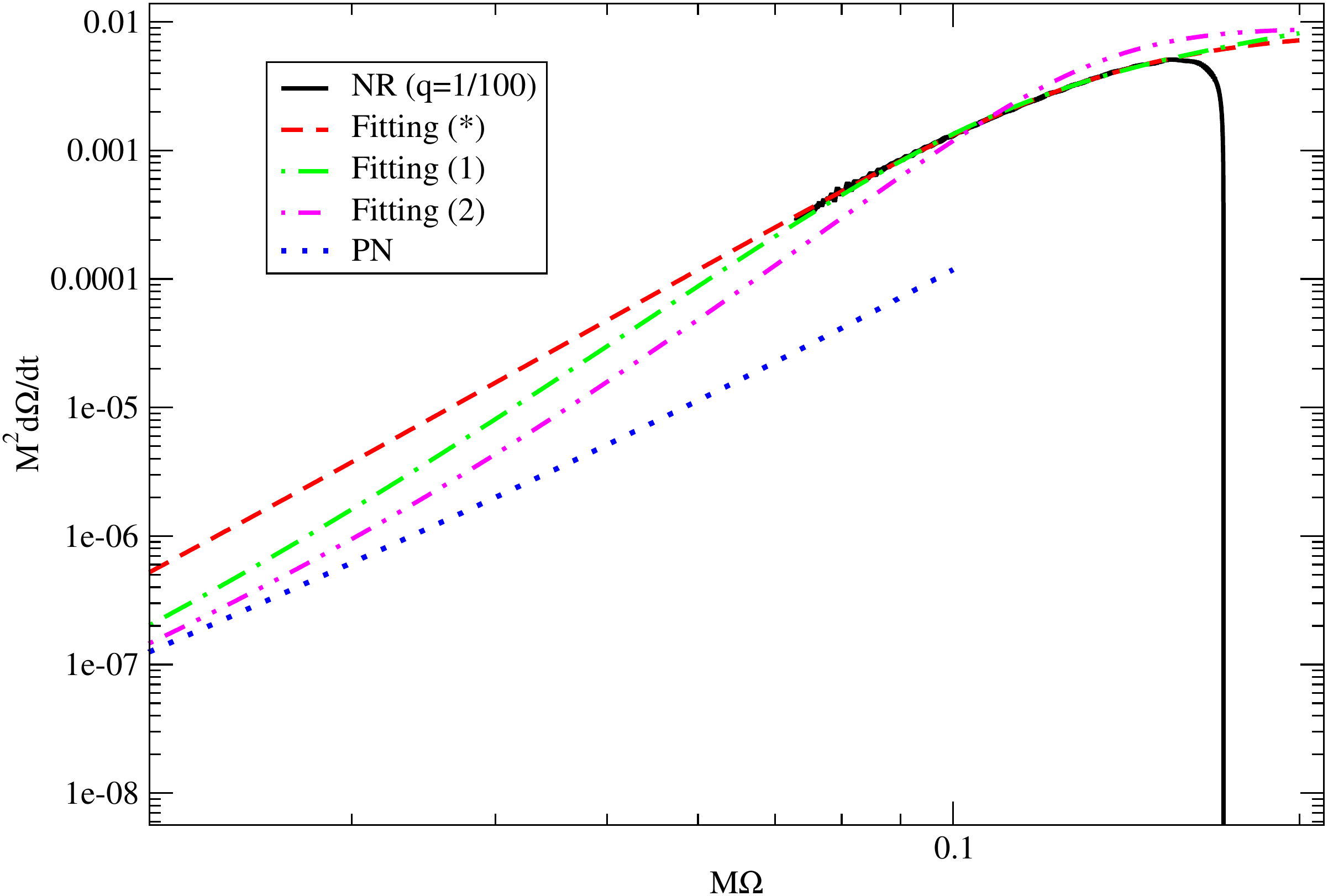}  
  \label{fig:OOt_100to1_PNP}
\end{figure}

For the $q=1/100$ case, which is shown in Fig.~\ref{fig:OOt_100to1_PNP},
we first found by a direct fit of all parameters that $\alpha<7$, which is not consistent with the PN approach. 
One possibility is that the trajectories are too short. If we were to
extend them to lower frequencies, we would expect that the slope
increases, leading to a larger $\alpha$.
Below, we try two alternate fitting methods in order to try to improve
the fit. 
This first  fit gives the best agreement with the numerical
trajectory so this is the one  we use in Sec.~\ref{sec:Results}
to calculate the gravitational waveforms. However, the second fit
reproduces the expected behavior at large separations, so we expect 
that it gives a better waveform in that regime.

Note that, as seen in Fig.~\ref{fig:OOt_100to1_PNP},
the end point of the NR trajectory, which corresponds to $R_{\rm Sch}=2M$,
has $d\Omega/dt<0$. This is a feature of Schwarzschild geodesics in
Schwarzschild coordinates (e.g., a quasi-circular
geodesic slightly inside the ISCO has $d\Omega/dt=-0.0113$).
Hence it appears that in the $q=1/100$ case the trajectories are much closer
 to geodesic motion than in the other cases.

Due to the fact that the $q=1/100$ simulations were very short,
obtaining accurate fitting parameters for the trajectory is 
error prone. We therefore consider several different techniques.
If we assume a simple fitting with respect to the mass ratio,
\bea
(A,\,\alpha,\,B,\,\beta) &=& c_0 \eta^{c_1} \,, 
\eea
for the fitting parameters
in the case of the $q=1/10$ and $q=1/15$ cases, 
we obtain
\bea
A &=& 0.797021 \, \eta^{-1.22855} \,,
\nonumber \\
\alpha &=& 5.26843\,  \eta^{-0.126379} \,,
\nonumber \\
B &=& 5.48252 \, \eta^{-0.160938} \,,
\nonumber \\
\beta &=& 6.80971 \, \eta^{-0.244245} \,. 
\label{eq:c0c1}
\eea
Then, inserting the symmetric mass ratio $\eta=10/121$ for 
$q=1/100$ gives
fitting parameters for $q=1/100$. We report these values on the
third line labeled $q=1/100$ in Table~\ref{tab:dOdt_AalpBbet_mT4}.
The curve obtained by these parameters is denoted by 
``Fitting (2)'' in Fig.~\ref{fig:OOt_100to1_PNP}. Note that these
parameters actually produce a relatively poor fit for the
trajectory.
On the other hand, if we assume $\alpha=8$ and $\beta=15$ and
fit the remaining parameters using the NR trajectory, we
obtain the ``Fitting (1)'' parameters (second line
labeled $q=1/100$ in Table~\ref{tab:dOdt_AalpBbet_mT4}). This fit
smoothly interpolates between the PN trajectory at small frequencies
and the start of the numerical trajectory at higher frequencies.
Since both the standard fitting procedure, and 
``Fitting(1)'' give reasonable fits, but different predictions,
in order to accurately determine the constant in Eqs.~(\ref{eq:c0c1}),
we need both more accurate and longer evolutions NR simulations of 
$q=1/100$.

\subsection{The orbital radius $R$ versus orbital frequency $\Omega$ in the NR coordinates}

In order to complete our description of the trajectories we need a
fitting function that will allow us to 
express the separation $R$ as a function of $\Omega$.
We use a function $R(\Omega)$ inspired
by the ADM-TT PN approach~\cite{Schafer:2009dq},
because the isotropic coordinates (used in the ADM-TT PN approach)
and the radial isotropic ``trumpet'' stationary $1+\log$ slices 
of the Schwarzschild spacetime 
are very similar~\cite{Kelly:2007uc,Tichy:2002ec}.

The fitting function based on the ADM-TT PN approach is given by 
\begin{widetext}
\bea
R &=& \frac{M}{(M\Omega)^{2/3}}
\,  \biggl( 1+ \left( -1+\frac{1}{3}\,\eta \right) (M\Omega)^{2/3}
+ \left( -\frac{1}{4}+{\frac {9}{8}}\,\eta+\frac{1}{9}\,{\eta}^{2} \right) (M\Omega)^{4/3}
\nonumber \\ && 
+ \left( 
-\frac{1}{4}-{\frac {1625}{144}}\,\eta
+{\frac {167}{192}}\,\eta\,{\pi }^{2}-\frac{3}{2}\,{\eta}^{2}
+{\frac {2}{81}}\,{\eta}^{3} \right) (M\Omega)^{2} \biggr)
/\left(1+a_0 (\Omega/\Omega_0)^{a_1}\right) + C  \,,
\label{eq:OmegaR_PNP}
\eea
\end{widetext}
where $R$ and $\Omega$ are in the NR coordinates,
and $a_0$, $a_1$ and $C$ are fitting parameters, and
 $M\Omega_0=(1/3)^{3/2}$.
Note that $a_1$ should be greater than $2$ to be consistent 
with the PN calculation.

Initially, we expected that $C$, which is the difference between
the NR radial coordinate and ``trumpet'' coordinate,
would be very small because this term is inconsistent
with the ADM-TT PN formula. However, we found that this term is
non-trivial for all three mass-ratio cases.

\begin{table}[t]
   \caption{
   The fitting parameters for the ``PN$+$'' (a modification of the ADM-TT PN calculation)
   in Eq.~(\ref{eq:OmegaR_PNP})
   for the relation between $R$ and $\Omega$ in the NR coordinates.
   }
   \label{tab:R_Ca0a1}
\begin{ruledtabular}
\begin{tabular}{c|ccc}
 Mass ratio & $C$ & $a_0$ & $a_1$  \\ \hline
 $q=1/10$ & $0.216953$ &  $0.513214$ & $4.68472$ 
 \\
 $q=1/15$ & $0.237427$ &  $0.600321$ & $4.57899$ 
 \\
 $q=1/100$ & $0.198137$ &  $0.923360$ & $5.29681$ 
\end{tabular}
\end{ruledtabular}
\end{table}

\begin{figure}[!h]
  \caption{
  The fitting of the orbital radius $R$ vs. orbital frequency $\Omega$in the NR coordinates
  for the $q=1/10$ case. 
  The solid (black), dashed (red) and dotted (blue)
  curves show the NR, fitting and ADM-TT PN evolutions, respectively.
  The fitting by using Eq.~(\ref{eq:OmegaR_PNP}) is valid up to $M\Omega = 0.175$.
  In the NR evolution, the end point of the solid (black) curve corresponds to $R_{\rm Sch}=2M$,
  and $M\Omega = 0.175$ is the NR orbital frequency around $R_{\rm Sch}=3M$,
assuming the NR coordinate system can be approximated by a ``trumpet'' 
  stationary $1+\log$ slice of the Schwarzschild spacetime.
  } 
  \includegraphics[width=3in]{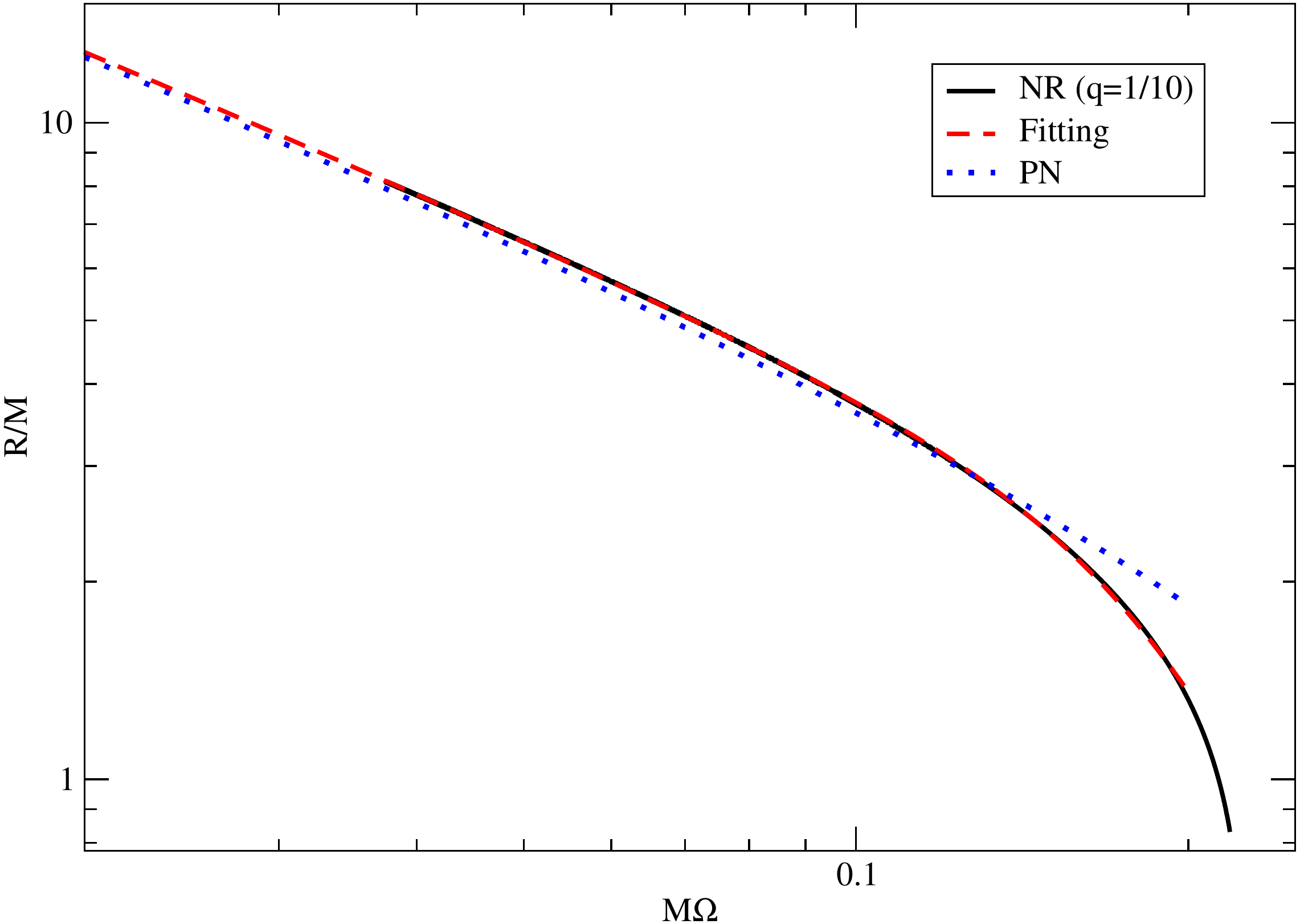}
  \label{fig:OmegaR_NR_10to1}
\end{figure}

\begin{figure}[!h]
  \caption{
  The fitting of the orbital radius $R$ vs. orbital frequency $\Omega$
  in the NR coordinates for the $q=1/15$ case. 
  The solid (black), dashed (red) and dotted (blue)
  curves show the NR, fitting and ADM-TT PN evolutions, respectively.
  The fitting by using Eq.~(\ref{eq:OmegaR_PNP}) is valid up to $M\Omega = 0.17$.
  In the NR evolution, the end point of the solid (black) curve corresponds to $R_{\rm Sch}=2M$,
  and $M\Omega = 0.17$ is the NR orbital frequency around $R_{\rm Sch}=3M$,
assuming the NR coordinate system can be approximated by a ``trumpet'' 
  stationary $1+\log$ slice of the Schwarzschild spacetime.
  } 
  \includegraphics[width=3in]{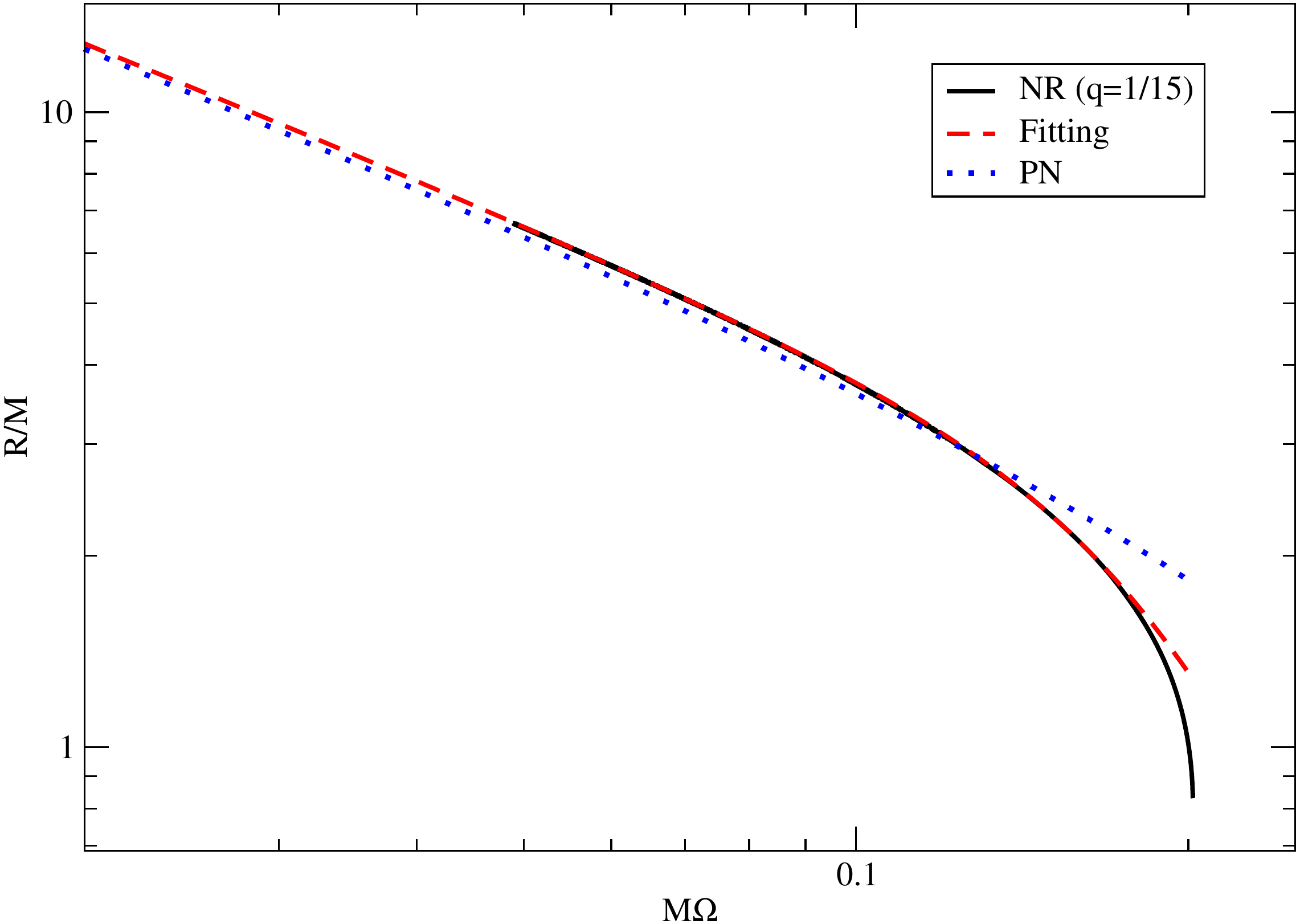}
  \label{fig:OmegaR_NR_15to1}
\end{figure}

\begin{figure}[!h]
  \caption{
  The fitting of the orbital radius $R$ vs. orbital frequency $\Omega$ in the NR coordinates
  for the $q=1/100$ case. 
  The solid (black), dashed (red) and dotted (blue)
  curves show the NR, fitting and ADM-TT PN evolutions, respectively.
  The fitting by using Eq.~(\ref{eq:OmegaR_PNP}) is valid up to $M\Omega = 0.15$.
  In the NR evolution, the end point of the solid (black) curve corresponds to $R_{\rm Sch}=2M$,
  and $M\Omega = 0.15$ is the NR orbital frequency around $R_{\rm Sch}=3M$,
  assuming the NR coordinate system can be approximated by a ``trumpet'' 
  stationary $1+\log$ slice of the Schwarzschild spacetime.
  } 
  \includegraphics[width=3in]{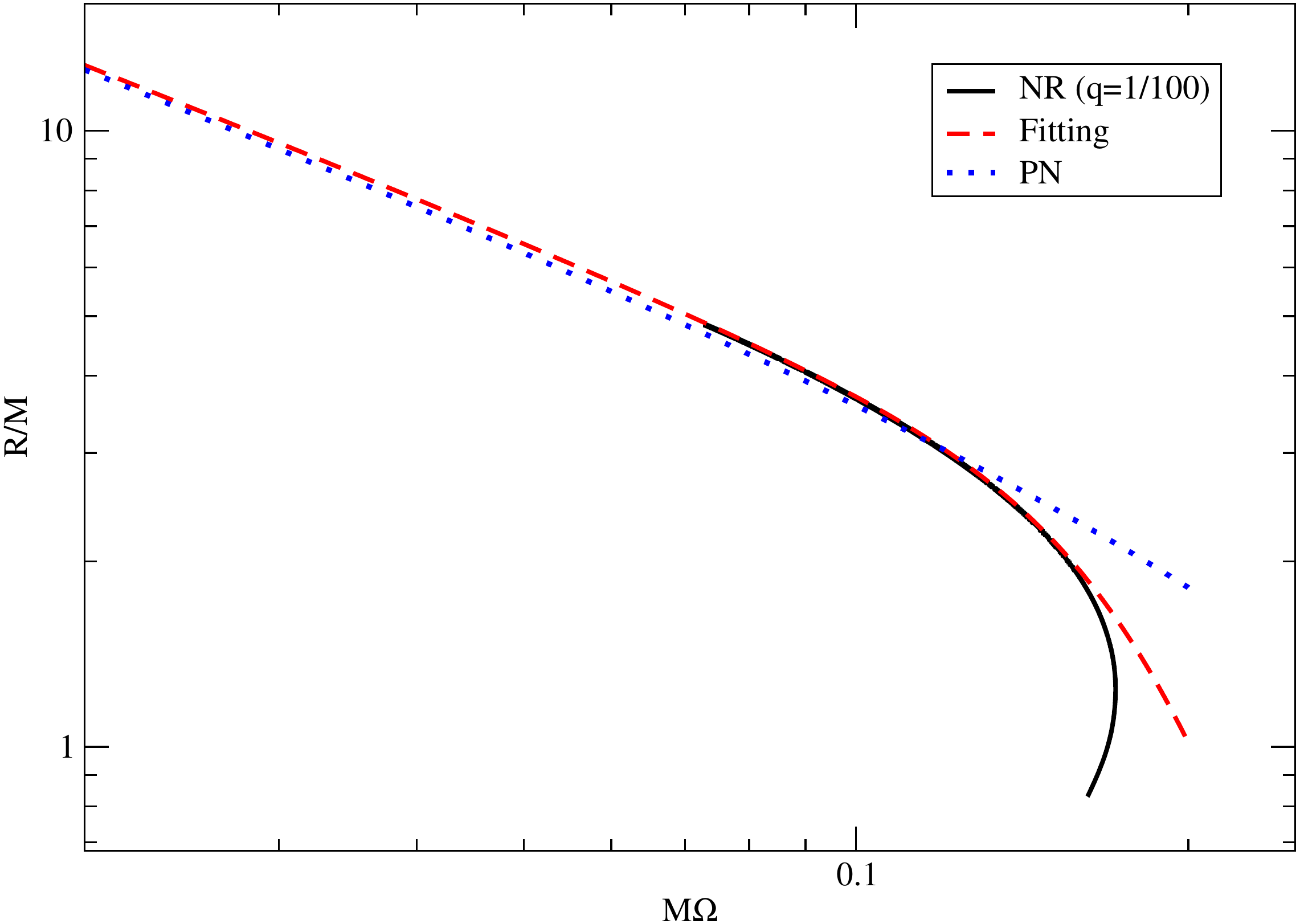}
  \label{fig:OmegaR_NR_100to1}
\end{figure}

The results of the fits are shown in
Figs.~\ref{fig:OmegaR_NR_10to1}-\ref{fig:OmegaR_NR_100to1} for
$q=1/10$, $q=1/15$, and $q=1/100$, respectively, and the
 fitting parameters are summarized in Table~\ref{tab:R_Ca0a1}.
We see  from Table~\ref{tab:R_Ca0a1} that the values of $a_1$ for all cases are
consistent with the current PN calculation ($a_1>2$), 
and find a non-negligible $C \sim 0.2$ for all three cases.
Due to this $C$ contribution, there is a finite difference between
the orbital radius evaluated in the ADM-TT PN approach
and that by the fitting function in Eq.~(\ref{eq:OmegaR_PNP}) 
even for smaller $M\Omega$.

As a result of the above fitting analysis, 
it is necessary to consider a radial transformation
to remove the offset $C$ between the NR coordinates 
($R_{\rm NR} = R$ for Eq.~(\ref{eq:OmegaR_PNP})) 
and the ``trumpet'' coordinates,
\bea
R_{\rm NR} \to R_{\rm Log}=R_{\rm NR}-C \,.
\label{eq:Ctrans}
\eea
After removing this offset, we then transform to the standard
Schwarzschild coordinates,
where the explicit time and radial coordinate transformations,
\bea
(T_{\rm Log},\,R_{\rm Log}) \to (T_{\rm Sch},\,R_{\rm Sch}) \,,
\eea
are given in Ref.~\cite{Brugmann:2009gc}.
Here, we assume $T_{\rm NR}=T_{\rm Log}$.
We note that after applying the $C$ transformation discussed above,
we find that the frequency when $R_{\rm Sch}=3M$ is given by
$M\Omega=0.161$, $0.158$ 
and $0.152$ for the $q=1/10$,  $q=1/15$
and $q=1/100$ cases, respectively.

Note that this procedure of fitting NR trajectories directly in the NR
coordinates and then
transforming to Schwarzschild coordinates, appears simpler
 than the alternative of evolving the perturbations directly in the
``trumpet'' coordinate system.
 We nevertheless
study this alternative in the Appendix \ref{app:Trumpet}, where we
provide the sourceless part of the perturbative equation and study its
characteristic speeds in ``trumpet'' coordinates.

\section{Results}\label{sec:Results}

We obtain an approximation to $\Omega(T_{\rm NR})$ by integrating
Eq.~(\ref{eq:OOt_PNP}). We then obtain $R(T_{\rm NR})$ using
Eq.~(\ref{eq:OmegaR_PNP}) and the approximate $\Omega(T_{\rm NR})$.
The initial orbital phase is not fixed, and we set it here such that
the matching between the perturbative and numerical waveforms
are maximized.

Note that part of the difference between the NR waveform
and the perturbative waveforms is due to eccentricity in the numerical
trajectories. 
This affects the time step of the orbital evolution through
\bea
dt &=& \left( \frac{d\Omega}{dt}\right)^{-1} d\Omega \,.
\eea
This may help explain why in the $q=1/10$ case, we observe a slower time evolution
in the fitting trajectory than the NR evolution.

In order to generate the final plunge trajectory and regularize
the
behavior of the source near the horizon in the SRWZ formalism~\cite{Lousto:2010qx},  we attach the model for the
trajectories to a plunging geodesic in a continuous way.
To do this, we choose a matching radius $R_M\sim 3M$ in Schwarzschild
coordinates (the light ring) and then determine the energy and angular
momentum of the trajectory at that point. We then complete the
trajectory from $R\sim R_M$ to $R\sim2M$ 
using the geodesic with the same energy and angular
momentum.

When we set the matching radius to $R_{\rm M} = 3M$, 
we find that the energy $E$ and angular momentum $L$
in the Schwarzschild coordinates are given by 
\bea
E 
&=& \frac{1}{3}\, u^t \,,
\nonumber \\ 
L
&=& 9\, M^2 \,u^{\phi} \,,
\label{eq:evalCons}
\eea
where $u^{\mu}=dx^{\mu}/d\tau$ is the four velocity,
and we focus on equatorial orbits. 
To evaluate $u^{t}$ in the above equation, we use
\bea
g_{\mu\nu} u^{\mu} u^{\nu} = -1 \,,
\label{eq:Ut}
\eea
where $g_{\mu\nu}$ denotes the Schwarzschild metric, 
and at $R_{\rm M} = 3M$ this becomes
\bea
u^{t} &=&  
\biggl[ 
 \frac{1}{3}
- 3 (\dot {R}_{\rm Sch}(T_{\rm Sch}))^2
\nonumber \\ && \quad 
- 9\,M^2 (\dot {\Phi}_{\rm Sch}(T_{\rm Sch}))^2 
\biggr]^{-1/2}
 \,.
\eea
Here, $\dot {R}_{\rm Sch}=u^r/u^t=dR_{\rm Sch}/dT_{\rm Sch}$ 
and $\dot {\Phi}_{\rm Sch}=u^\phi/u^t=d\Phi_{\rm Sch}/dT_{\rm Sch}$
are the velocity components
obtained from the numerical data
after the coordinate transformation from the ``trumpet'' coordinates
to the Schwarzschild coordinates.
We also use Eq.~(\ref{eq:Ut}) to calculate
$u^{t}$ in the source term of the SRWZ formalism.

In the SRWZ formalism, the gravitational waveforms 
at a sufficiently distant location $R_{\rm Obs}$ is given by 
\bea
&& \frac{R_{\rm Obs}}{M}
\left(
h_{+} - i\,h_{\times} \right)
=
\sum_{\ell m} \frac{\sqrt{(\ell-1)\ell(\ell+1)(\ell+2)}}{2M} 
\nonumber \\ && \qquad \qquad \qquad \times 
\left(
\Psi_{\ell m}^{\rm (even)} - i\, \Psi_{\ell m}^{\rm (odd)}
\right)
{}_{-2}Y_{\ell m}
\,.
\label{eq:SRWZwaves}
\eea
We use the normalized waveform, i.e.,
the expression in the right hand side of the above equation to
calculate
the wave amplitude.

On the other hand, to convert the NR $\psi_4$, which is extrapolated to infinity 
via Eq.~(\ref{eq:asymtpsi4ext}), to the waveform $h$,
we use the ``pyGWAnalysis'' code in {\sc EinsteinToolkit}~\cite{einsteintoolkit}
(see Ref.~\cite{Reisswig:2010di} for details).
Then, we calculate the match between the NR and SRWZ waveforms 
in the advanced LIGO (Zero Det, High Power) noise curve~\cite{AdvLIGONC}. 
In Table~\ref{tab:match_allcases}, we summarize the match 
for the three mass ratios.
In the following subsections, we treat each mass ratio case separately.

\begin{table} [ht]
\centering 
  \caption{The match between the NR and SRWZ waveforms
  for various total masses using the advanced LIGO noise curve.
  For the smaller masses, the entire waveform is in the advanced
  LIGO band, while for the larger masses, only the final ringdown part
  of the waveform is in band. For these larger masses,
  we integrate over frequencies $f \geq 10$Hz.
  Here we only used the $(\ell=2,\,m=2)$ mode to calculate the match.
}
\label{tab:match_allcases}
  \begin{ruledtabular}
  \begin{tabular}{c|ccc}
   Total mass ($M_{\odot}$) & $q=1/10$ & $q=1/15$ & $q=1/100$ \\
   \hline
      100 & 0.983019 & 0.982900 & 0.995162 \\
   \hline
      200 & 0.992766 & 0.992704 & 0.995216 \\
   \hline
      300 & 0.996219 & 0.996514 & 0.995486 \\
   \hline
      400 & 0.996781 & 0.997734 & 0.995513 \\
   \hline
      500 & 0.996844 & 0.997973 & 0.995474
  \end{tabular}
  \end{ruledtabular}
\end{table}

\subsection{The $q=1/100$ case}

For the $q=1/100$ case,
the fitting formulas in Eqs.~(\ref{eq:OOt_PNP}) and (\ref{eq:OmegaR_PNP}) 
are valid for orbital frequencies up to  $M\Omega = 0.15$ in the ``trumpet'' coordinates. 
Here, we extend them beyond $M\Omega = 0.15$ by attaching these
trajectories to a plunging geodesic at $R_M$ (as explained above).

Ideally, we would like to use as much of the fitting trajectory
as possible. However, when
attempting to use the  fitting trajectory to calculate the 4-velocity at small
radii,
we can not evaluate $u^t$ 
for $R_{\rm Sch} \leq 2.12M$ from Eq.~(\ref{eq:Ut}) because the
approximation that the background metric is Schwarzschild in
``trumpet'' coordinates breaks
down and the vector $(1, \dot {R}_{\rm Sch}, 0, \dot {\Phi}_{\rm Sch})$ becomes spacelike
in the background metric (but not in the  numerical
metric). This breakdown can be thought to arise from the interaction
of the singular part of the conformal factor due to the smaller BH
(when it is close to horizon of the larger one) 
with the singular part of the conformal factor due to the larger BH.
This, in turn, changes $\tilde \Gamma^i$, and therefore the gauge,
when
the two BHs approach each other.
Also, since we see some delay in the phase evolution
for a small matching radius near the horizon,
we set $R_{\rm M} \sim 3M$ for the start of the geodesic approximation.
At this matching radius, $E$ and $L$ for the Schwarzschild geodesic 
are evaluated from the orbital separation and three velocity 
using Eq.~(\ref{eq:evalCons}). We find
\bea
E &=& 0.93942436 \,, 
\nonumber \\ 
L/M &=& 3.46184226 \,.
\eea
In the SRWZ formalism, we set the non-dimensional spin parameter
$\alpha=0.033$ (see Table~\ref{tab:remnant}).
Although this value is obtained in the NR simulation,
we can also obtain it from an empirical formula~\cite{Lousto:2009mf}.

\begin{figure}[!h]
  \caption{
  The amplitude (TOP) and phase (BOTTOM) of the $(\ell=2,\,m=2)$ mode of $h$ for the $q=1/100$ case.
  The waveforms were translated in time such that the maximum in the
  amplitude occurs at $t=0$ and the phases were adjusted by a constant
  offset such that $\phi=0$ at $t=-75M$.
  The solid (black) and dashed (red) curves show the NR and SRWZ waveforms, respectively.
  } 
  \includegraphics[width=3in]{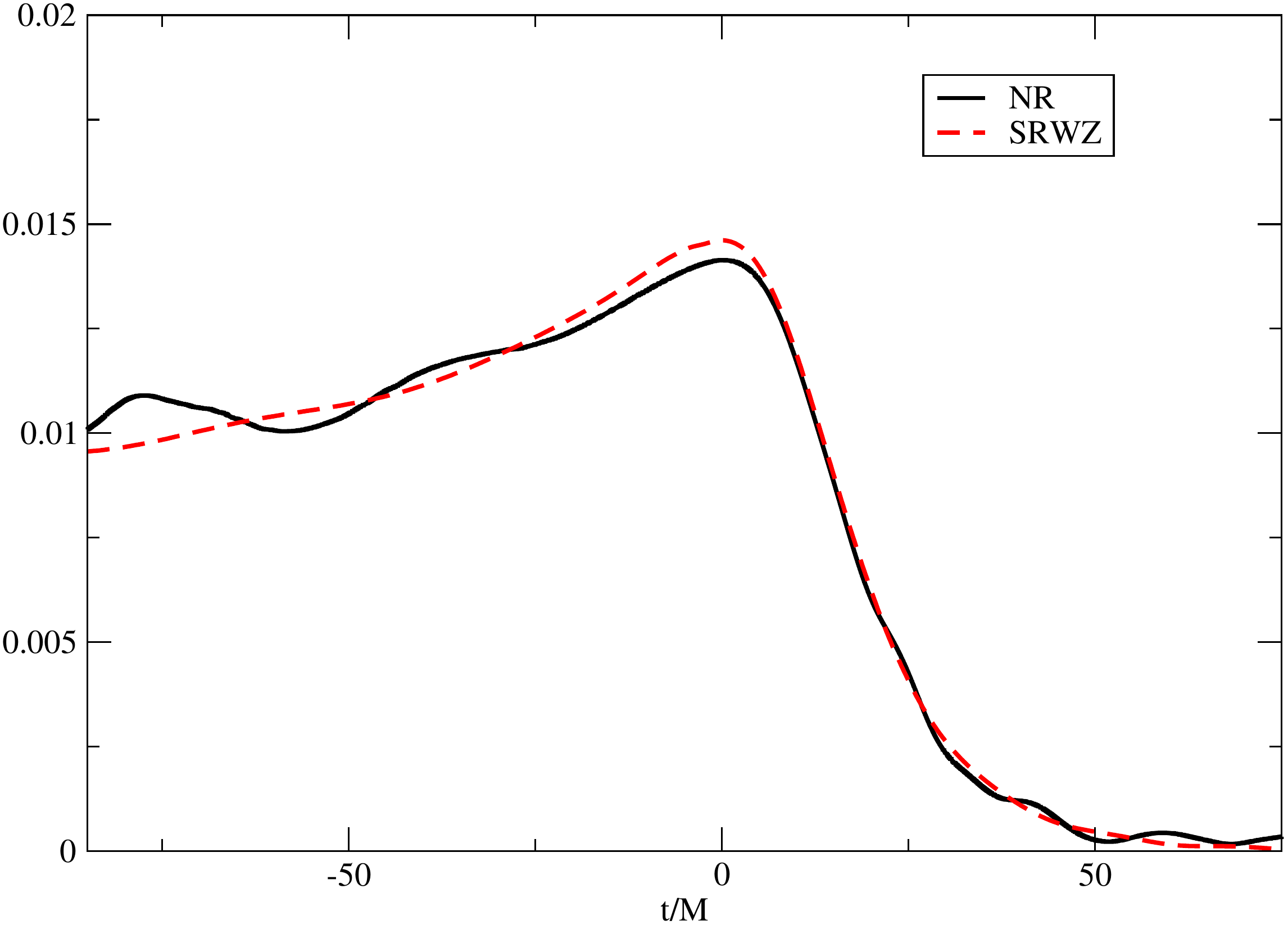}
  \includegraphics[width=3in]{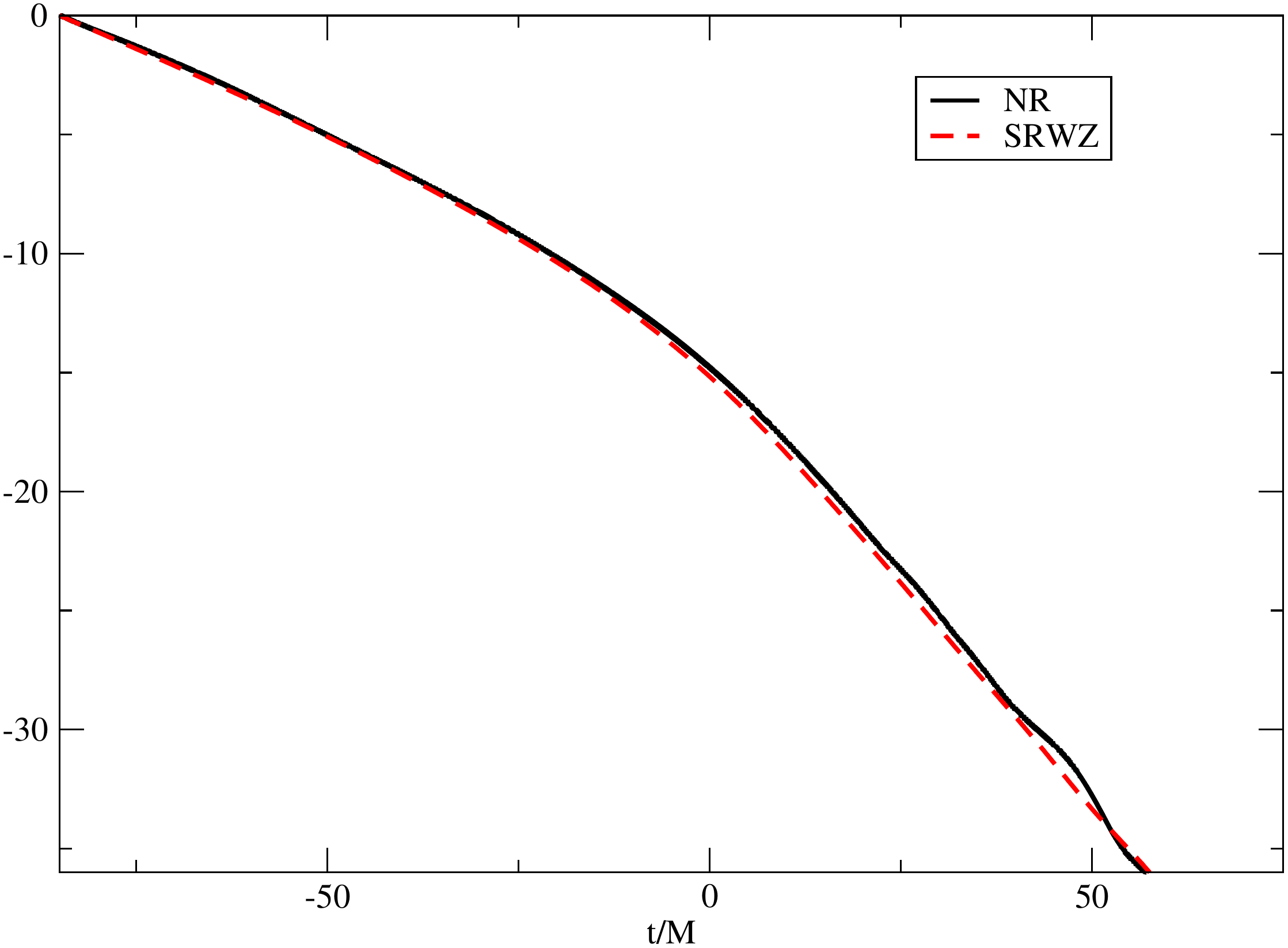}
  \label{fig:comparison_AP_100to1}
\end{figure}

We next calculate the match in the advanced LIGO noise curve.
For the integration in the frequency domain,
we consider gravitational wave frequencies $M\Omega_{22} \geq 0.15$.
For a total mass $M=484M_{\odot}$,
where we have the lower frequency cut off (corresponding to the start
of the numerical waveform) at $f_{\rm low}=10.01$Hz, 
the match between the NR and SRWZ waveforms is 
\bea
{\cal M}_{22} &=& 0.995477 \,,
\eea
for the $(\ell=2,\,m=2)$ mode.
The amplitude and phase evolutions for the NR and SRWZ waveforms 
are shown in Fig.~\ref{fig:comparison_AP_100to1}.
To obtain this figure, we translated the waveforms so that
the maximum in the amplitude occurs at $t=0$ and
used the freedom in the phase to set it to zero
at $t=-75M$.

\begin{figure}[!h]
  \caption{
  The amplitude (TOP) and phase (BOTTOM) of the $(\ell=2,\,m=1)$ mode of $h$ for the $q=1/100$ case.
  The waveforms were translated in time such that the maximum in the
  amplitude occurs at $t=0$ and the phases were adjusted by a constant
  offset such that $\phi=0$ at $t=-75M$.
  The solid (black) and dashed (red) curves show the NR and SRWZ waveforms, respectively.
  } 
  \includegraphics[width=3in]{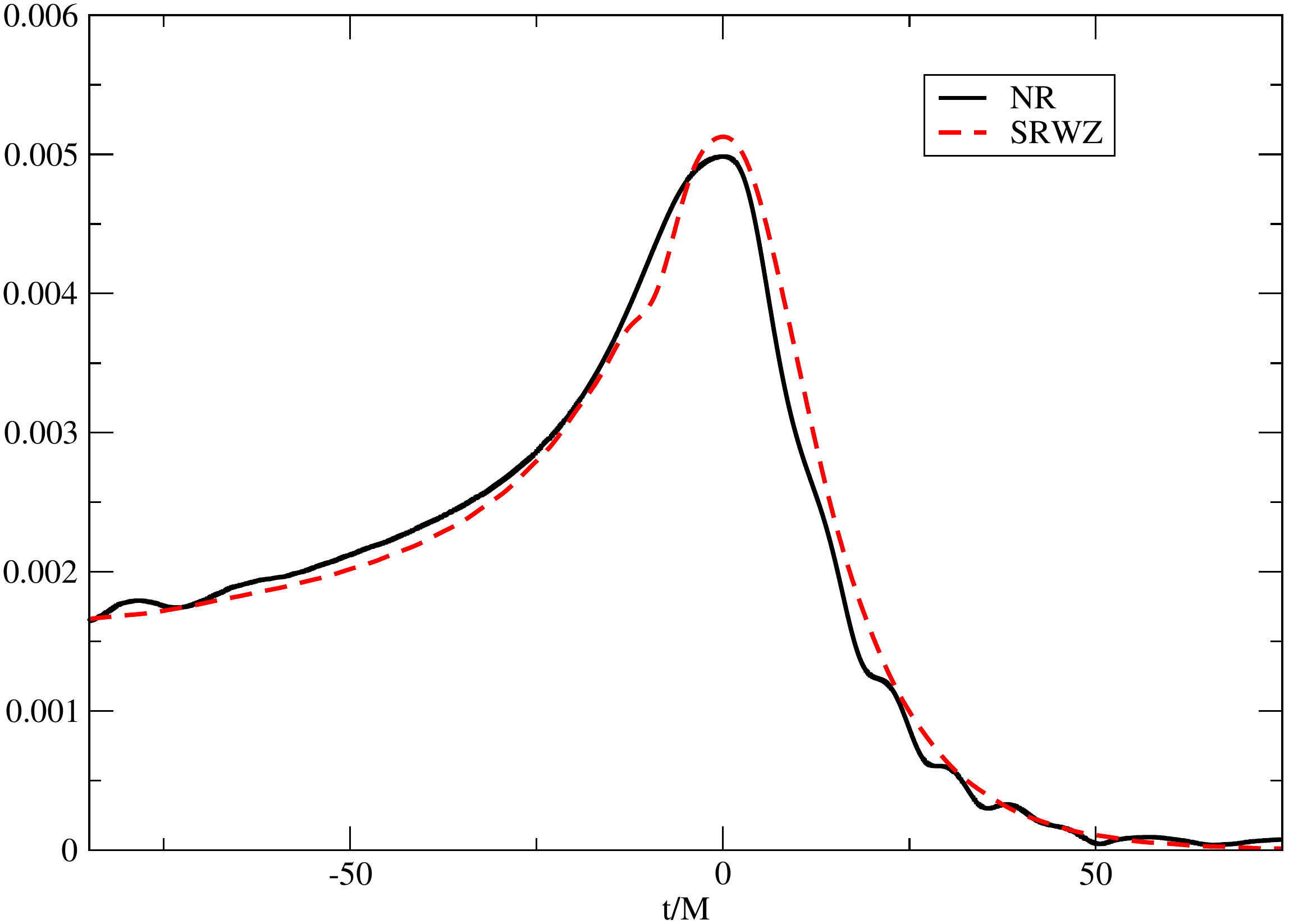}
  \includegraphics[width=3in]{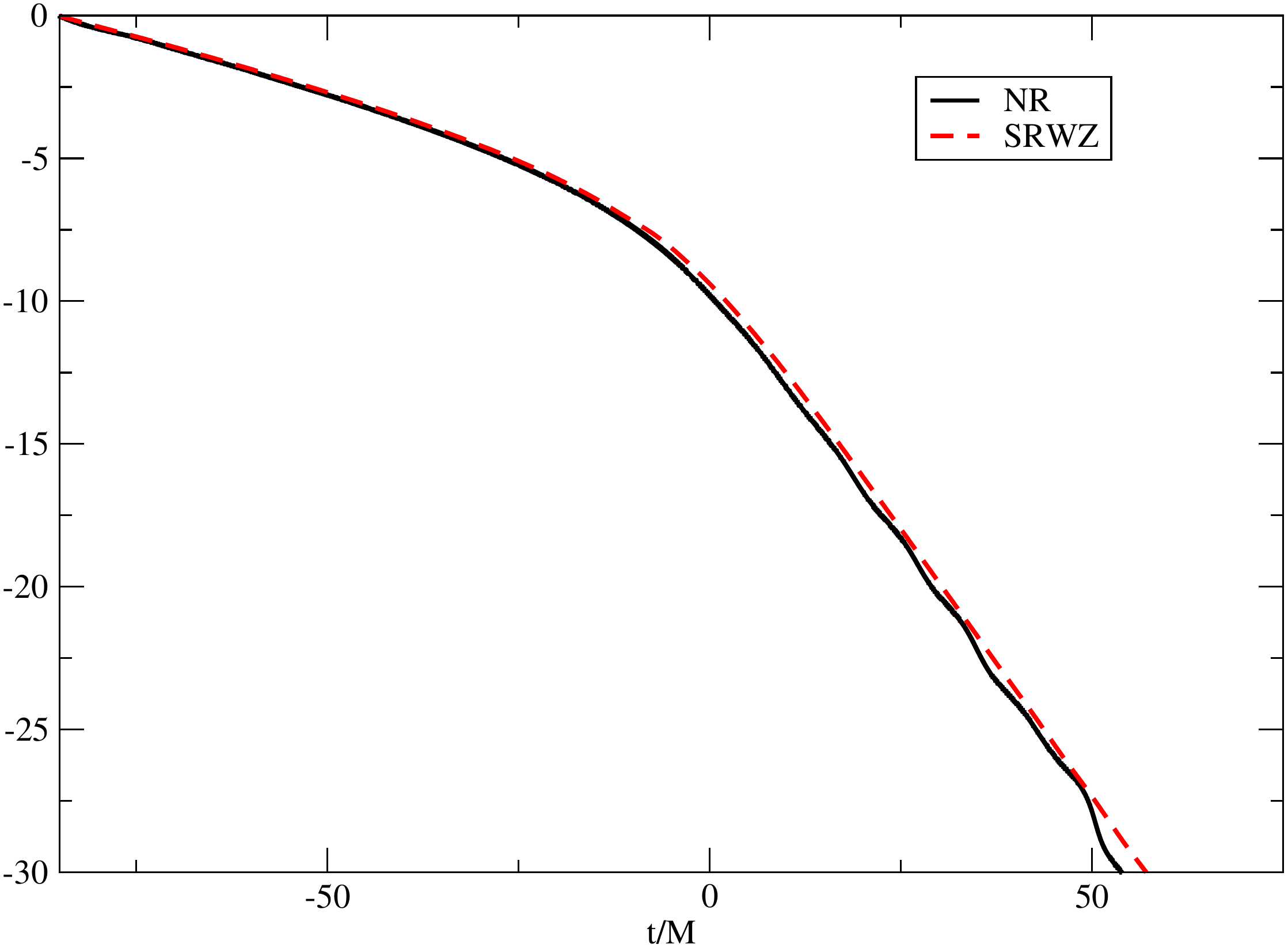}
  \label{fig:comparison_AP_100to1_21}
\end{figure}

The $(\ell=2,\,m=1)$ mode of $h$ is the leading 
contribution from the odd parity in the RWZ calculation.
The amplitude and phase evolutions for this mode 
are shown in Fig.~\ref{fig:comparison_AP_100to1_21}.
After adding a time translation so that the maximum in the amplitude
occurs at $t=0$ and a phase translation so that the phase $\phi=0$ is
zero at $t=-75M$, 
we still see a small difference between the phases.
For a total mass of $M=242M_{\odot}$ (this mass is chosen
such that the lowest frequency is just in the advanced LIGO band 
$f_{\rm low}=10.01$Hz), 
the match between the NR and SRWZ waveforms for this mode is given by
\bea
{\cal M}_{21} &=& 0.957966 \,.
\eea

\begin{figure}[!h]
  \caption{
  The amplitude (TOP) and phase (BOTTOM) of the $(\ell=3,\,m=3)$ mode of $h$ for the $q=1/100$ case.
  The waveforms were translated in time such that the maximum in the
  amplitude occurs at $t=0$ and the phases were adjusted by a constant
  offset such that $\phi=0$ at $t=-75M$.
  The solid (black) and dashed (red) curves show the NR and SRWZ waveforms, respectively.
  } 
  \includegraphics[width=3in]{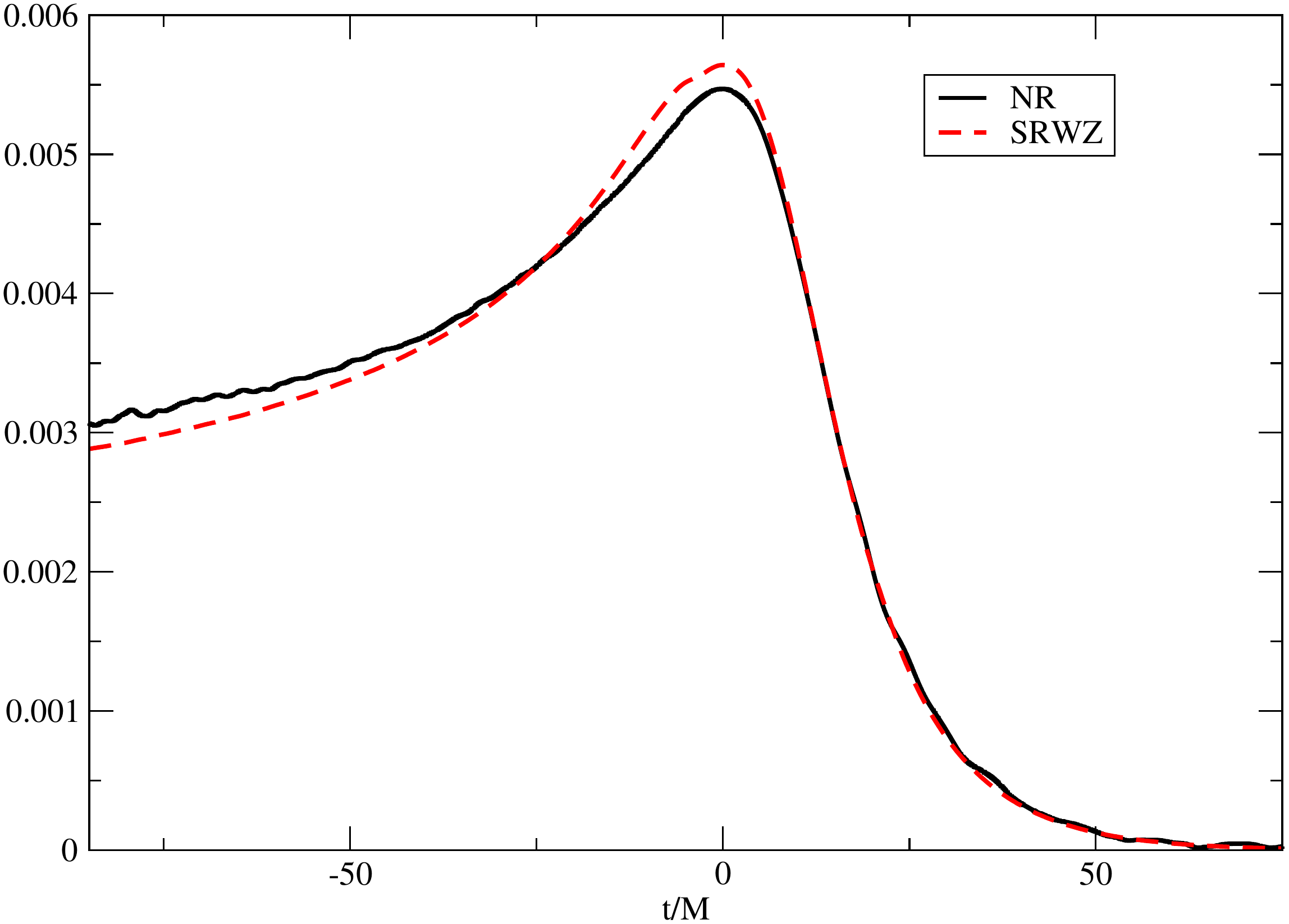}
  \includegraphics[width=3in]{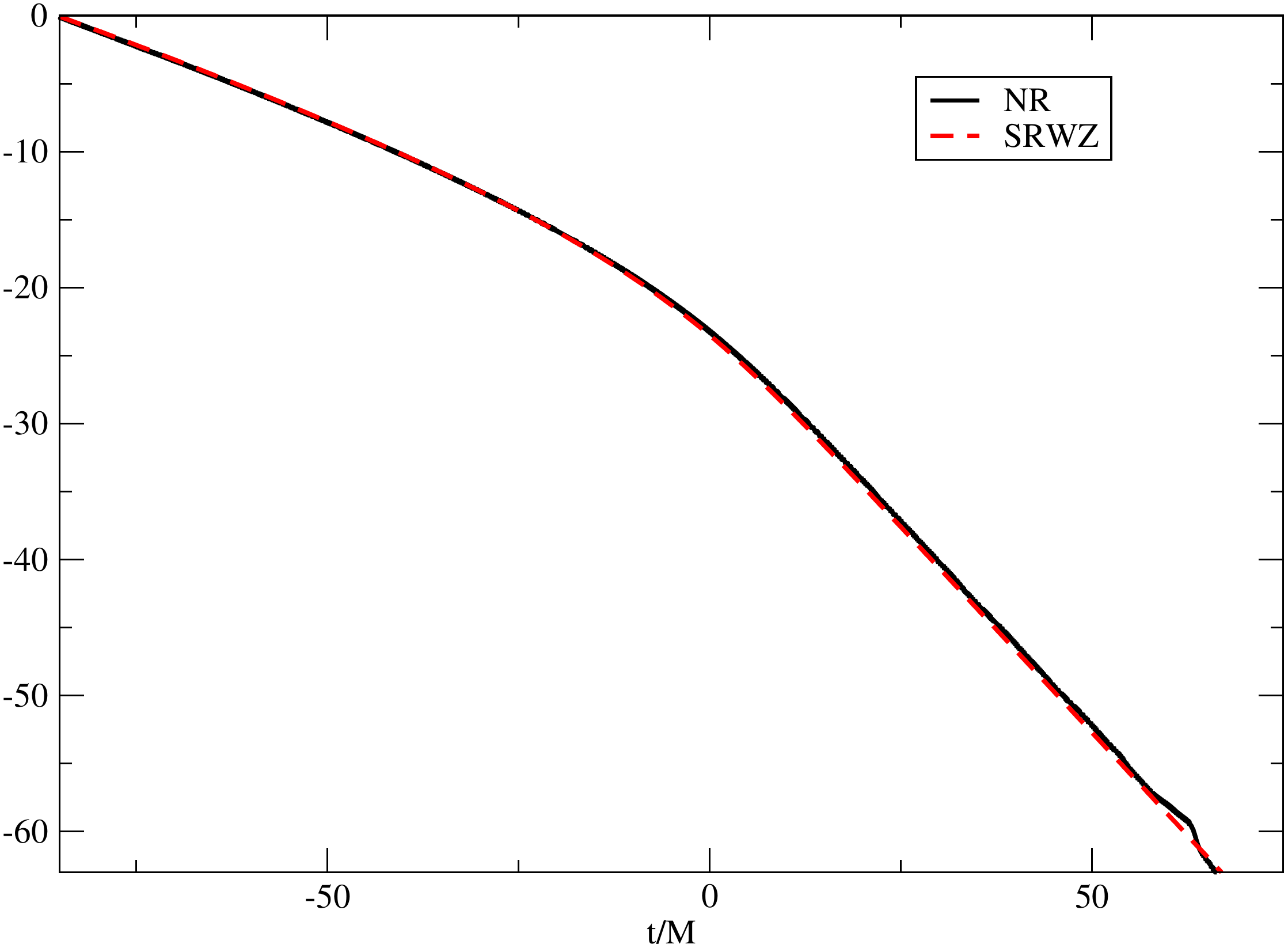}
  \label{fig:comparison_AP_100to1_33}
\end{figure}

For the $(\ell=3,\,m=3)$ mode of $h$,
the amplitude and phase evolutions are shown in Fig.~\ref{fig:comparison_AP_100to1_33}.
For $M=726M_{\odot}$ (mass chosen so that $f_{\rm low}=10.01$Hz),
 the match for this
mode is
\bea
{\cal M}_{33} &=& 0.993895 \,.
\eea

\subsection{The $q=1/10$ case}

For the $q=1/10$ case, 
the fitting formulae is valid up to larger orbital frequencies
than for the $q=1/100$ case.
Nevertheless, $u^t$ can not be evaluated 
for $R_{\rm Sch} \leq 2.26M$ from Eq.~(\ref{eq:Ut}).
Therefore, we need to start the geodesic approximation at
radii larger than this,
and again we set the matching radius $R_{\rm M} \sim 3M$.
We find that the energy and angular momentum for the Schwarzschild geodesic 
are given by
\bea
E &=& 1.02546489 \,, 
\nonumber \\ 
L/M &=& 3.95926146 \,,
\eea
at the matching radius.
Not that the energy is $E>1$ because
we simply project the orbital quantities on the Schwarzschild spacetime
by using Eqs.~(\ref{eq:evalCons}) and (\ref{eq:Ut}) (this is a strong
indication that the motion of the $q=1/10$ binary is non-geodesic
even close to the larger BH).
In the SRWZ formalism, we set the non-dimensional spin parameter to
$\alpha=0.26$, consistent with the measured spin from the numerical
simulation (see Table~\ref{tab:remnant}).

\begin{figure}[!h]
  \caption{
  The amplitude (TOP) and phase (BOTTOM) of the $(\ell=2,\,m=2)$ mode of $h$ for the $q=1/10$ case.
  The waveforms were translated in time such that the maximum in the
  amplitude occurs at $t=0$ and the phases were adjusted by a constant
  offset such that $\phi=0$ at $t=-830M$.
  The solid (black) and dashed (red) curves show the NR and SRWZ waveforms, respectively.
  } 
  \includegraphics[width=3in]{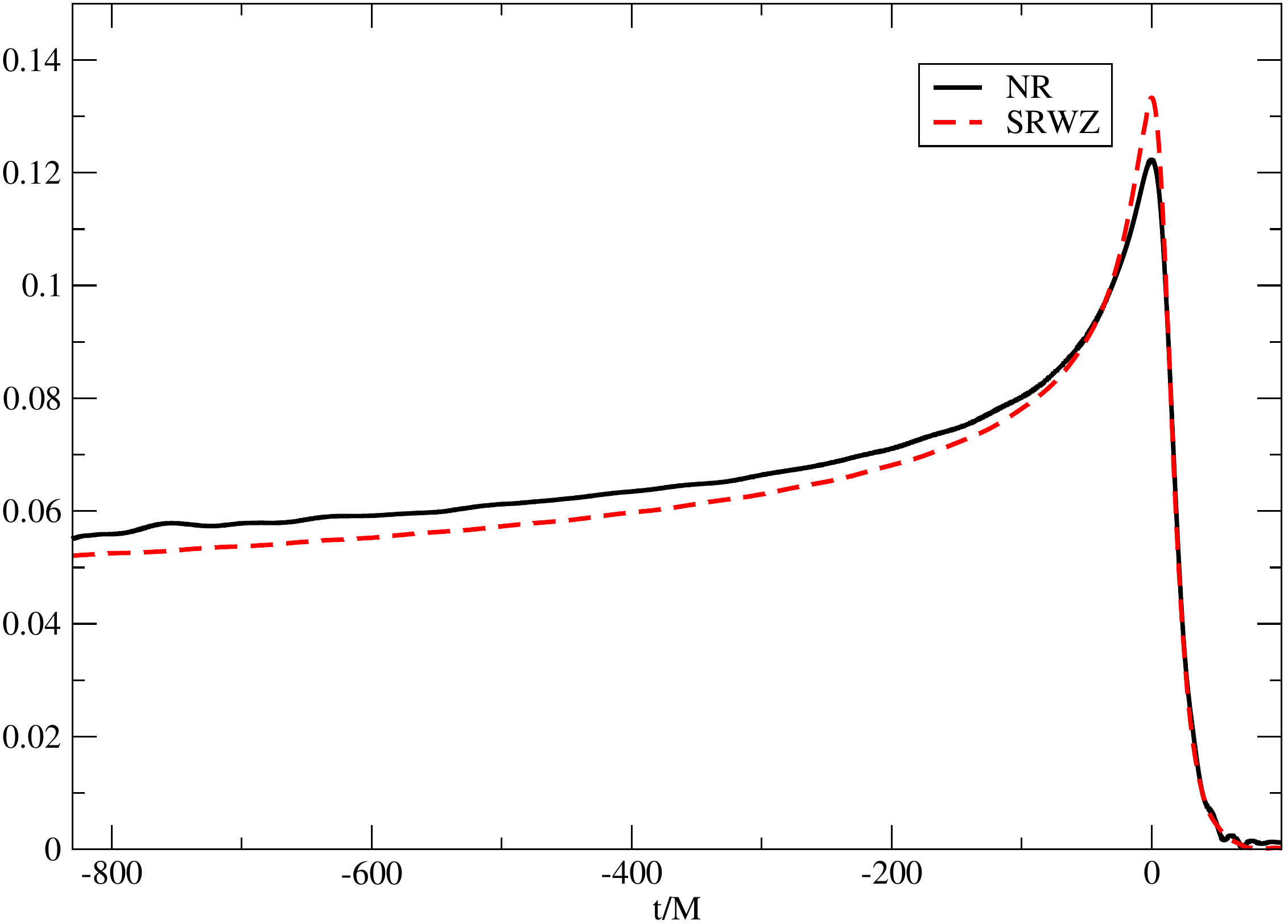}
  \includegraphics[width=3in]{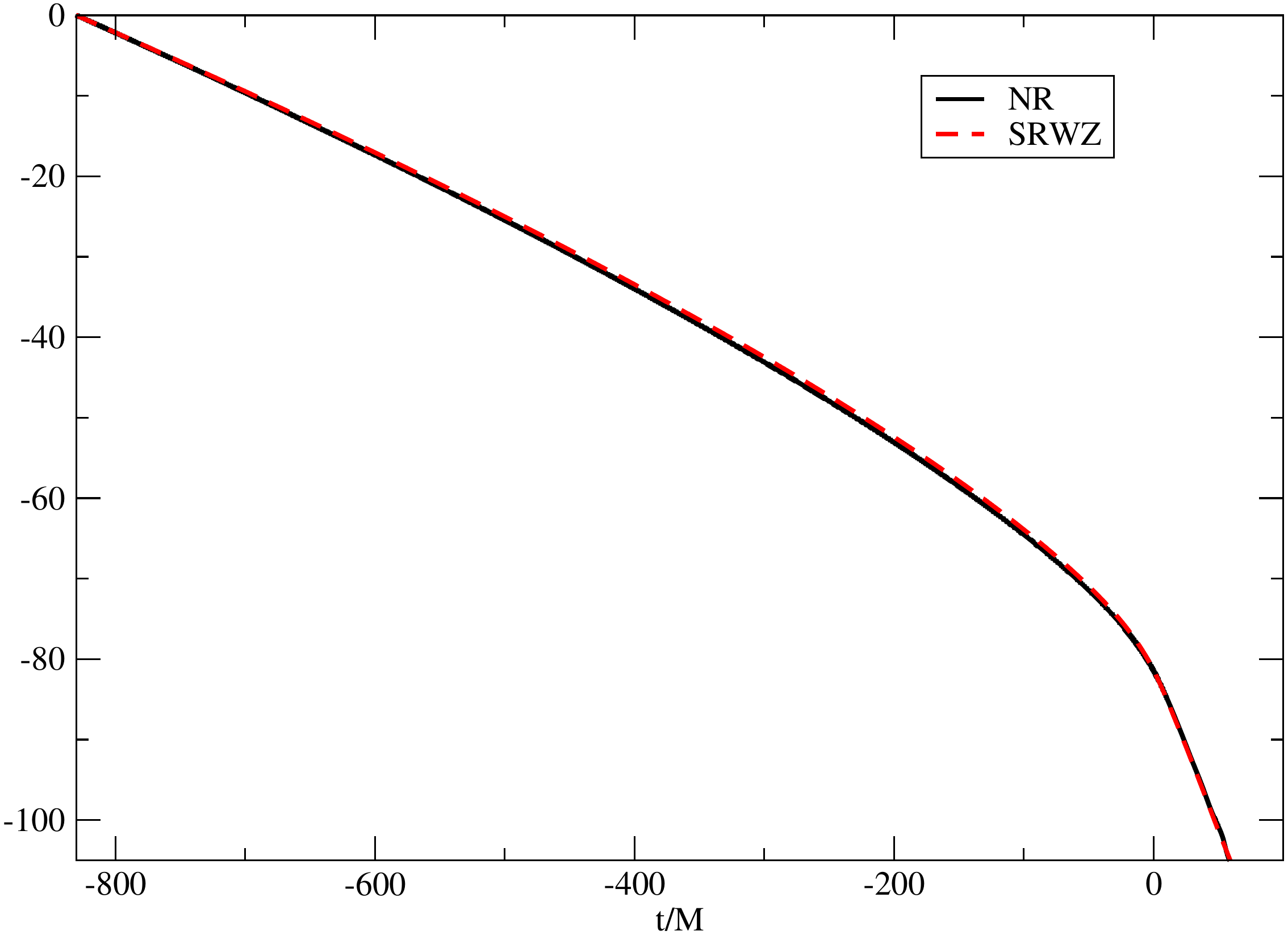}
  \label{fig:comparison_AP_10to1}
\end{figure}

The amplitude and phase evolutions of the $(\ell=2,\,m=2)$ mode of $h$
are shown in Fig.~\ref{fig:comparison_AP_10to1}.
While the phases of the NR and SRWZ waveforms
are in good agreement, there are relatively
large differences in the amplitudes both for the inspiral and merger phases.
We will discuss the amplitude differences in more detail  in Sec.~\ref{subsec:amp_diff}.

We consider the match calculation for $M\Omega_{22} \geq 0.075$ 
and total mass $M=242M_{\odot}$, i.e., 
the lower frequency cut off at $f_{\rm low}=10.01$Hz.
In this case the match between the NR and SRWZ waveforms is 
\bea
{\cal M}_{22} &=& 0.994669 \,,
\eea
for the $(\ell=2,\,m=2)$ mode in the advanced LIGO noise curve.

\begin{figure}[!h]
  \caption{
  The amplitude (TOP) and phase (BOTTOM) of the $(\ell=2,\,m=1)$ mode of $h$ for the $q=1/10$ case.
  The waveforms were translated in time such that the maximum in the
  amplitude occurs at $t=0$ and the phases were adjusted by a constant
  offset such that $\phi=0$ at $t=-830M$.
  The solid (black) and dashed (red) curves show the NR and SRWZ waveforms, respectively.
  } 
  \includegraphics[width=3in]{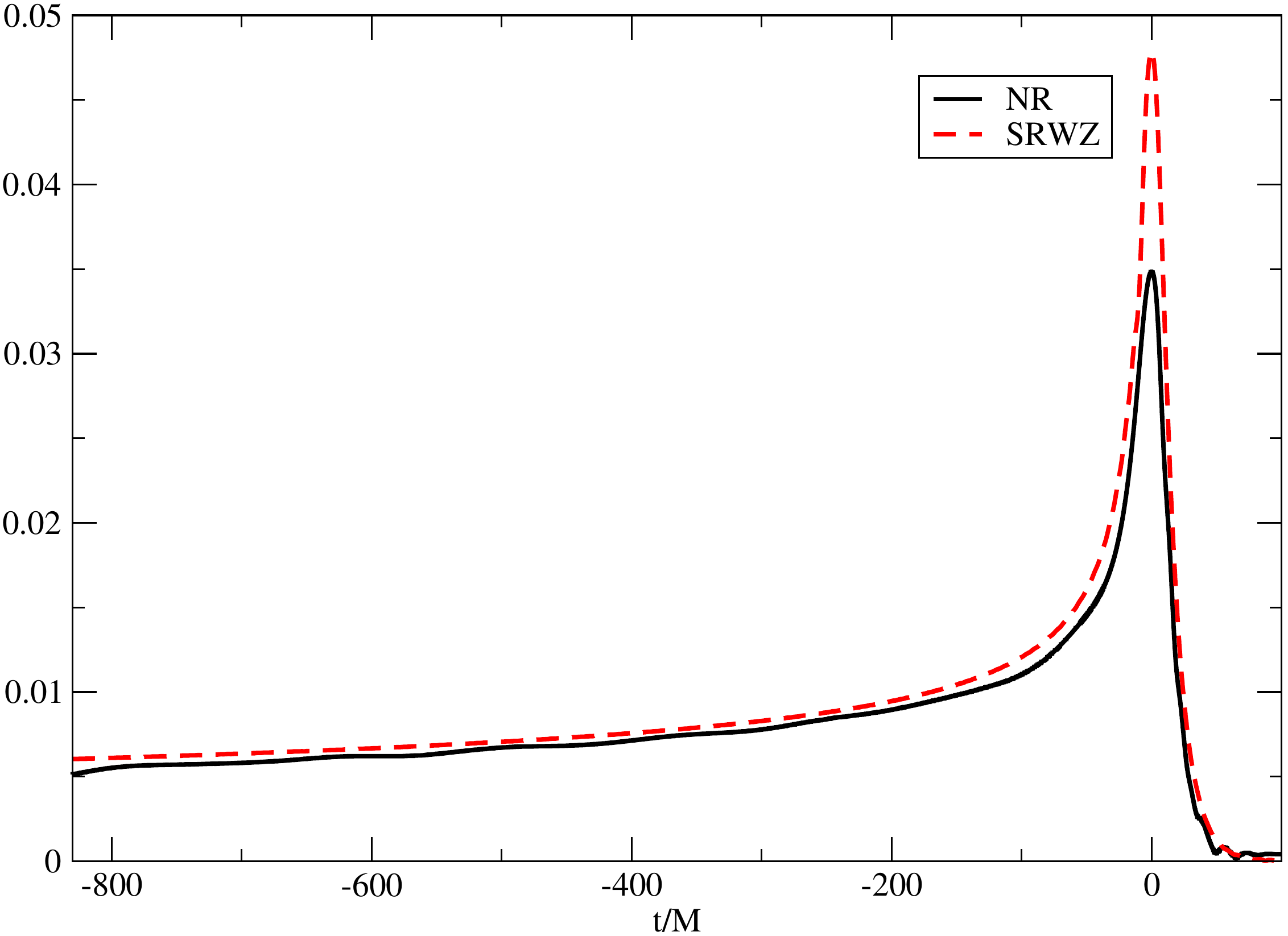}
  \includegraphics[width=3in]{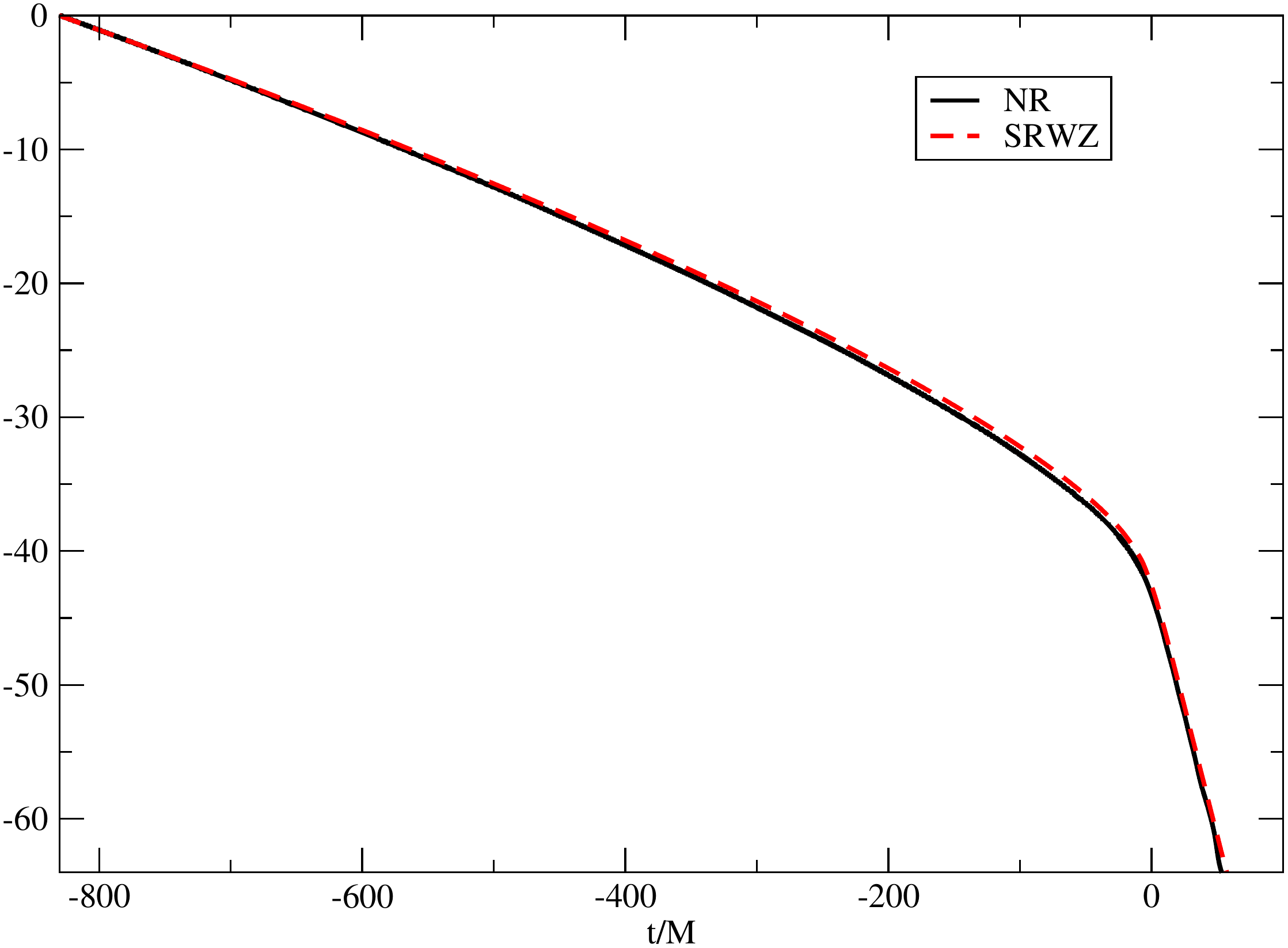}
  \label{fig:comparison_AP_10to1_21}
\end{figure}

\begin{figure}[!h]
  \caption{
  The amplitude (TOP) and phase (BOTTOM) of the $(\ell=3,\,m=3)$ mode of $h$ for the $q=1/10$ case.
  The waveforms were translated in time such that the maximum in the
  amplitude occurs at $t=0$ and the phases were adjusted by a constant
  offset such that $\phi=0$ at $t=-830M$.
  The solid (black) and dashed (red) curves show the NR and SRWZ waveforms, respectively.
  } 
  \includegraphics[width=3in]{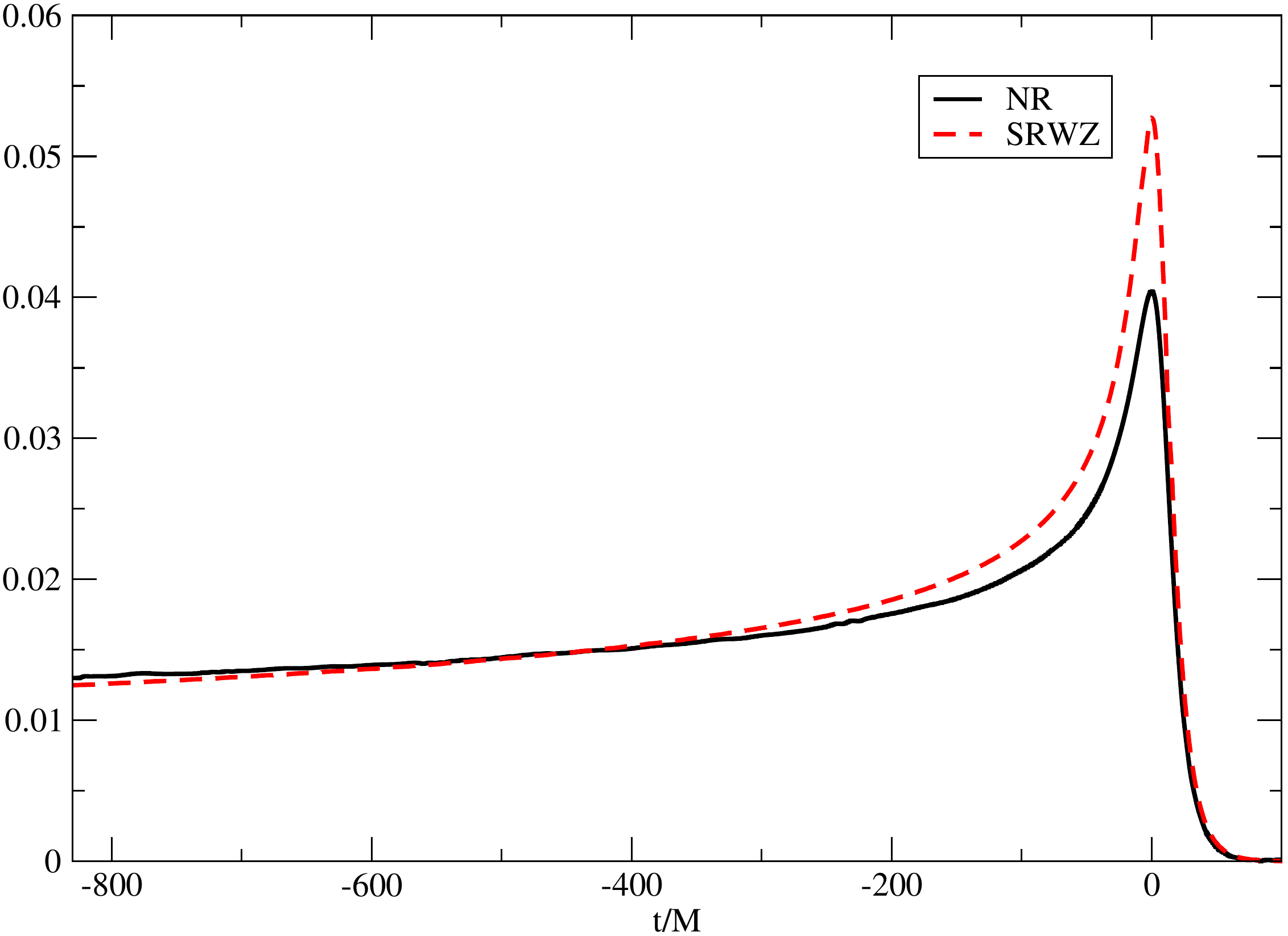}
  \includegraphics[width=3in]{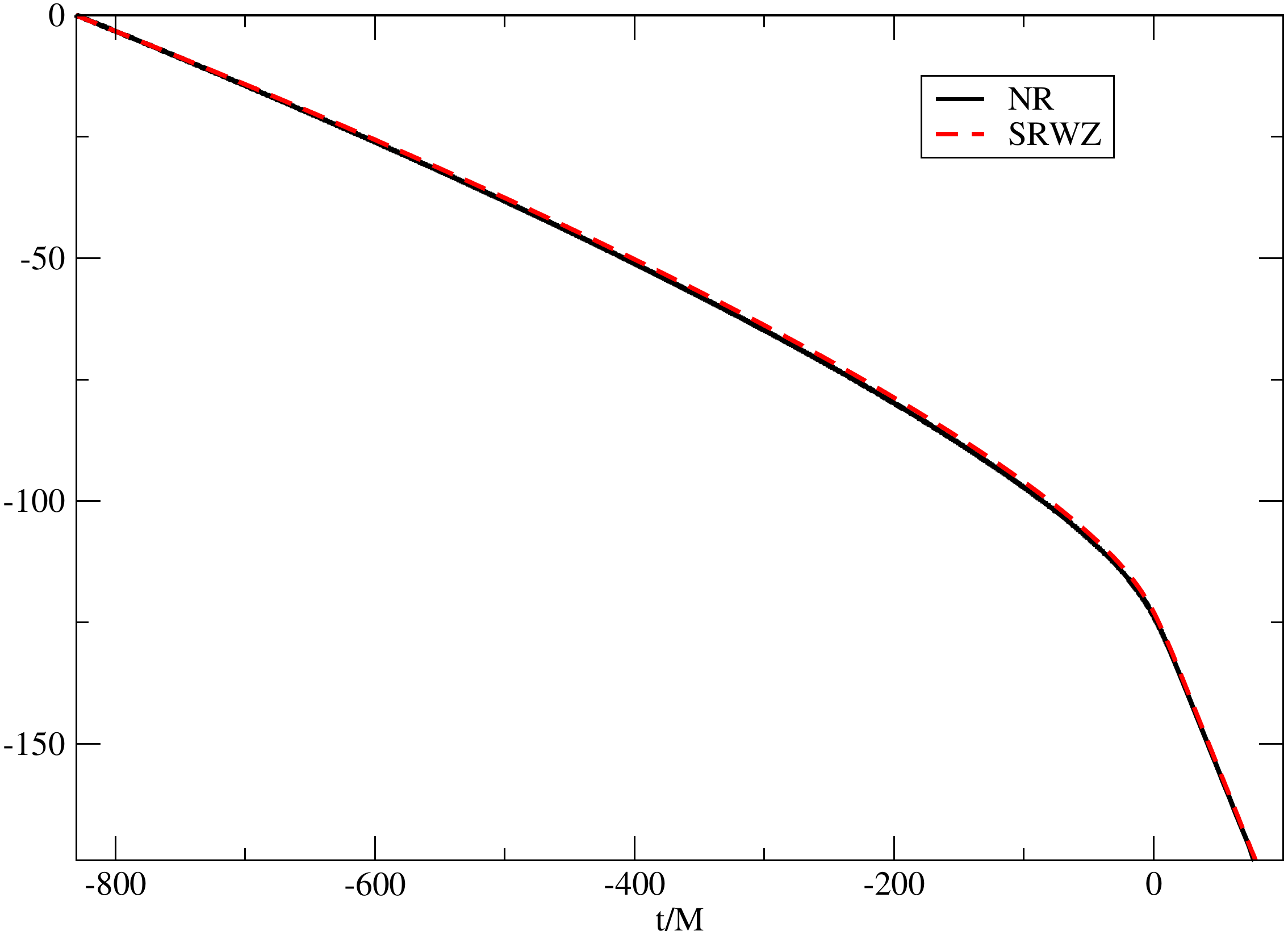}
  \label{fig:comparison_AP_10to1_33}
\end{figure}

The amplitude and phase evolutions for the $(\ell=2,\,m=1)$ mode of $h$
are shown in Fig.~\ref{fig:comparison_AP_10to1_21}.,
For $M=121M_{\odot}$ ($f_{\rm low}=10.01$Hz), the match for this mode
is 
\bea
{\cal M}_{21} &=& 0.989382 \,.
\eea

The amplitude and phase evolutions for the $(\ell=3,\,m=3)$ mode of $h$
are shown in Fig.~\ref{fig:comparison_AP_10to1_33}.
For $M=363M_{\odot}$ ($f_{\rm low}=10.01$Hz), the match for this mode is 
\bea
{\cal M}_{33} &=& 0.987414 \,.
\eea
We have a similar mismatch in the $(\ell=2,\,m=1)$ and 
$(\ell=3,\,m=3)$ modes.

We note that although 
the $(\ell=2,\,m=1)$ and $(\ell=3,\,m=3)$ modes have a slightly large 
mismatch ($1-{\cal M}$), these modes are 
the sub-dominant contributions to the gravitational wave.
The maximum amplitudes are $\sim 0.12$, $0.035$ and 
$0.04$ for the $(\ell=2,\,m=2)$, $(\ell=2,\,m=1)$ and 
$(\ell=3,\,m=3)$ modes, respectively.
Therefore, we find that from a rough estimation
the contribution of the mismatch for each mode are almost same
 in the gravitational wave.

Also, as discussed in Ref.~\cite{Lousto:2010qx}, unlike the case
of $q=1/100$, here
the spin corrections in the SRWZ formalism
are important for  obtaining the correct
 ringdown phase.
We see that the correct ringdown frequency was obtained by
comparing the  
slope of the phase evolution in Figs.~\ref{fig:comparison_AP_10to1},
\ref{fig:comparison_AP_10to1_21} and \ref{fig:comparison_AP_10to1_33}.

\subsection{The $q=1/15$ case}

For the $q=1/15$ case, 
the fitting formula is valid up to larger orbital frequencies
than for the $q=1/100$ case.
Here too, we can not calculate $u^t$ 
for $R_{\rm Sch} \leq 2.21M$ from Eq.~(\ref{eq:Ut}),
and again we use the geodesic approximation
and a matching radius  $R_{\rm M} \sim 3M$.
We find that the energy and angular momentum are given by 
\bea
E &=&  0.985372666 \,, 
\nonumber \\ 
L/M &=&  3.74699937 \,.
\eea
In the SRWZ formalism, we set the non-dimensional spin parameter
$\alpha=0.19$, consistent with the numerical results (see
Table~\ref{tab:remnant}).

\begin{figure}[h!]
  \caption{
  The amplitude (TOP) and phase (BOTTOM) of the $(\ell=2,\,m=2)$ mode of $h$ for the $q=1/15$ case.
  The waveforms were translated in time such that the maximum in the
  amplitude occurs at $t=0$ and the phases were adjusted by a constant
  offset such that $\phi=0$ at $t=-600M$.
  The solid (black) and dashed (red) curves show the NR and SRWZ waveforms, respectively.
  } 
  \includegraphics[width=3in]{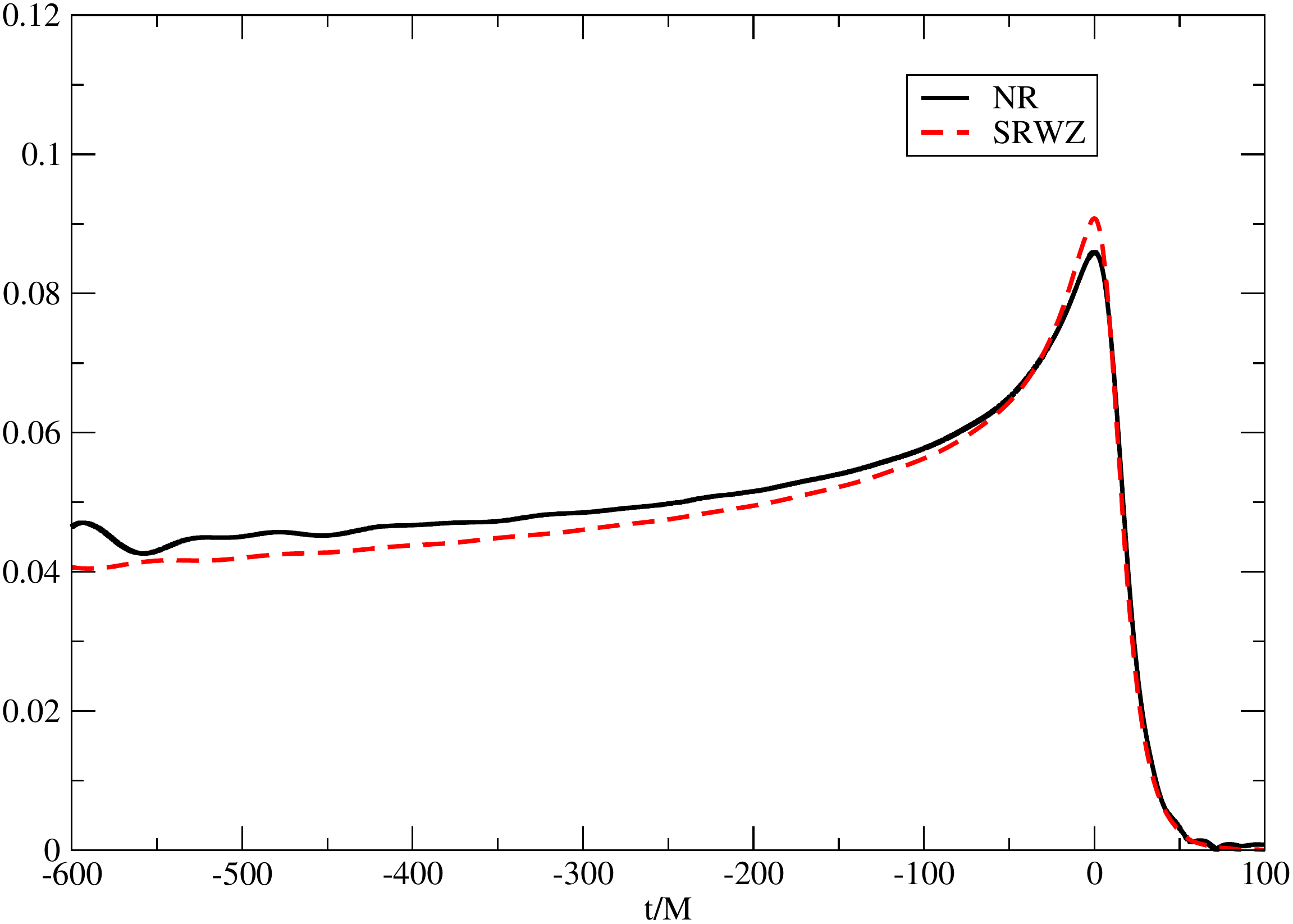}
  \includegraphics[width=3in]{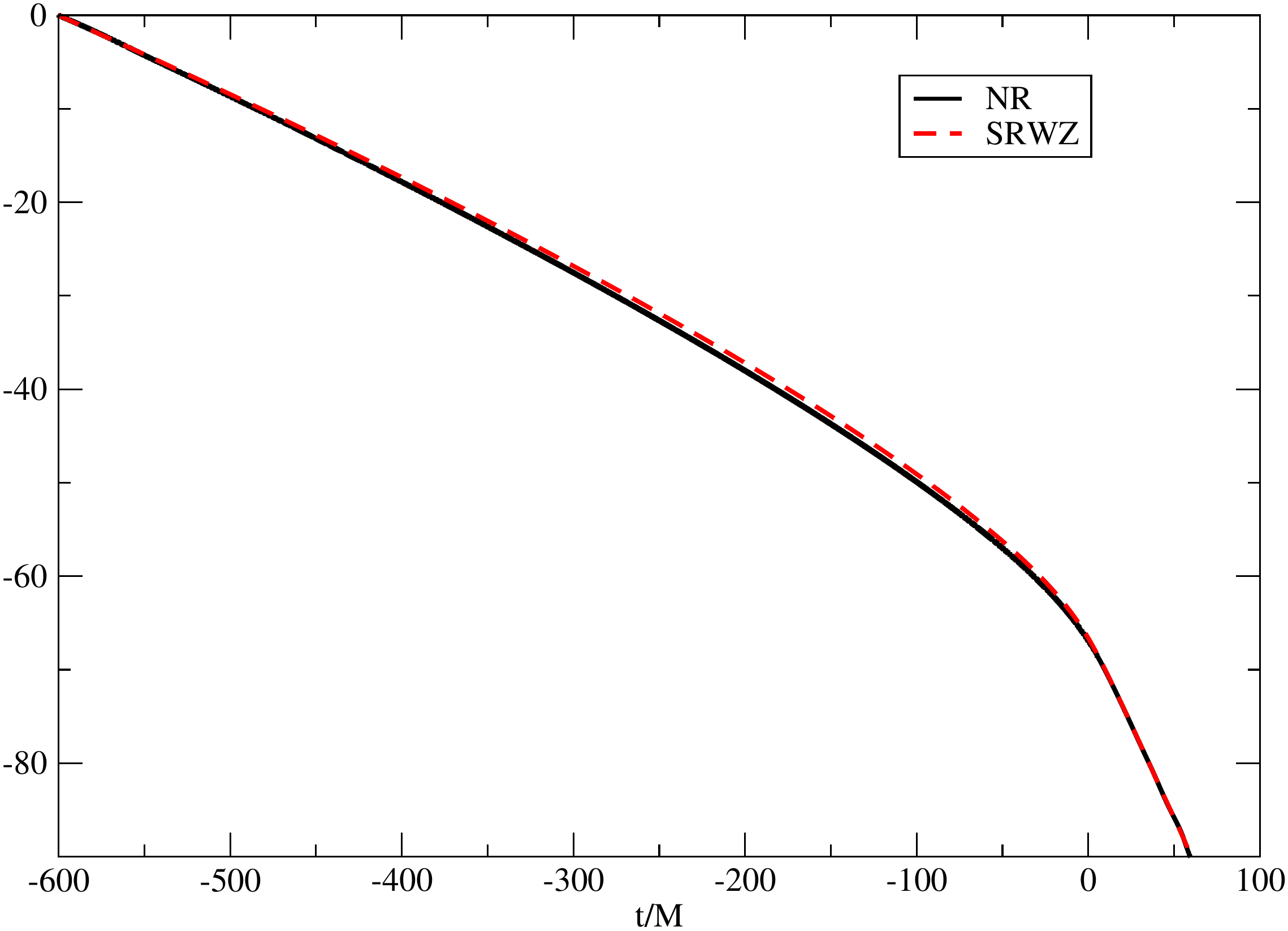}
  \label{fig:comparison_AP_15to1}
\end{figure}

The amplitude and phase evolutions of the $(\ell=2,\,m=2)$ mode of $h$
are shown in Fig.~\ref{fig:comparison_AP_15to1}.
The match for $M\Omega_{22} \geq 0.09$ 
with a total mass of  $M=290M_{\odot}$ 
($f_{\rm low}=10.02$Hz) is
\bea
{\cal M}_{22} &=& 0.996039 \,,
\eea
using the advanced LIGO noise curve.

\begin{figure}[h!]
  \caption{
  The amplitude (TOP) and phase (BOTTOM) of the $(\ell=2,\,m=1)$ mode of $h$ for the $q=1/15$ case.
  The waveforms were translated in time such that the maximum in the
  amplitude occurs at $t=0$ and the phases were adjusted by a constant
  offset such that $\phi=0$ at $t=-600M$.
  The solid (black) and dashed (red) curves show the NR and SRWZ waveforms, respectively.
  } 
  \includegraphics[width=3in]{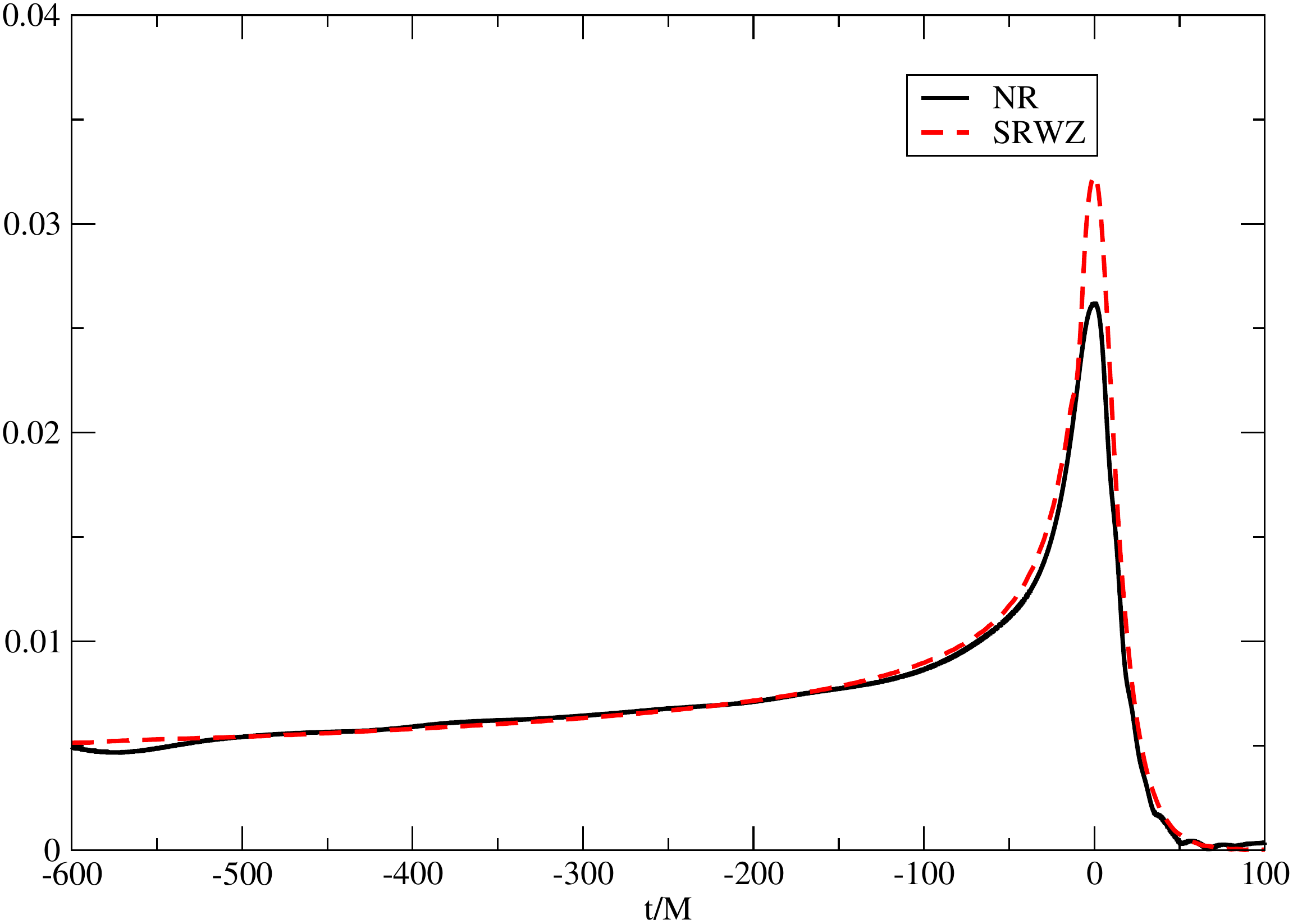}
  \includegraphics[width=3in]{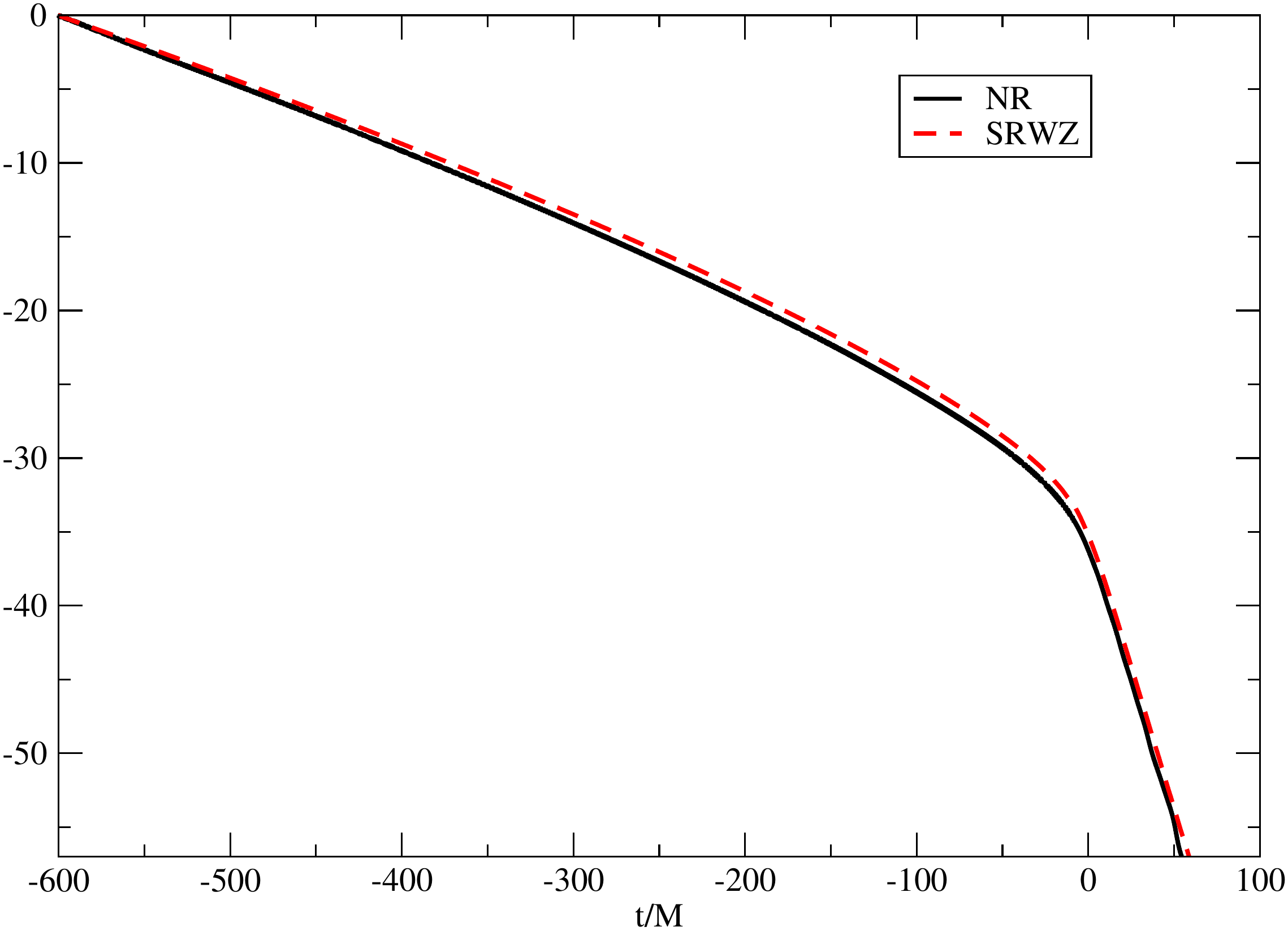}
  \label{fig:comparison_AP_15to1_21}
\end{figure}

The amplitude and phase evolutions for the $(\ell=2,\,m=1)$ mode of $h$
are shown in Fig.~\ref{fig:comparison_AP_15to1_21}.
For $M=145M_{\odot}$ ($f_{\rm low}=10.02$Hz), the match for this mode
is
\bea
{\cal M}_{21} &=& 0.989005 \,.
\eea

\begin{figure}[!h]
  \caption{
  The amplitude (TOP) and phase (BOTTOM) of the $(\ell=3,\,m=3)$ mode of $h$ for the $q=1/15$ case.
  The waveforms were translated in time such that the maximum in the
  amplitude occurs at $t=0$ and the phases were adjusted by a constant
  offset such that $\phi=0$ at $t=-600M$.
  The solid (black) and dashed (red) curves show the NR and SRWZ waveforms, respectively.
  } 
  \includegraphics[width=3in]{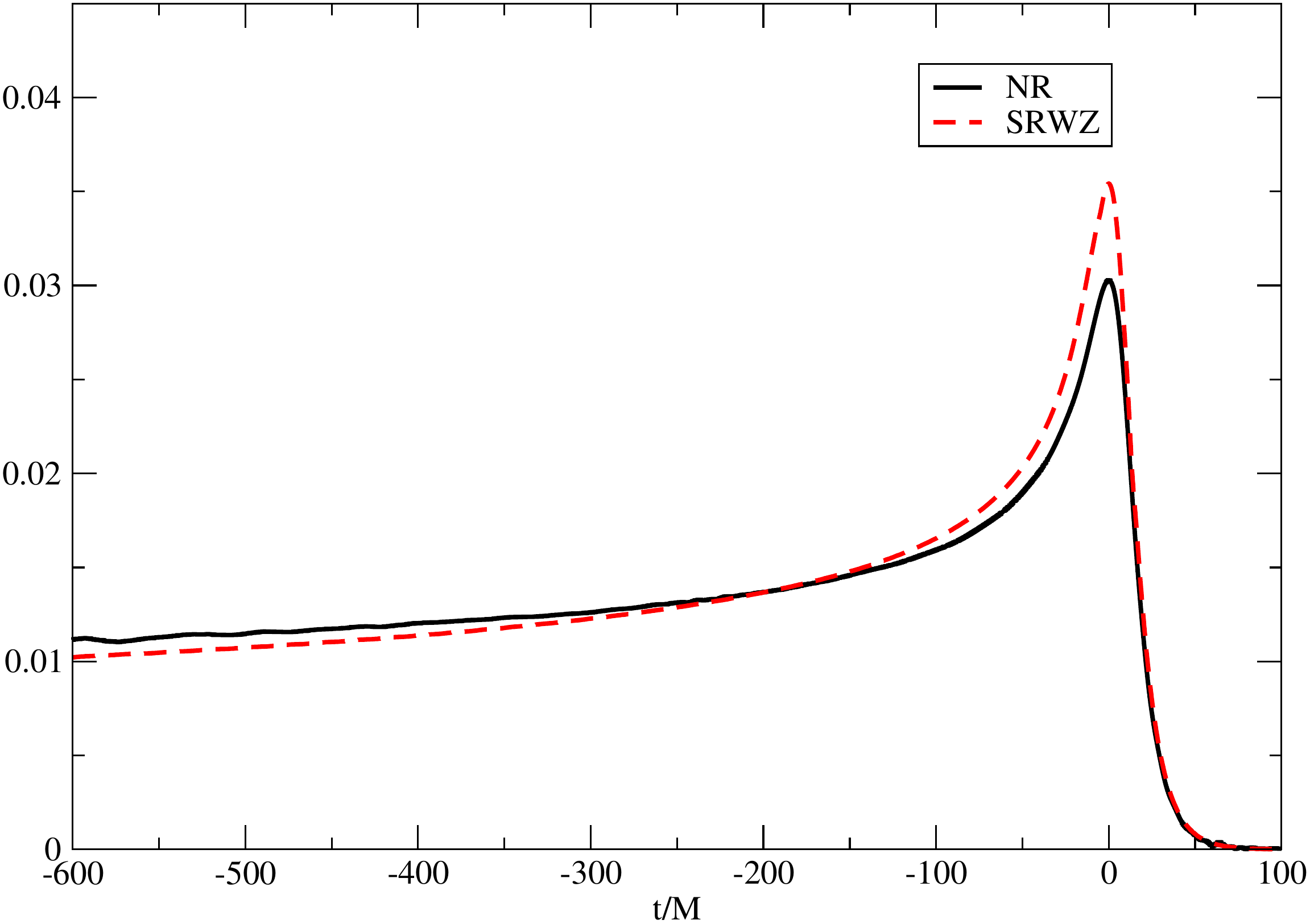}
  \includegraphics[width=3in]{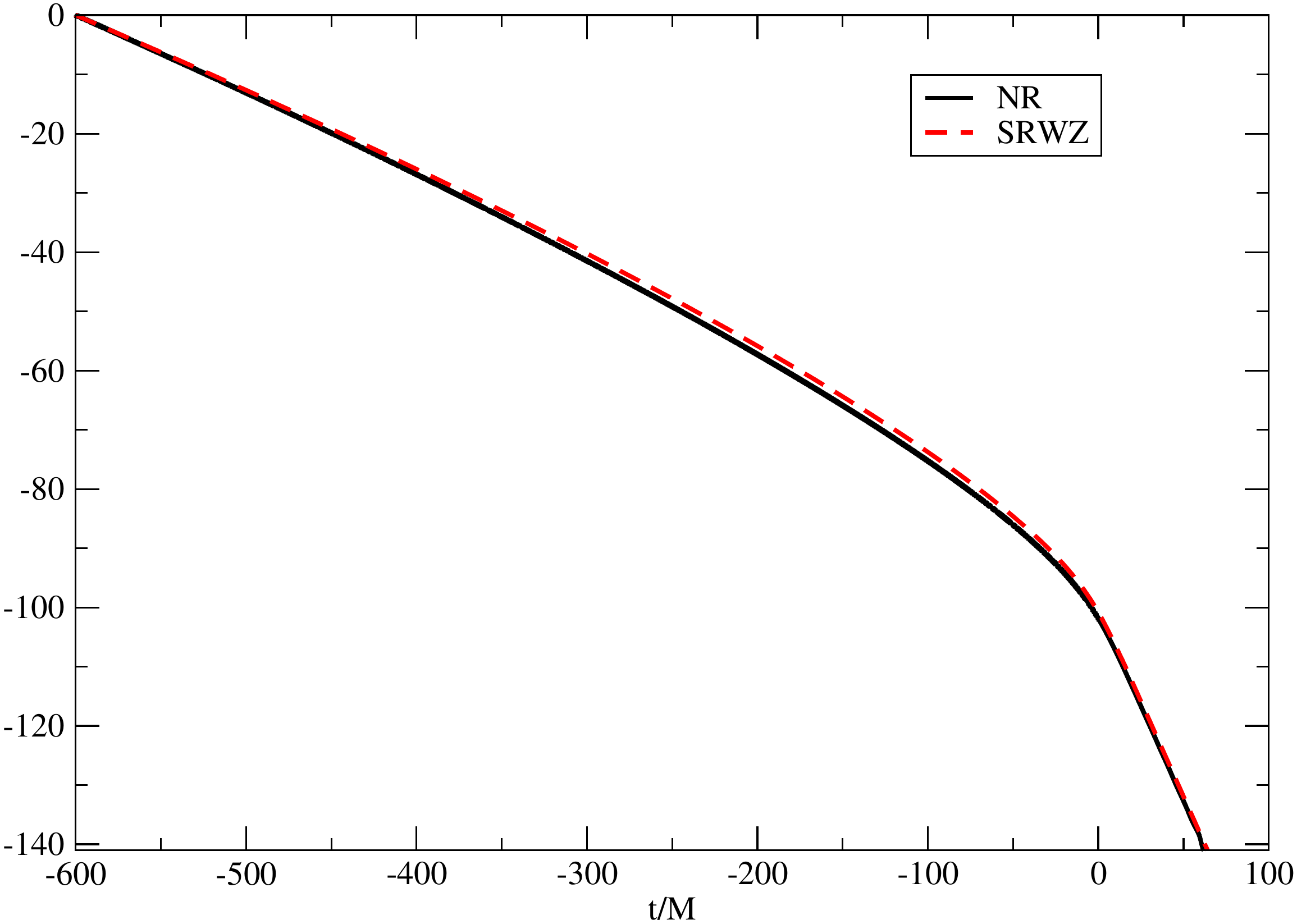}
  \label{fig:comparison_AP_15to1_33}
\end{figure}

The amplitude and phase evolutions for the $(\ell=3,\,m=3)$ mode of $h$
are shown in Fig.~\ref{fig:comparison_AP_15to1_33}.
For $M=435M_{\odot}$ ($f_{\rm low}=10.01$Hz), the match for this is
\bea
{\cal M}_{33} &=& 0.986921 \,.
\eea

As was seen in the $q=1/10$ case, 
we have a somewhat large mismatch 
for the sub-dominant $(\ell=2,\,m=1)$ and $(\ell=3,\,m=3)$ modes.

\subsection{Amplitude differences}\label{subsec:amp_diff}

\begin{figure}[!h]
  \caption{
  The amplitude difference of the $(\ell=2,\,m=2)$ mode of $h$ for the $q=1/10$ case.
  Here, the difference is normalized by the NR amplitude. We see $\sim 10\%$ difference
  in the inspiral and merger phases.
  } 
  \includegraphics[width=3in]{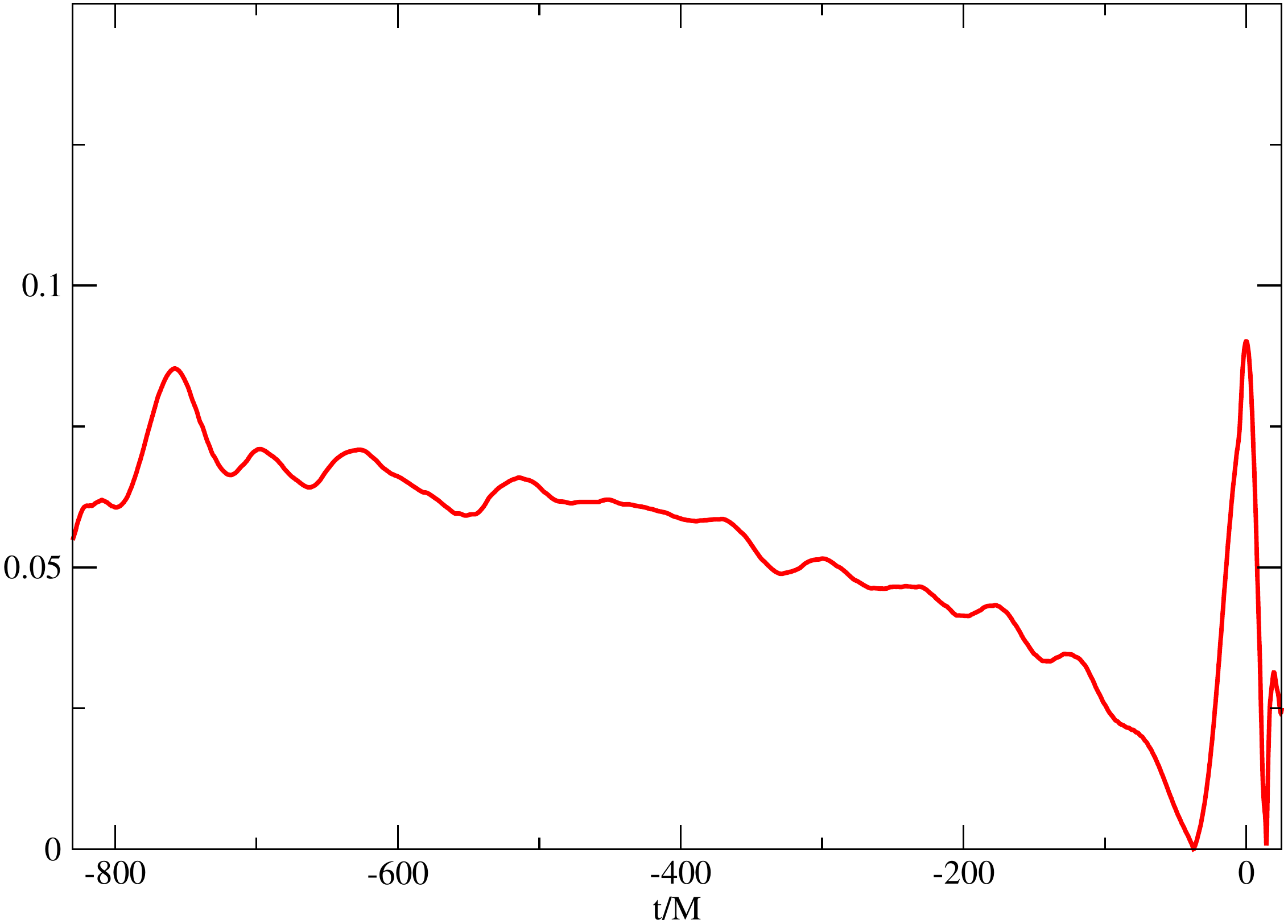}
  \label{fig:comparison_amp_diff_10}
\end{figure}

Here, we focus on the amplitude of the $(\ell=2,\,m=2)$ mode of $h$
obtained using NR and the SRWZ formalism.
In Fig.~\ref{fig:comparison_amp_diff_10}, we show the amplitude difference between
 the NR and SRWZ waveforms
for the $q=1/10$ case.
There is a large ($\sim 10\%$) difference in the amplitudes both in the inspiral and merger phases.

To understand this difference, we compare the amplitude differences 
between the NR and SRWZ waveforms for the other mass ratio cases.
Using the following definition for the amplitude difference,
\bea
\delta A_{22}=|A_{22}^{\rm (SRWZ)}-A_{22}^{\rm (NR)}| \,,
\eea
we find that the amplitude differences for the
$q=1/10$ and $q=1/15$ cases
around the maximum amplitude obey 
\bea
\delta A_{22}^{(q=1/10)} &\sim& 
2.3 \, \times \, \delta A_{22}^{(q=1/15)}
\nonumber \\ 
&\sim& 1.41^{2.42} \, \times \,\delta A_{22}^{(q=1/15)} \,,
\eea
where $1.41$ is the ratio of the symmetric mass ratios $\eta$.

\begin{figure}[!h]
  \caption{
  The amplitude difference $\delta A_{22}$
  of the $(\ell=2,\,m=2)$ mode of $h$ for the $q=1/10$ and $1/15$ cases.
  } 
  \includegraphics[width=3in]{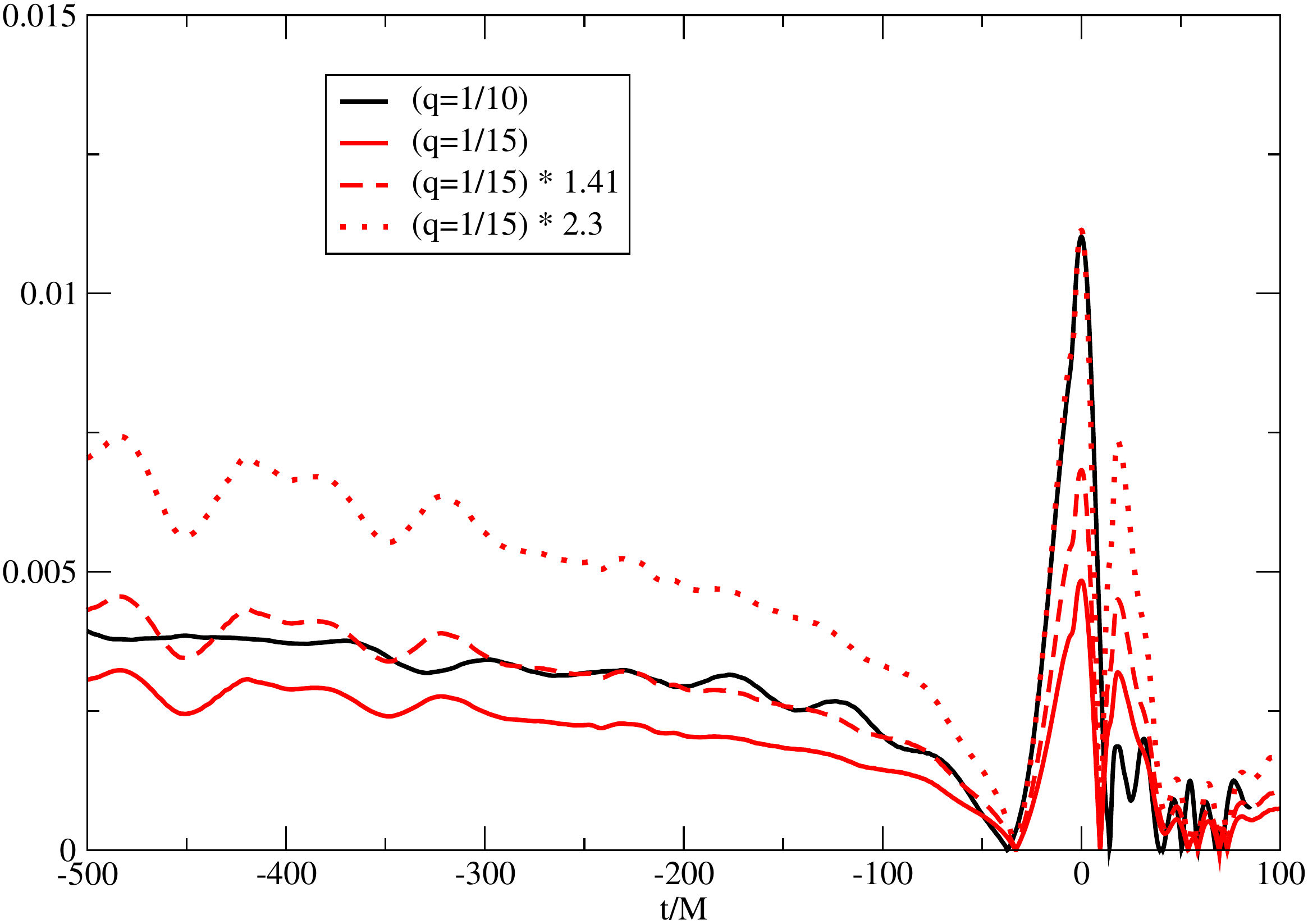}
  \label{fig:comparison_amp_diff}
\end{figure}

Figure~\ref{fig:comparison_amp_diff} shows the amplitude differences
(note that $t=0$ denotes the time of the maximum amplitude) for
$q=1/10$ and $q=1/15$.
We see that the amplitude difference before merger 
is proportional to the ratio of the symmetric mass ratio.
This behavior is different from our previous analysis in Ref.~\cite{Lousto:2010qx}, 
where we directly used the NR trajectories in the perturbative calculation of the waveforms,
and found that 
the amplitude difference in the inspiral phase scales like $\eta^2$.
Since the fitting formula is based on the quasicircular evolution,
any eccentricity of the orbit may create amplitude differences during the inspiral phase.

\begin{figure}[!h]
  \caption{
  The amplitude difference $\delta A_{22}$
  of the $(\ell=2,\,m=2)$ mode of $h$ for the $q=1/10$ and $1/100$ cases.
  } 
  \includegraphics[width=3in]{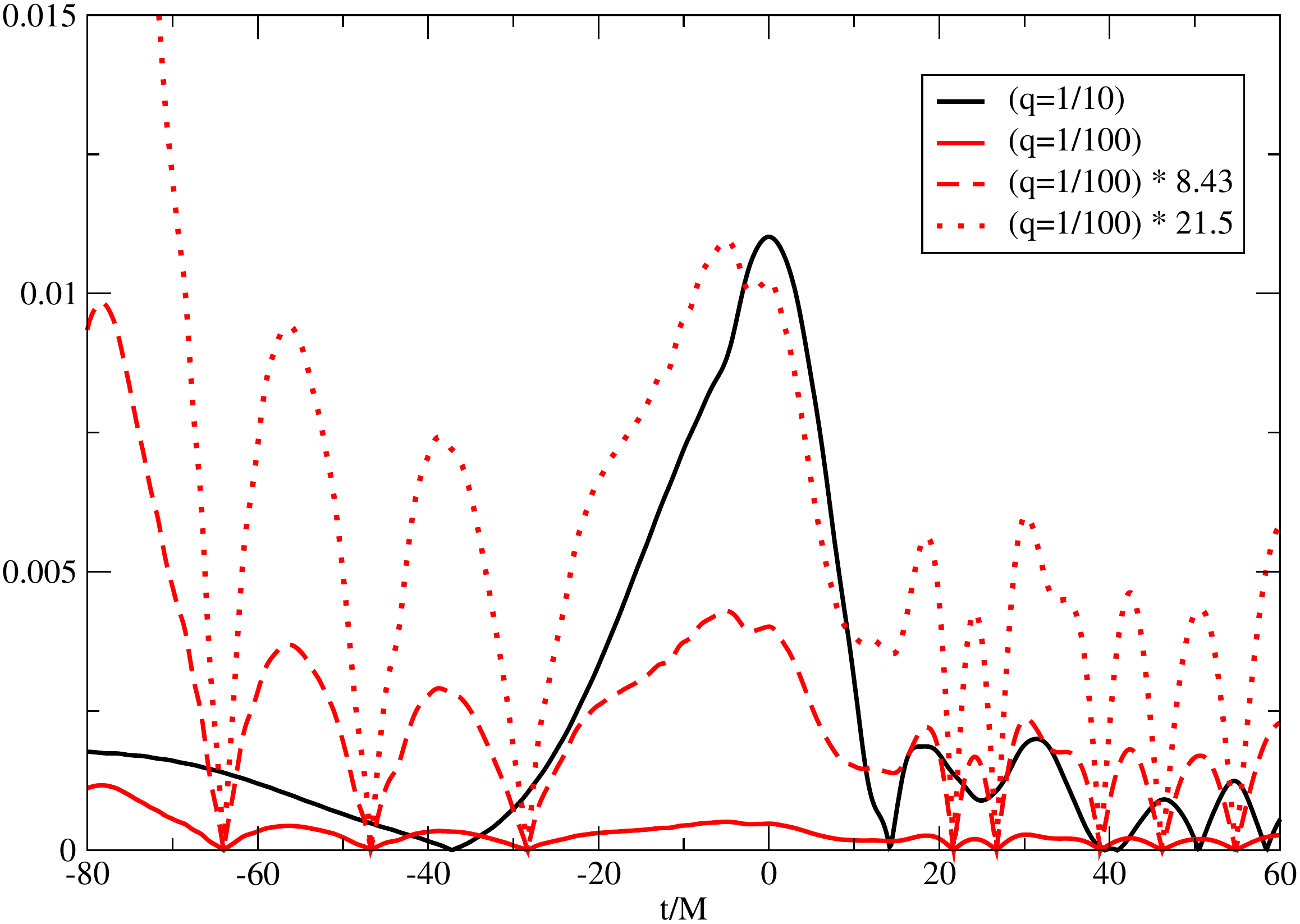}
  \label{fig:comparison_amp_diff2}
\end{figure}

Figure~\ref{fig:comparison_amp_diff2} shows the amplitude differences
both for the $q=1/10$ and $q=1/100$ waveforms.
Near the maximum amplitude, we find that the amplitude differences
scale like 
\bea
\delta A_{22}^{(q=1/10)} &\sim& 
21.5 \, \times \,\delta A_{22}^{(q=1/100)}
\nonumber \\ 
&\sim& 8.43^{1.44} \, \times \,\delta A_{22}^{(q=1/100)} \,,
\eea
where $8.43$ is the ratio of the symmetric mass ratio.
Note that this rescaling does not capture the behavior of the
amplitude differences during the inspiral phase.
Although it is difficult to obtain any exact relation between the
amplitude differences during the inspiral (possibly due to effects of
eccentricity), 
we do observe nonlinear effects in the amplitude difference between the NR and SRWZ waveforms 
around the maximum amplitude, which we infer from the fact that
the amplitude difference scales nonlinearly with the mass ratio.

\begin{figure}[!h]
  \caption{
  The amplitude difference $\delta A_{22}$
  of the $(\ell=2,\,m=2)$ mode of $h$ for the $q=1/15$ and $1/100$ cases.
  } 
  \includegraphics[width=3in]{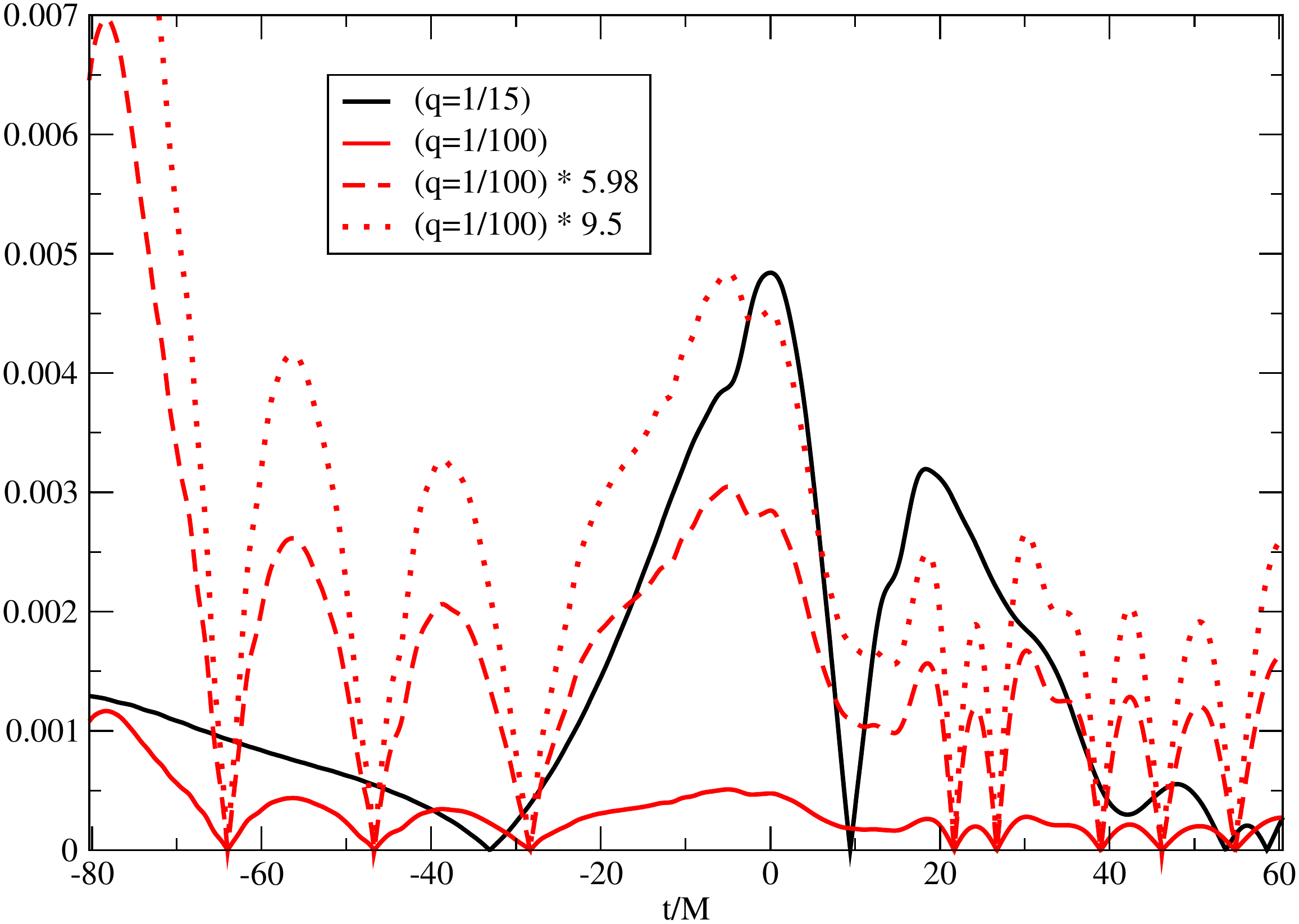}
  \label{fig:comparison_amp_diff3}
\end{figure}

Figure~\ref{fig:comparison_amp_diff3} shows the amplitude differences
both for the $q=1/15$ and $q=1/100$ cases.
Near the maximum amplitude, we find that the amplitude differences
scale like 
\bea
\delta A_{22}^{(q=1/15)} &\sim& 
9.5 \, \times \,\delta A_{22}^{(q=1/100)}
\nonumber \\ 
&\sim& 5.98^{1.26} \, \times \,\delta A_{22}^{(q=1/100)} \,,
\eea
where $5.98$ is the ratio of the symmetric mass ratios.
Again, we see the similar behavior to $q=1/10$ and
$q=1/100$ comparison.

\begin{figure}[!h]
  \caption{
  The amplitude of the $(\ell=2,\,m=2)$ mode of $h$ for the $q=1/10$ case.
  We have set the maximum amplitudes at $t=0$ by the time translation.
  The solid (black), dashed (red) and dotted (blue) curves show 
  the waveforms from the NR, SRWZ and SRWZ without the $C$ transformation
  in Eq.~(\ref{eq:OmegaR_PNP}), respectively.
  } 
  \includegraphics[width=3in]{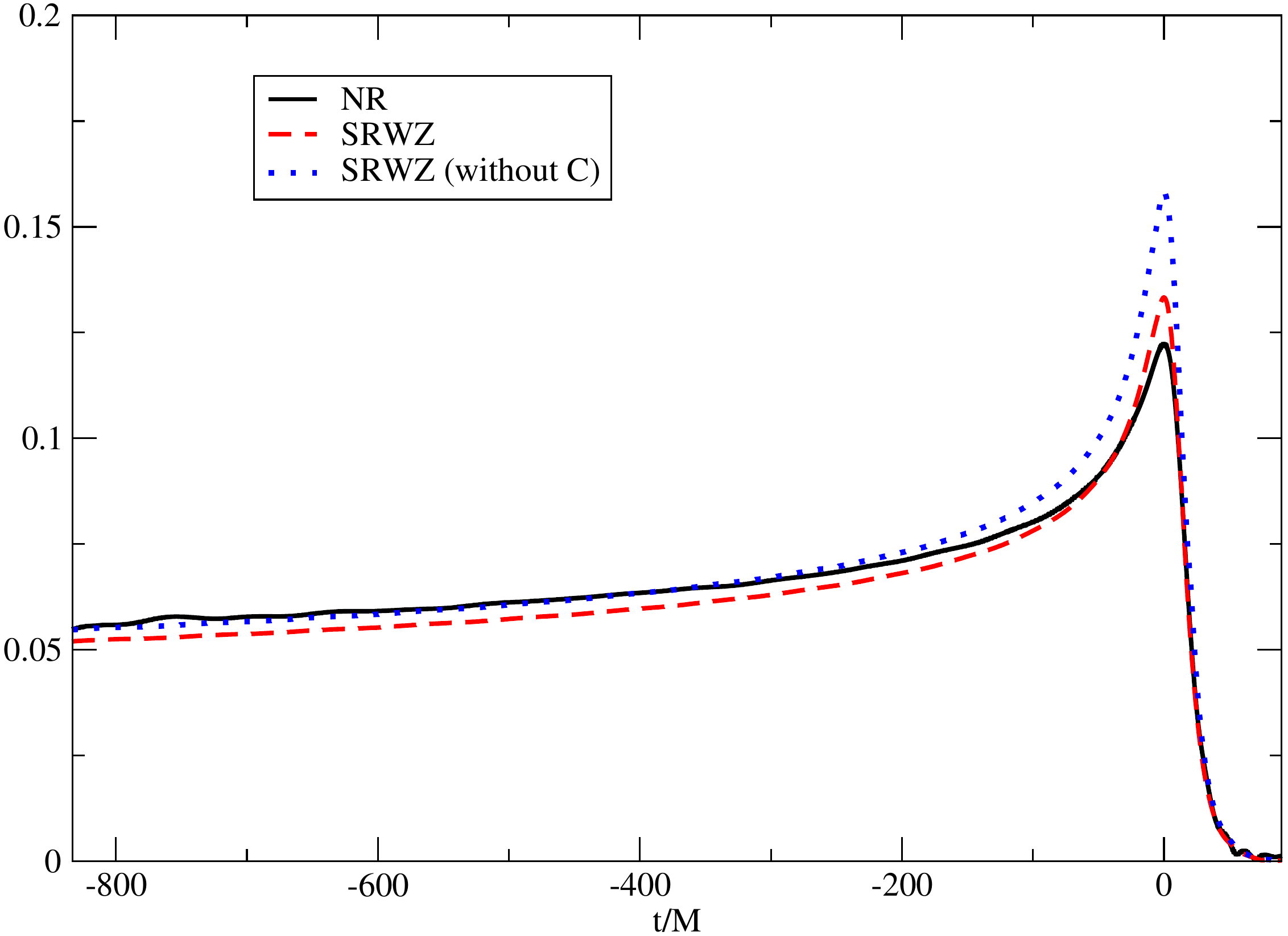}
  \label{fig:comparison_10to1_amp2}
\end{figure}

We note that the $C$ transformation in Eq.~(\ref{eq:Ctrans})
also affects the amplitude difference.
As seen in Fig.~\ref{fig:comparison_10to1_amp2},  
if we directly use the orbital radius $R_{\rm NR}$ to calculate the waveforms, 
we obtain an even larger amplitude than that obtained using the
corrected orbital radius $R_{\rm Log}$.
This behavior is expected based on the fact that while the orbital
radius is larger, the orbital frequency remains the same.
Although the orbital radius $R_{\rm NR}$ gives
a similar amplitude to the NR waveform in the initial inspiral part,
we have a much larger amplitude in the merger phase.

\section{Discussion}\label{sec:Discussion}

In the first paper of this series~\cite{Lousto:2010qx} we established that
perturbation theory of black holes in the extreme-mass-ratio expansion
$q=m_1/m_2\ll1$, as in Eq.~(\ref{eq:SRWZeqs}), can faithfully 
represent an approximation to the full
numerical simulations, if we provide 
the full numerical relative trajectory
of the holes as a function of time. In this second paper we proceed to
model these trajectories using full numerical simulations and a
parametrized extension of 3.5 PN quasi-adiabatic evolutions,
Eqs.~(\ref{eq:OOt_PNP}) and (\ref{eq:OmegaR_PNP}). This form of the
expansion allows us to incorporate the PN trajectory at large distances
and smoothly fit to the full numerical ones up to separations as small
as
$R_{\rm Sch}=3M$, at which point we can describe the subsequent trajectory by
a Schwarzschild plunging geodesic. In this approximation, 
we set the values of the energy $E$ and angular momentum $L$ of the geodesic
to those of the  previously fitted trajectory
evaluated at $R_{\rm Sch}=3M$.

Once we verified the procedure for each of the test runs, i.e.\ for
$q=1/10,\,1/15$ and $1/100$, we  obtain an expression for the fitting parameters
as a function of $q$, as given in Eqs.~(\ref{eq:c0c1}).
The other piece of information needed by the perturbative approach, in
addition to the relative trajectory, are the mass and spin of the
final black hole formed after merger to provide the background metric.
The spin and mass parameters for the SRWZ equation (\ref{eq:SRWZeqs}) can be 
estimated from empirical formulae~\cite{Lousto:2009mf,Rezzolla:2008sd}.
Note that this formula correctly predicted
the remnant  black hole's mass and spin~\cite{Lousto:2010ut}
for the three cases of mass ratio  studied here.

The current study can be considered a successful proof of principle, and
needs to be supplemented by a thorough set of runs in the intermediate
mass ratio regime $0.01\lesssim q\lesssim 0.1$ that start at separations
where the post-Newtonian approximation is still valid, i.e.\ 
$R\gtrsim10M$ and go through the merger of the holes until the formation
of a common horizon. The fitting formulae 
Eqs.~(\ref{eq:OOt_PNP}) and (\ref{eq:OmegaR_PNP}) with Eqs.~(\ref{eq:c0c1})
should provide a good approximation within the fitting interval, but
extrapolation to both the comparable masses or extreme mass ratio
could
quickly lose accuracy. The fitted tracks from our work can be compared
to alternative approaches to the problem such as the EOB~\cite{Damour:2009kr}
and EOBNR~\cite{Pan:2009wj,Yunes:2010zj} and the final
waveforms to direct phenomenological
ones as given in Refs.~\cite{Santamaria:2010yb,Sturani:2010ju}.
Our approach provides a complementary model for the intermediate mass-
ratio regime to the above approaches, which  mostly apply to the
comparable-mass regime.

The extension of our approach to initially highly-spinning black holes
can be done essentially through the same lines depicted here. 
Numerical simulations of highly spinning holes with small mass ratio
are feasible~\cite{Lousto:2008dn}, and the trajectories would already
include the effect of the spins of both holes. The perturbative
waveform calculation would use the Teukolsky
formalism~\cite{Teukolsky:1973ha} 
and its second order extension~\cite{Campanelli:1998jv}.
Fitting formulae based on the extension of
spinning post-Newtonian trajectories with free parameter dependence
on the mass ratio and effective spins should be used together with
the final matching to plunging geodesics on an spinning background.
The final remnant formulae for spinning binaries have already been proposed
and fitted to available simulations~\cite{Lousto:2009mf,
Rezzolla:2008sd}.

\acknowledgments 

We thank Marcelo Ponce for providing the method to reduce the
horizon mass loss required to make accurate simulations in the
small mass ratio regime.
We gratefully acknowledge the NSF for financial support from Grants
No. PHY-0722315, No. PHY-0653303, No. PHY-0714388, No. PHY-0722703,
No. DMS-0820923, No. PHY-0929114, PHY-0969855, PHY-0903782,
and No. CDI-1028087; and NASA for financial support from NASA Grants 
No. 07-ATFP07-0158 and No. HST-AR-11763.  Computational resources were
provided by the Ranger cluster at TACC (Teragrid allocation TG-PHY060027N) 
and by NewHorizons at RIT.

\appendix

\section{Perturbative Equations in 1+log ``trumpet'' coordinates}\label{app:Trumpet}

In this work we used standard Schwarzschild coordinates to describe
the background geometry and trajectory of the particle. To obtain the
latter, we transformed the particle's location from the NR coordinates
(assuming they were represented by $1+\log$ trumpet coordinates of a
Schwarzschild spacetime) into the standard Schwarzschild coordinates.
It is also useful to see the form of the perturbative equations in
these $1+\log$ trumpet coordinates. This will help to assess potential
regions where this approach might have limitations.  In particular,
since the trumpet coordinates are obtained in the late-time,
stationary limit, wherever there still are strong dynamics, the
approach could be inaccurate.

The generalized Regge-Wheeler-Zerilli equations for an arbitrary 
spherically symmetric
background were found in Ref.~\cite{Sarbach:2001qq}.
Here, for the sake of simplicity, we use only the generalized 
Regge-Wheeler equation given in Ref.~\cite{Pazos:2006kz}. 
Formally, this equation can be written as 
\bea
\ddot{\Phi}_{RW} &=& c_1 \,\dot{\Phi'}_{RW}+  c_2 \,\Phi^{''}_{RW} +
c_3 \,\dot{\Phi}_{RW} 
\nonumber \\ && 
+ c_4 \,\Phi'_{RW} + c_5 \,\Phi_{RW} \,,
\eea
where the dot and dash denote the time and radial derivatives, and
the coefficients
$c_i$ ($i=1,\,2,\,3,\,4,\,5$) are determined by the background metric. 

For the background metric, we use the notation
of~\cite{Brugmann:2009gc}, which has the form
\bea
ds^2 &=& - (\alpha^2 - \beta^2) dt^2 + \frac{2\beta}{f} dt dr
\nonumber \\ &&
+ \frac{1}{f^2} dr^2 + R^2 (d\theta^2 + \sin^2\theta d\phi^2) \,.
\eea
where $\alpha$, $\beta$, $f$ and $R$ are functions of the radial 
coordinate $r$ only. 
The coefficients $c_i$ are then given by
\bea
c_1 &=& 2 \,f(r) \,\beta(r) \,,
\nonumber \\ 
c_2 &=& (\alpha(r)^2-\beta(r)^2)\,f(r)^2 \,,
\nonumber \\ 
c_3 &=& -(\beta(r) \,\alpha'(r)-\alpha(r) \,\beta'(r))\, 
\frac{f(r)}{\alpha(r)} \,,
\nonumber \\ 
c_4 &=& \frac{f(r)^2}{\alpha(r)}\,
(-2 \,\alpha(r) \,\beta(r) \,\beta'(r)+\alpha(r)^2 \,\alpha'(r)
\nonumber \\ && 
+\beta(r)^2 \,\alpha'(r))
+ f(r) \,f'(r) \,(\alpha(r)^2-\beta(r)^2) 
\,,
\nonumber \\ 
c_5 &=& - \alpha(r)^2 \,V(r) \,,
\label{eq:C}
\eea
where $V$ denotes the Regge-Wheeler potential,
\bea
V(r) &=& \frac{1}{R(r)^2} \, 
\left(\ell \,(\ell+1)- \frac{6\,M}{R(r)} \right) \,.
\eea
When using the standard Schwarzschild coordinates, the above equation
becomes the standard Regge-Wheeler equation. 
We note that for the Zerilli equation,
we may set~\cite{Sarbach:2001qq}
\bea
V(r) &=& \frac{2}{R(r)^3 (\lambda R(r)+3M)^2}
\left[
\lambda^2(\lambda+1)R(r)^3
\right. \nonumber \\ &&  \left.
+3\lambda^2MR(r)^2+9\lambda M^2R(r)+9M^3
\right] \,,
\eea
where $\lambda=(\ell-1)(\ell+2)/2$.

We now consider the stationary $1+\log$ slices of the 
Schwarzschild spacetime in isotropic coordinates 
as an approximation to the NR coordinate system. 
As discussed in Ref.~\cite{Brugmann:2009gc}, the metric in 
this coordinate system can be obtained numerically. 

Since the Schwarzschild metric is a solution of the vacuum General 
Relativity field equations, and since we are using a Killing lapse and
shift~\cite{Brugmann:2009gc}, we have
\bea
\alpha(r) &=& f(r) R'(r) \,,
\nonumber \\ 
\alpha(r)^2 - \beta(r)^2 &=& 1 - \frac{2M}{R(r)} \,.
\label{eq:SSEE}
\eea

From the stationary $1+\log$ slice condition, we have
\bea
C e^\alpha &=& \left(\alpha^2 - 1 + \frac{2M}{R}\right) R^4 \,,
\eea
where the constant $C$ is determined  by requiring that
 $\alpha'$ be regular at the ``critical'' point  (where the
denominator vanishes).
\bea
\frac{C}{M^4} &=& \frac{1}{128}(3+\sqrt{10})^3 \exp(3-\sqrt{10}) \,.
\eea

In isotropic coordinates we obtain
\bea
\alpha(r) &=& \frac{r \,R'(r)}{R(r)} \,.
\eea

Since the above equations are transcendental,
we have to find numerical
solutions. However,  it is also useful to study their asymptotic behavior 
for large $r$. At large $r$ we have,
\bea
R(r) &=& r + M + \frac{1}{4}\,\frac{M^2}{r}
-\frac{1}{8}\,\frac{C\,e}{r^3}+ O(1/r^4) \,,
\nonumber \\
\alpha(r) &=& 1- \frac{M}{r}+\frac{1}{2}\,\frac{M^2}{r^2}
-\frac{1}{4}\,\frac{M^3}{r^3}
\nonumber \\ &&
+\left(\frac{1}{8}+\frac{1}{2} \,\frac{C}{M^4} \,e \right) \,\frac{M^4}{r^4} 
+ O(1/r^5) \,,
\eea
and the other two functions $\beta$ and $f$ can be found
from Eqs.~(\ref{eq:SSEE}). 

Inserting these functions into Eqs.~(\ref{eq:C}), 
we find that the coefficients $c_i$ are given by
\begin{widetext}
\bea
c_1 &=& \frac{2\,\sqrt{C\,e}}{r^2} + O(1/r^3) \,,
\nonumber \\ 
c_2 &=& 
1-{\frac {4\,M}{r}}+\frac{17}{2}\,{\frac {{M}^{2}}{{r}^{2}}}
-{\frac {{13\,M}^{3}}{{r}^{3}}} 
+ \left( {\frac {259}{16}}+\frac{1}{4}\,\frac{C}{M^4}\,e \right) 
\,{\frac {{M}^{4}}{{r}^{4}}}
+ O(1/r^5) \,,
\nonumber \\ 
c_3 &=& -2\,{\frac {\sqrt {C\,e}}{{r}^{3}}} + O(1/r^4)
\,,
\nonumber \\ 
c_4 &=& {\frac {2\,M}{{r}^{2}}}-\frac{17}{2}\,{\frac {{M}^{2}}{{r}^{3}}}
+{\frac {39}{2}}\,{\frac {{M}^{3}}{{r}^{4}}}
+ \left( -{\frac {259}{8}}+\frac{3}{2}\,\frac{C}{M^4}\,e  \right) 
\,{\frac {{M}^{4}}{{r}^{5}}} + O(1/r^6)
\,,
\nonumber \\ 
c_5 &=& 
{\frac {\ell \left( \ell+1 \right) }{{r}^{2}}}
-{\frac {6\,M \left( 1+\ell \left( \ell+1 \right) \right) }{{r}^{3}}}
+\frac{1}{2}\,{\frac {{M}^{2} \left( 84+37\,\ell \left( \ell+1 \right) \right) }{{r}^{4}}}
-\frac{1}{2}\,{\frac {{M}^{3} \left( 303+79\,\ell \left( \ell+1 \right) \right) }{{r}^{5}}}
\nonumber \\ && 
+ \left( 378+{\frac {1059}{16}}\,\ell \left( \ell+1 \right) 
+\frac{9}{4}\,\frac{C}{M^4}\,e\,\ell \left( \ell+1 \right) \right) 
\,\frac{M^4}{r^6} + O(1/r^7)
\,.
\eea
\end{widetext}
If we set $C=0$, the above coefficients are the same as those derived 
in the standard isotropic coordinates 
($r_{\rm Sch}=r_{\rm Iso}(1+M/(2r_{\rm Iso}))^2$) 
for the Schwarzschild spacetime. 

\begin{figure}[ht]
\begin{center}
\caption{
  The solid (black), dashed (blue), and
  dotted (red) 
  curves show the standard Regge-Wheeler (TOP) and Zerilli (BOTTOM) 
  potentials in the standard
  the Schwarzschild coordinates,  isotropic 
  coordinates, and ``trumpet'' coordinates (which are the stationary
  $1+\log$ slices of the
  Schwarzschild spacetime in isotropic coordinates), respectively, for the $\ell=2$ mode.
  Note that the location of the horizon is different
  in each system. $r_{\rm Sch}<2M$, $r_{\rm Iso}<M/2$, 
  $r_{\rm Log}<0.8304M$  are 
  inside the horizon. 
  Note that the potential is finite at the horizon,
  and is always positive.
}
\includegraphics[width=3.4in]{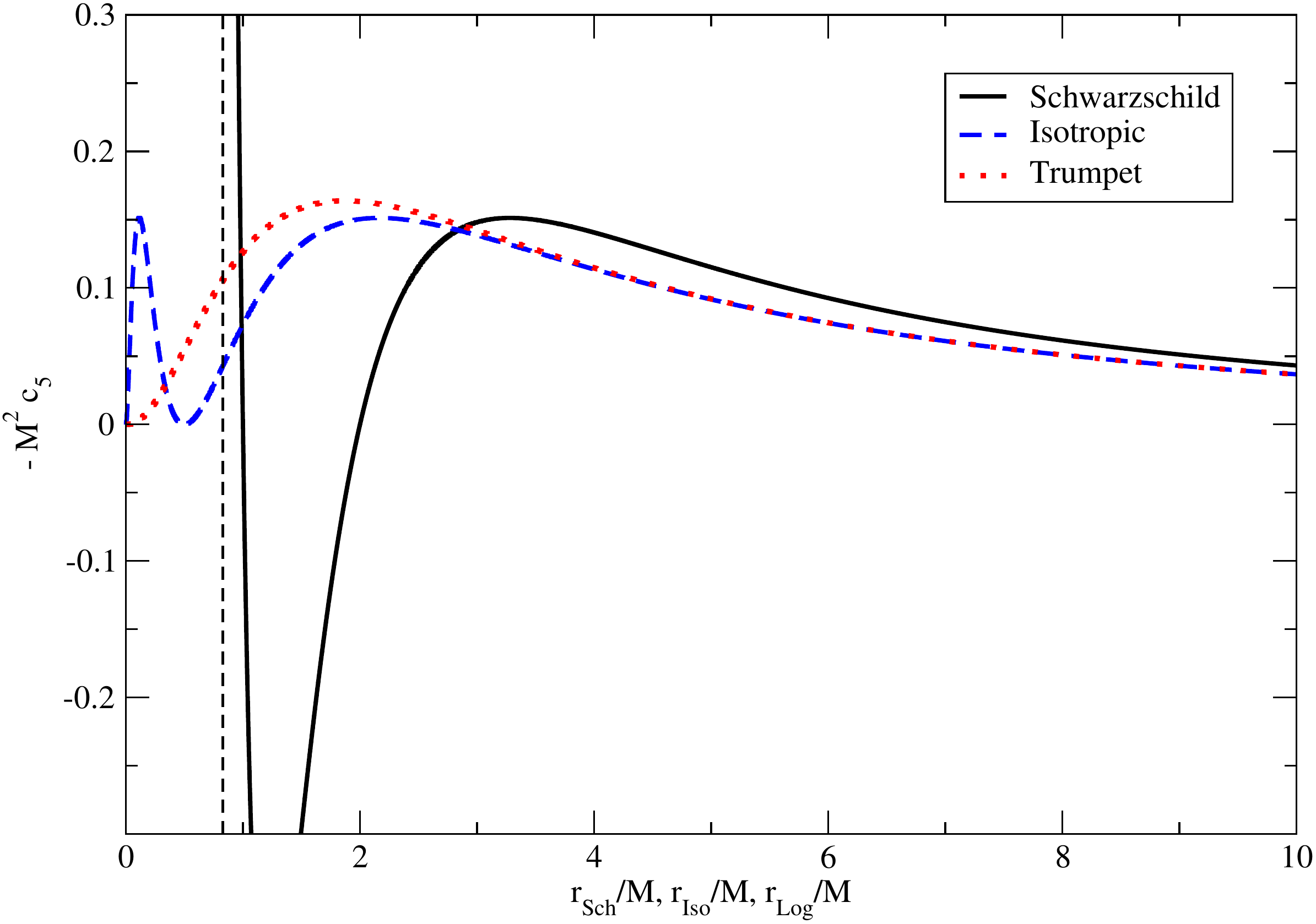}
\includegraphics[width=3.4in]{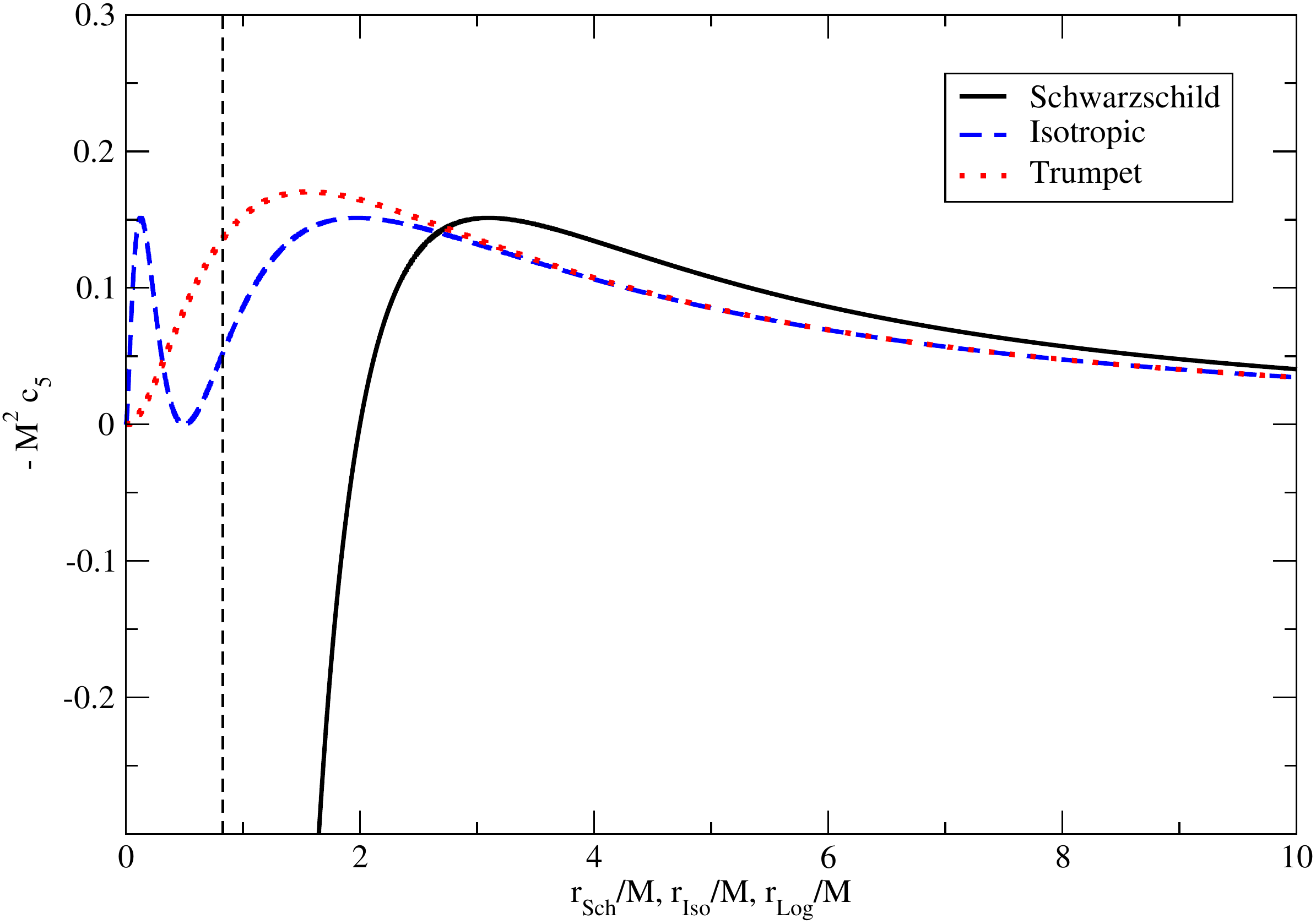}
\label{fig:BruegV}
\end{center}
\end{figure}

\begin{figure}[ht]
\begin{center}
\caption{
  The solid (black), dashed (blue), and dotted (red) curves show 
  the radial derivative 
  of the standard Regge-Wheeler (TOP) and Zerilli (BOTTOM) potentials
  in the standard Schwarzschild coordinates, isotropic coordinates, 
  and ``trumpet'' coordinates, respectively, 
  for the $\ell=2$ mode.
  $r_{\rm Sch}<2M$, $r_{\rm Iso}<M/2$, and $r_{\rm Log}<0.8304M$ are 
  inside the horizon. 
  The vertical dashed lines denotes the location of the
  horizon in the ``trumpet'' coordinates.
}
\includegraphics[width=3.4in]{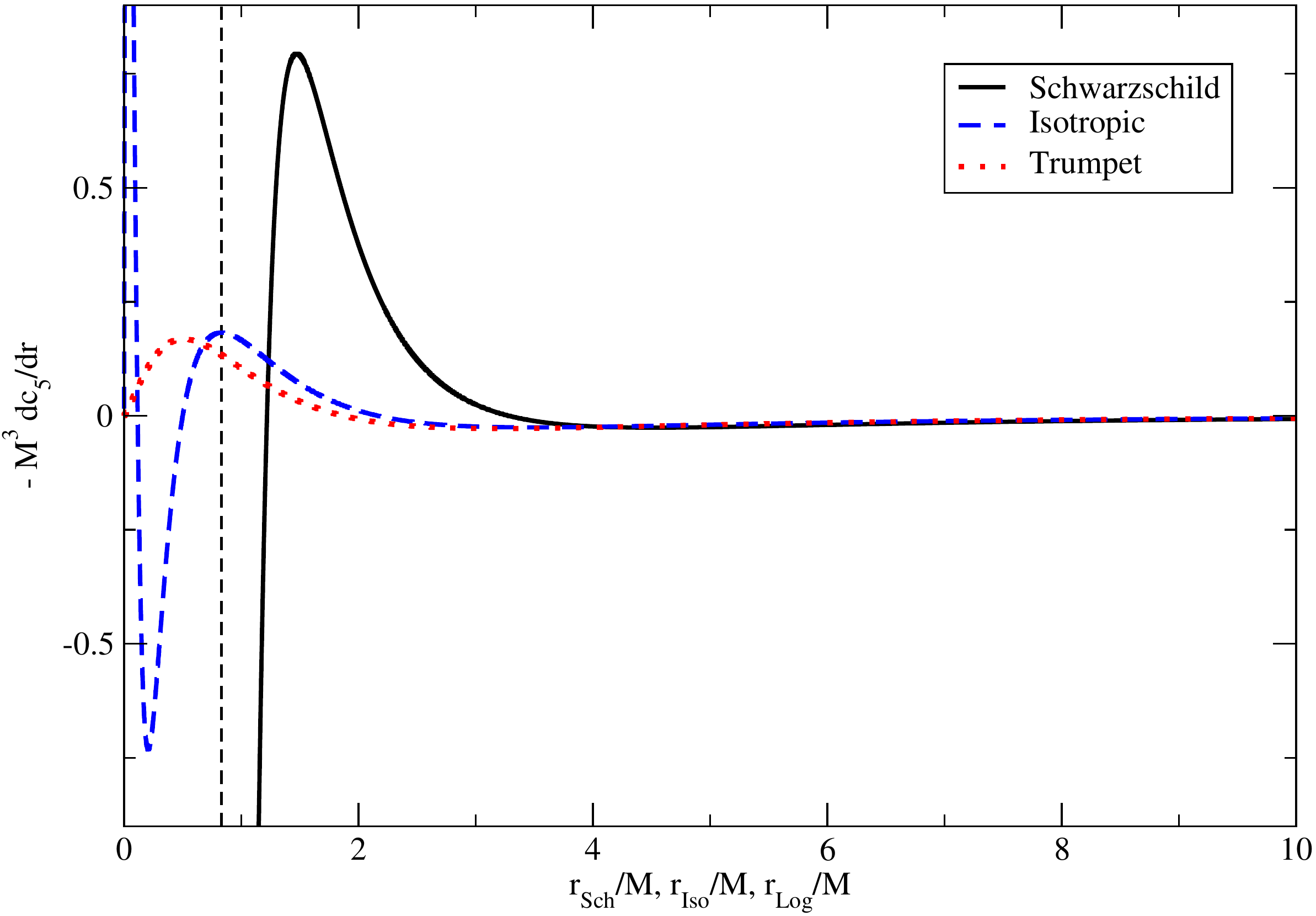}
\includegraphics[width=3.4in]{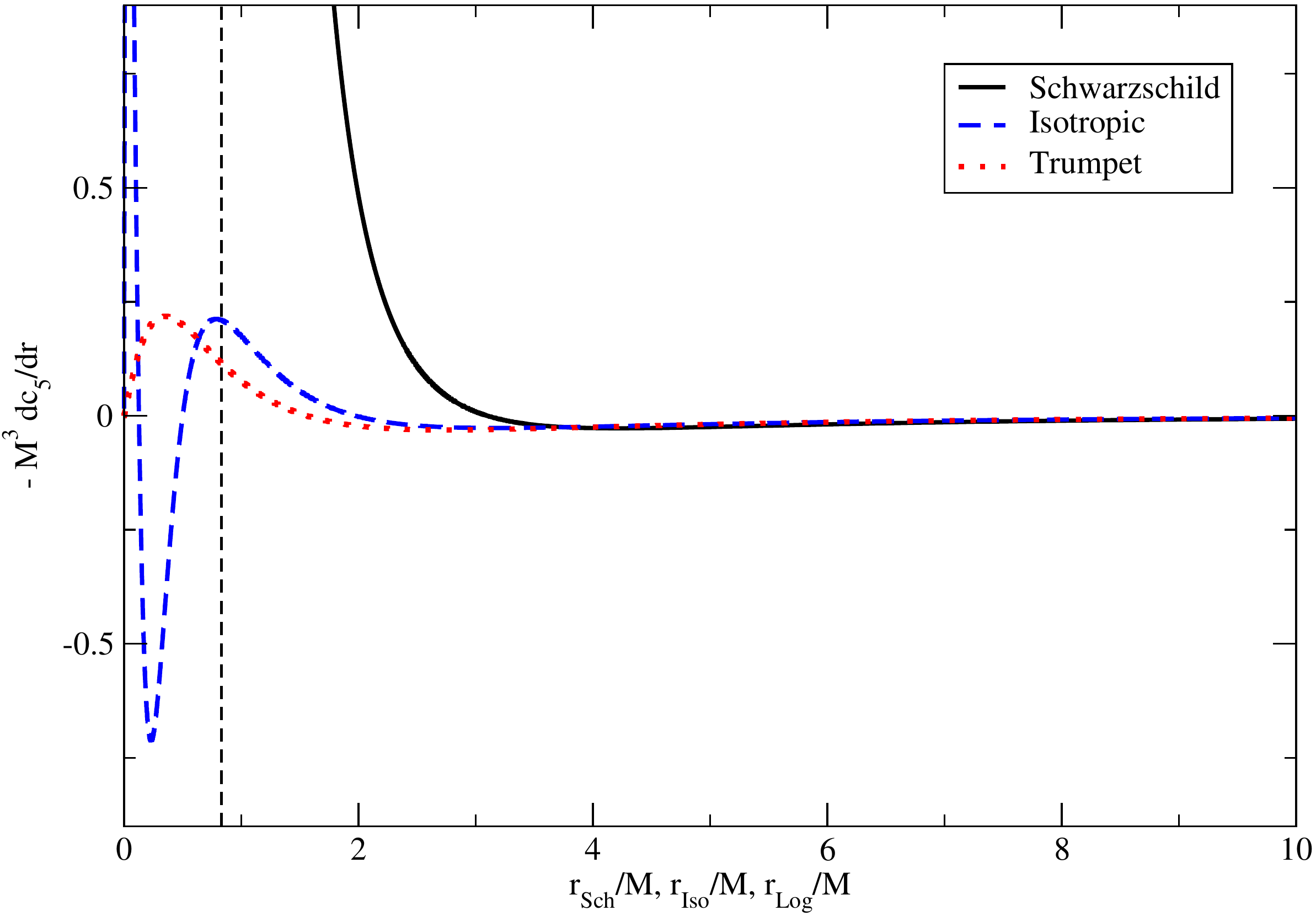}
\label{fig:BruegdV}
\end{center}
\end{figure}

An important piece of information in setting up the grid refinements,
particularly for the smaller black hole is the perturbative 
potential~\cite{Lousto:2010ut}.
Using the numerical method given in Ref.~\cite{Brugmann:2009gc}, 
we can obtain the potential term $c_5$. 
The ``Regge-Wheeler'' potentials for the standard 
Schwarzschild ($r_{\rm Sch}$), isotropic ($r_{\rm Iso}$),
and the trumpet radial coordinates ($r_{\rm Log}$)
(associated to the NR coordinates) are shown in Fig.~\ref{fig:BruegV},
while the radial derivatives of these potentials
are shown in Fig.~\ref{fig:BruegdV}. 
In the trumpet coordinates, the radial coordinate covers 
up to $r_{\rm Log}=0$, and $\alpha$ and $R$ around $r_{\rm Log}=0$ 
have the following forms~\cite{Brugmann:2009gc}. 
\bea
\alpha(r_{\rm Log}) &\sim& r_{\rm Log}^{1.091} \,,
\nonumber \\ 
R(r_{\rm Log}) &\sim& R_0 + \left(\frac{r_{\rm Log}}{1.181} \right)^{1.092} \,,
\eea
where $R_0 \simeq 1.312M$. 
From the above equation, we see that the sphere corresponding to
$r_{\rm Log}=0$ has a finite circumference.

We observe that the differences between the ``trumpet'' and isotropic
coordinates (outside the horizon), as determined by differences in the
potentials, occurs in a region between the horizon and the maximum of
the potential. This lends credence to the postulate that the main
differences between the numerical coordinates and the ``trumpet''
coordinates lie in this region, as well. Indeed, as the two BHs
approach each other, the conformal function $W$ near the larger BH must
change (since $W=0$ at the location of the smaller BH, and varies
smoothly outside the two punctures). Thus, as the small BH falls
through the maximum of the potential, the coordinate system must
become distorted in this important region. Therefore, we may expect
that the assumption that the trajectory in the background spacetime is
well described by assuming the coordinate radius is the trumpet radial
coordinate breaks down.  This may explain the need to match
trajectories to plunging geodesics with slightly larger $E$ and $L$
values than one might expect (i.e.\ $E>1$). 

Another use of the perturbative potentials shown in Fig.~\ref{fig:BruegV}
and, in particular, its radial derivative shown in Fig.~\ref{fig:BruegdV},
is to guide the set up of the numerical grid structure near the
black holes \cite{Lousto:2010ut}. To set up the mesh refinement levels,
we chose grids that cover the region between the zeros of the
derivative of the potential. 
The idea is that we need to model the curvature and
the gravitational radiation emitted by the small BH (which drives
the merger, and hence the physics). At the zeros of the
derivative of the potentials, the variations are minimized.
Furthermore
the separations between zeros increases, naturally leading to a choice
of small-width, high resolution grids between the first zeros, one
step lower in resolution between the second two, followed by a
sequence of coarser grids.

\begin{figure}[ht]
\begin{center}
\caption{
  The ingoing and outgoing characteristic speeds. 
  We see that $c_+=0$ but $c_-$ is finite at the horizon
  (denoted by the vertical dashed line,
  $r_{\rm Log}=0.8304M$), 
  and that both speeds are negative inside the horizon.
  The thin curves show the characteristic speeds for the
  perturbation in  the isotropic coordinates. Note that in
  this latter case, both speeds are zero on the horizon
  ($r_{\rm Iso}=M/2$).
}
\includegraphics[width=3.4in]{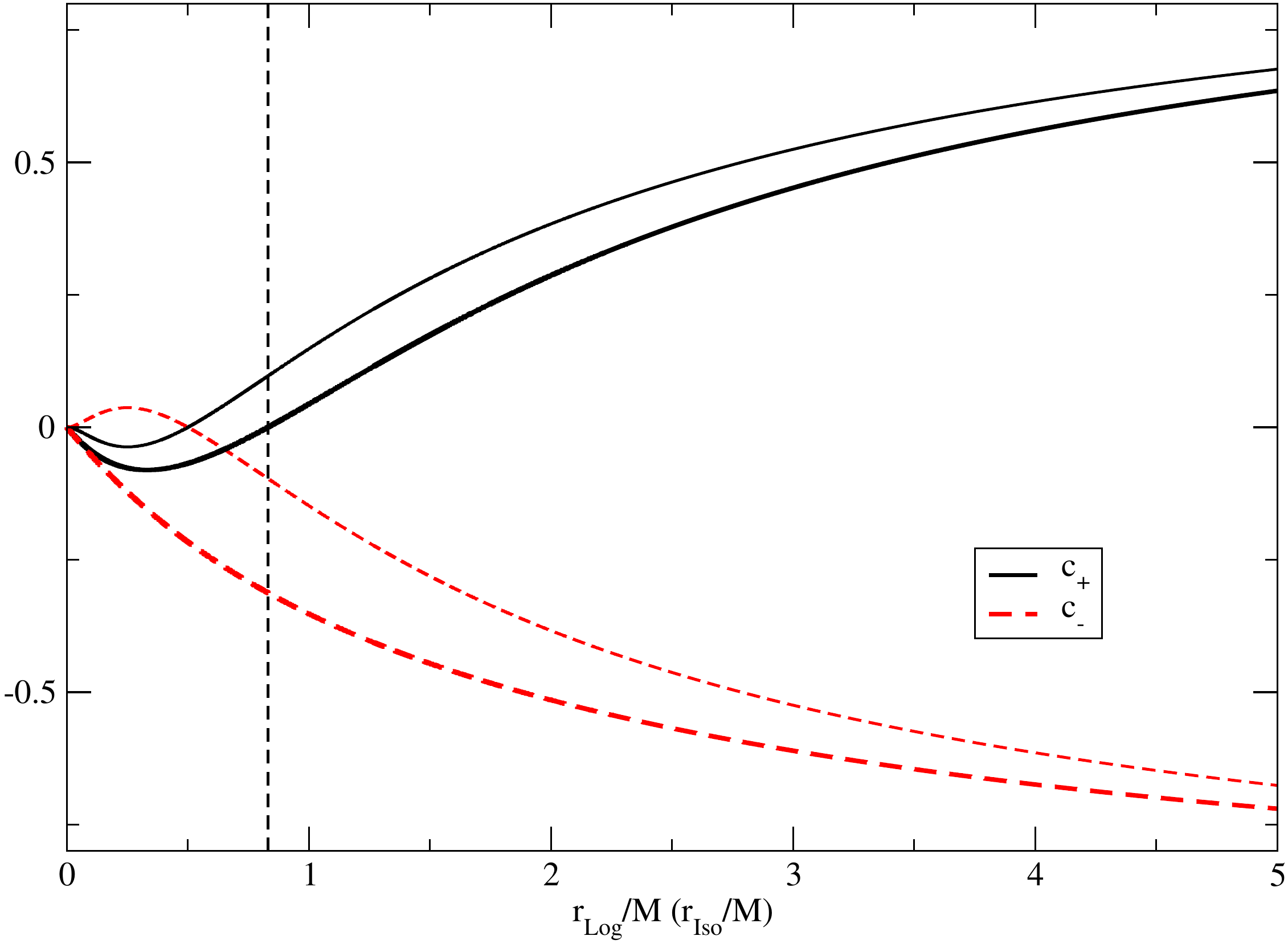}
\label{fig:characteristic}
\end{center}
\end{figure}

In order to understand one of the main features of the wave propagation, 
we also calculate the ingoing and outgoing characteristic speeds 
for this system~\cite{Zenginoglu:2009ey}. 
Using our definition of the metric, these are given by 
\bea
c_+ &=& f(r)\,\left(\alpha(r)-\beta(r)\right) \,,
\nonumber \\ 
c_- &=& -f(r)\,\left(\alpha(r)+\beta(r)\right) \,.
\eea 
Figure~\ref{fig:characteristic} shows
the radial dependence in the trumpet coordinates. Note that inside the horizon,
($r_{\rm Log}=0.8304M$), both speeds remain negative and that the outgoing
characteristic speed vanishes precisely at the horizon. Both modes reach
their asymptotic speeds $(\pm1)$, at much larger radii.

It is interesting to observe the differences between the
usual isotropic coordinates in the standard Schwarzschild slicing and
the ``trumpet'' coordinates. The initial data are in isotropic
coordinates, therefore these differences are indicative of how
the background has to change from the initial slice
to the eventual ``trumpet'' slice.

\bibliographystyle{apsrev}
\bibliography{../../../Bibtex/references}

\end{document}